\documentclass[letterpaper,twocolumn,prl,aps,superscriptaddress,amsmath,amssymb,floatfix]{revtex4-2}
\usepackage{mathptmx}
\DeclareMathAlphabet{\mathcal}{OMS}{cmsy}{m}{n}

\usepackage[latin9]{inputenc}
\usepackage{color}
\usepackage{amsmath}
\usepackage{amssymb}
\usepackage{graphicx}
\usepackage{esint}
\usepackage{makecell}
\usepackage[unicode=true,
bookmarks=true,bookmarksnumbered=false,bookmarksopen=false,
breaklinks=false,pdfborder={0 0 1},backref=false,colorlinks=true]
{hyperref}
\hypersetup{
linkcolor=magenta,urlcolor=blue,citecolor=blue,pdfstartview={FitH},hyperfootnotes=false}

\usepackage{braket}
\usepackage{physics}

\makeatletter

\usepackage{textcomp}
\usepackage{epstopdf}

\pdfpageheight\paperheight
\pdfpagewidth\paperwidth

\@ifundefined{textcolor}{}{%
\definecolor{BLACK}{gray}{0}
\definecolor{WHITE}{gray}{1}
\definecolor{RED}{rgb}{1,0,0}
\definecolor{GREEN}{rgb}{0,1,0}
\definecolor{BLUE}{rgb}{0,0,1}
\definecolor{CYAN}{cmyk}{1,0,0,0}
\definecolor{MAGENTA}{cmyk}{0,1,0,0}
\definecolor{YELLOW}{cmyk}{0,0,1,0}
}

\usepackage{xcolor}\usepackage{soul}
\definecolor{blue}{rgb}{0,0,1}
\definecolor{red}{rgb}{1,0,0}
\definecolor{green}{rgb}{0,1,0}

\usepackage{soul}

\makeatother

\begin{document}

\title{Engineering the Nonlinearity of Bosonic Modes with a Multi-loop SQUID}

\author{Ziyue Hua}
\affiliation{Center for Quantum Information, Institute for Interdisciplinary Information Sciences, Tsinghua University, Beijing 100084, China}

\author{Yifang Xu}
\affiliation{Center for Quantum Information, Institute for Interdisciplinary Information Sciences, Tsinghua University, Beijing 100084, China}

\author{Weiting Wang}
\affiliation{Center for Quantum Information, Institute for Interdisciplinary Information Sciences, Tsinghua University, Beijing 100084, China}

\author{Yuwei Ma}
\affiliation{Center for Quantum Information, Institute for Interdisciplinary Information Sciences, Tsinghua University, Beijing 100084, China}

\author{Jie Zhou}
\affiliation{Center for Quantum Information, Institute for Interdisciplinary Information Sciences, Tsinghua University, Beijing 100084, China}

\author{Weizhou Cai}
\affiliation{CAS Key Laboratory of Quantum Information, University of Science and Technology of China, Hefei 230026, China}

\author{Hao Ai}
\affiliation{School of Integrated Circuits, Tsinghua University, Beijing 100084, China}

\author{Yu-xi Liu}
\affiliation{School of Integrated Circuits, Tsinghua University, Beijing 100084, China}

\author{Ming Li}
\email{lmwin@ustc.edu.cn}
\affiliation{CAS Key Laboratory of Quantum Information, University of Science and Technology of China, Hefei 230026, China}
\affiliation{Hefei National Laboratory, Hefei 230088, China}

\author{Chang-Ling Zou}
\email{clzou321@ustc.edu.cn}
\affiliation{CAS Key Laboratory of Quantum Information, University of Science and Technology of China, Hefei 230026, China}
\affiliation{Hefei National Laboratory, Hefei 230088, China}

\author{Luyan Sun}
\email{luyansun@tsinghua.edu.cn}
\affiliation{Center for Quantum Information, Institute for Interdisciplinary Information Sciences, Tsinghua University, Beijing 100084, China}
\affiliation{Hefei National Laboratory, Hefei 230088, China}


\begin{abstract}
Engineering high-order nonlinearities while suppressing lower-order terms is crucial for quantum error correction and state control in bosonic systems, yet it remains an outstanding challenge. Here, we introduce a general framework of Nonlinearity-Engineered Multi-loop SQUID (NEMS) device, enabling the realization of arbitrary nonlinearities by tuning fluxes in multiple loops within superconducting circuits. We demonstrate specific examples of NEMS devices that selectively engineer pure cubic, quartic, and quintic interactions with suppressed parasitic couplings, showing great promise for realizing Kerr-cat bias-preserving {\scshape cnot} gates and stabilizing four-leg cat qubits. By opening new avenues for tailoring nonlinear Hamiltonians of superconducting devices, this work enables sophisticated and precise manipulation of bosonic modes, with potential applications in quantum computation, simulation, and sensing. 
    
\end{abstract}

\maketitle

\section{Introduction}
Bosonic modes, such as those of superconducting resonators and mechanical oscillators, have emerged as a promising platform for quantum information processing~\cite{Joshi2021bosonicqubits,Cai2021Bosonic,Copetudo2024ReviewBosonicQEC}. 
In particular, bosonic modes offer unique advantages for quantum error correction (QEC), a crucial ingredient for fault-tolerant quantum computation~\cite{Devoret2013SQEC}. 
By encoding quantum information in the large Hilbert space of an oscillator, bosonic QEC codes can achieve high performance with reduced hardware overhead compared to conventional qubit-based codes, {providing improvement in quantum computing~\cite{Grimsmo2020PRXCatandBinomial,Gouzien2023Performance,Liu2024HybridOscillatorQubit,Crane2024HybridOscillatorQubit}, simulation~\cite{Flurin2017Observing,Hu2018Simulation,Wang2020Efficient}, and sensing~\cite{Wang2019Heisenberglimited,Deng2024metrology100Fock}}. 
Prominent bosonic coding approaches, including the binomial codes~\cite{Michael2016NewClass}, {four-component cat (4-cat) codes}~\cite{Leghtas2013Hardware-Efficient}, and GKP codes~\cite{GKP2001Encoding}, have successfully enhanced the lifetime of quantum information~\cite{Hu2019binomial,Ma2020PASS,Gertler2021BosonicAQEC,Reinhold2020ErrorcorrectedGate} or even surpassed the break-even point~\cite{Ofek2016Extending,Ni2023Beating,Sivak2023Realtime}. While lower-order bosonic codes like the 2-cat codes~\cite{Mirrahimi2014Dynamically} and dual-rail codes~\cite{Teoh2023Dualrail} may not provide enough space for quantum error correction, they can serve as possible physical qubits for cascaded coding strategies. 

Specifically, the stabilized 2-cat qubit exhibits biased-noise characteristics, where the bit-flip errors are exponentially suppressed with an increase in the average photon number~\cite{Grimm2020Stabilization,Lescanne2020Exponential,Berdou2023OneHundred,Reglade2024NatureDCat10s}. Moreover, by leveraging the continuous nature of the phase space, it is possible to construct {\scshape x}, {\scshape cnot}, and Toffoli gates that maintain the biased-noise property~\cite{Guillaud2019Repetition,Puri2020Bias-preserving}. The existence of bias-preserving gates reduces the resource overhead for concatenating 2-cat qubits with binary QEC codes, such as the repetition-cat codes~\cite{Guillaud2019Repetition,Guillaud2021Error,Gouzien2023Performance,Putterman2024ConcatenatedCat} and surface-cat codes~\cite{Darmawan2021Practical,Chamberland2022Building}. The noise-biased gate operation gives the 2-cat qubit a unique advantage in cascade encoding over other noise-biased two-level systems~\cite{Aliferis2008Fault-tolerent,Tuckett2018Ultrahigh,Bonilla2021XZZX}, offering an efficient method for implementing fault-tolerant quantum computation.

However, unleashing the full potential of bosonic QEC requires precise control over the nonlinear interactions within bosonic modes. 
{Although various bosonic operations can be realized with the help of an ancillary qubit~\cite{Blais2004CavityQED,Khaneja2005GEAPE,Eickbusch2022ECD}, specially designed nonlinear elements are still required to construct stable Hamiltonians and nonlinear processes~\cite{Ma2021QuantumControl,Chapman2023SNAILBeamSplitter,Yao2023Highfidelity}.}
For example, the stabilization and manipulation of cat states typically necessitate the integration of auxiliary nonlinear components. 
Transmons~\cite{Koch2007Transmon} possess a static four-wave mixing processes and have been used in the stabilization of 2-cat states~\cite{Leghtas2015Confining,Touzard2018Coherent}. 
SQUID-based nonlinear elements, such as SQUID arrays or SQUID resonators, exhibit a reduced static nonlinearity but generate even-order nonlinear drives for three-wave mixing. Recently, these SQUID-based elements have been used in investigating the nonlinear dynamics of two-photon driven Kerr parametric oscillators (KPOs)~\cite{Wang2019Quantum,Andersen2020Quantum,Yamaji2022Spectroscopic,Yamaji2023Correlated,Iyama2024Observation,Hoshi2024EntanglingCat}. 
The superconducting nonlinear asymmetric inductive element (SNAIL)~\cite{Frattini2017SNAIL} offers a static 3rd-order nonlinearity while effectively suppressing the 4th-order nonlinearity by fine-tuning the operational points. This feature has made SNAIL a suitable platform for implementing single Kerr-cat qubits~\cite{Grimm2020Stabilization,Frattini2024Observation,Hajr2024TaKerrCat,Venkatraman2024DetunedKerrCat}. 
The asymmetrically threaded SQUID (ATS)~\cite{Lescanne2020Exponential} exhibits minimal static nonlinearity and generates strong odd-order nonlinear interactions under specific magnetic driving conditions. This property has positioned ATS as a controller for dissipative stabilization of cat qubits~\cite{Lescanne2020Exponential,Berdou2023OneHundred,Reglade2024NatureDCat10s}. 
The evolution of nonlinear devices within the cat-code experimental frameworks is towards reducing static nonlinearity alongside enhancing the strength and selectivity of the driven nonlinearities. Maximizing the desired driven nonlinearity facilitates rapid and complex operations, whereas minimizing undesired static or driven nonlinearity ensures operation fidelity and system coherence for further applications.

In this paper, we propose the Nonlinearity-engineered Multi-loop SQUID (NEMS), a class of novel superconducting nonlinear elements, for manipulating quantum states of bosonic modes. We construct an inductive element consisting of an inductively shunted multi-loop SQUID, then show that the static part and driven part of the potential can be engineered separately. For the driven part, we discuss different cases for odd-order and even-order terms and present the {\textit{symmetric-double-branch}} method for engineering the even-order terms. We demonstrate three NEMS devices as examples: NEMS-3 generates pure $ g_3^\text{driven} $, NEMS-4 generates pure $ g_4^\text{driven} $, and NEMS-5 generates pure $ g_5^\text{driven} $. Furthermore, We discuss the potential applications for these NEMS devices, such as implementing Kerr-cat bias-preserving {\scshape cnot} (BPCNOT) gates and stabilizing 4-cat states. Our designs underscore the potential of engineering high-order interactions of bosonic modes using Josephson junctions (JJs), which could significantly advance cat qubits and bosonic quantum error correction by exploring qudit codes and high-order codes with longer distances.

The remainder of this paper is organized as follows. In Sec. II, we introduce the general structure and working principles of the NEMS devices. Sections III and IV are dedicated to the design methodology for engineering odd-order and even-order nonlinearities, respectively, along with concrete examples of NEMS-3, NEMS-4, and NEMS-5. In Secs. V, we discuss the application of NEMS devices in realizing BPCNOT gates for Kerr-cat qubits and stabilizing 4-cat codes. Finally, we conclude in Sec. VI with an outlook on future directions and the potential impact of NEMS on bosonic quantum information processing.

\section{Principle of NEMS}

\begin{figure}
\includegraphics{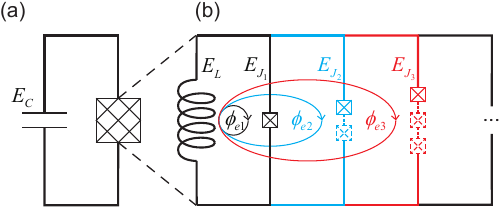}
\caption{Nonlinearity-Engineered Multi-loop SQUID (NEMS) circuit structure. (a) An anharmonic oscillator with designed nonlinearity is realized by connecting a NEMS with a large shunting capacitor with a charging energy $ E_\text{C} $ in parallel. The NEMS serves as a weakly nonlinear inductive element. (b) Circuit diagram of the NEMS, which consists of several parallel Josephson junction (JJ) branches forming multiple loops. The left-most branch, referred to as the inductor branch, contains several large JJs that provide a binding potential and act as a linear inductance. The branches on the right, referred to as Josephson branches, provide nonlinear driven Hamiltonians when subjected to an alternating magnetic field.}
\label{Fig1}
\end{figure}

Our target is to design a weakly nonlinear resonator that possesses controllable nonlinearity while also generating strong and tailorable nonlinear driving terms. Since most capacitors in superconducting systems are linear, we construct an LC oscillator consisting of a linear capacitor and a NEMS that serves as a nonlinear inductor, as schematically shown in Fig.~\ref{Fig1}(a). The key idea behind the NEMS design is to leverage a multi-loop SQUID design with multiple JJ branches, where the JJs provide the essential nonlinearity and the loops offer the geometry parameters for high degree of design flexibility, inspired by the architecture of ATS~\cite{Lescanne2020Exponential,Berdou2023OneHundred,Reglade2024NatureDCat10s}.
As depicted in Fig.~\ref{Fig1}(b), the NEMS comprises multiple JJ branches arranged in parallel, forming a multi-loop structure. As will shown below, by carefully engineering the sizes and numbers of JJs in each branch, as well as the magnetic flux threading each loop, we can precisely control the static and driven nonlinearities of the device. 

In our model, we neglect the capacitance within the JJs, {because the shunting capacitance is much larger than the JJ capacitance. Thus the NEMS reduces to a weakly nonlinear inductor, and the low-frequency oscillating mode is equivalent to an LC oscillator~\cite{Deppe2004JJCapacitances,Yao2023Highfidelity}.} 
The leftmost branch, termed the inductor branch, contains multiple large JJs that collectively provide a binding potential resembling that of a linear inductor. {This branch ensures {a weakly anharmonic oscillator (WAO)} behavior of the NEMS for a single-well potential with other branches providing perturbative anharmonicity.} The branches on the right, referred to as the Josephson branches, contain smaller JJs and are responsible for introducing nonlinear interactions to the device. By applying an alternating magnetic field to NEMS, we can selectively stimulate and control the strength of these nonlinear terms. 

\subsection{Hamiltonian representation of NEMS}

The Lagrangian describing the circuit schematically shown in  Fig.~\ref{Fig1} is given by
\begin{equation}\label{equ: L_total of NEMS}
    \begin{split}
        \mathcal{L}_\text{NEMS} & = \frac{1}{2}C_\text{S} \dot{\varphi}^2 - U_\text{NEMS}(\varphi, \vec{\phi}_{e}). \\ 
    \end{split}
\end{equation}
Here, $\varphi$ represents the {generalized variable of the phase difference across the leftmost inductor branch}, { $ C_\text{S} $ represents the capacitance of the large linear capacitor with a corresponding charging energy $ E_\text{C} \equiv e^2 / 2 C_\text{S} $,} and $ U_\text{NEMS}(\varphi, \vec{\phi}_{e}) $ represents the inductive potential energy of the NEMS, with the vector  $ \vec{\phi}_{e}=\{\phi_{e1},\,\phi_{e2},\,\phi_{e2},\,...\}$ denoting a set of control parameters due to external bias flux and drive acting on individual loops. 
For the inductor branch contains $ n_\text{L}$ large JJs with equal junction size $ E_{\text{J}_\text{L}} $, the inductor branch forms as an effective large super-inductor {$ E_\text{L} \equiv E_{\text{J}_\text{L}} / n_\text{L} $}. For the Josephson branches, each branch comprises several small JJs characterized by varying junction sizes defined as $ E_{\text{J}_i}=r_i E_\text{L}$ {(with $ r_i$ standing for the JJ energy ratio)} and the number of junctions $ n_{i} $. Then, the total inductive energy of NEMS can be derived as 
\begin{equation}\label{equ: U_tot of NEMS general}
    \begin{split}
        U_\text{tot}\left(\varphi\right) &= U_\text{NEMS}(\varphi, \vec{\phi}_{e}) \\
        &= E_\text{L} \frac{\varphi^2}{2}  - \sum_i  n_{i} E_{\text{J}_i} \cos\left( \frac{ \varphi + \phi_{ei} }{ n_{i} } \right).
    \end{split}
\end{equation}
By treating both the nonlinear terms and  the external drives as perturbations, the anharmonic LC oscillator can be equivalently described by the Hamiltonian of a quantized harmonic oscillator as
\begin{equation}\label{equ: H_lin of NEMS}
        \hat{\mathcal{H}}_\text{lin} = 4 E_\text{C} \hat{n}^2 +  E_\text{L} c_2^\text{static} \frac{\hat{\varphi}^2}{2}.
\end{equation}
Here, $\hat{n}$ and $\hat{\varphi}$ are the conjugate operators of {charge number and phase variables}, respectively, and $c_2^\text{static}$ is a coefficient for the harmonic energy and will be explained later. By representing the flux and charge fluctuation of the oscillator with bosonic ladder operators ($a$ and $a^\dagger$)
\begin{eqnarray}\label{equ: ZPF of NEMS}
\hat{\varphi}& \rightarrow \varphi_{\mathrm{zpf}}\left( \hat{a}^\dagger + \hat{a} \right),\\
\hat{n}& \rightarrow \mathrm{i} n_{\mathrm{zpf}} \left( \hat{a}^\dagger - \hat{a} \right),
\end{eqnarray}
we have 
\begin{equation}\label{equ: H_lin of NEMS oscillator}
\hat{\mathcal{H}}_\text{lin} = \omega_\text{static} \left( \hat{a}^\dagger \hat{a} + \frac{1}{2} \right),
\end{equation}
where the bosonic mode frequency is $\omega_\text{static}\equiv\sqrt{8 c_{2}^\text{static} E_\text{J} E_\text{C}}$, and $ \varphi_{\mathrm{zpf}} = \left({2E_\text{C}}/{c_{2}^\text{static} E_\text{J}}\right)^{{1}/{4}}$ and $n_{\mathrm{zpf}} = \frac{1}{2} \left({c_{2}^\text{static} E_\text{J}}/{2E_\text{C}}\right)^{{1}/{4}}$ represent the characteristic magnitudes of the zero-point fluctuations.

\subsection{Static and driven nonlinearities}

In the following, we discuss the anharmonicity of the system and the external drives. {Under a time-dependent magnetic driving field, }the potential in Eq.~\eqref{equ: U_tot of NEMS general} can be rewritten as 
\begin{equation}\label{equ: U_tot t of NEMS general}
        U_\text{tot}(\varphi,t) = U_\text{static}(\varphi) + U_\text{driven}(\varphi,t),
\end{equation} 
where the first term arises from the contributions of intrinsic nonlinear interactions and DC bias fields $\overline{\phi}_{ei}$ on each loop, while the second term results from the dynamic modulation of the system under time-dependent AC drive fields $\delta \phi_{ei}(t)=\phi_{ei}-\overline{\phi}_{ei}$.

The static part of the potential can be expanded by the contributions of individual loops as
\begin{equation}\label{equ: U_static of NEMS general}
        \frac{U_\text{static}\left(\varphi\right)}{E_\text{L}} = \frac{\varphi^2}{2} - \sum_{i} n_i r_i \cos\left( \frac{ \varphi + \overline{\phi_{ei}} }{ n_{i} }  \right). 
\end{equation}
{Generally, the static potential function has one minimum position $\varphi^*$ if the system behaves as a WAO.}
Considering the Taylor expansion of the potential energy with respect to the harmonic oscillator, i.e., { $ U_\text{static}(\varphi) / E_\text{L}  = c_0^\text{static} + \sum_{n\geq2} \frac{1}{n!} c_{n}^\text{static} \left(\varphi - \varphi^* \right)^n$ } with $c_{n}^{\mathrm{static}}$ being coefficients for the $n$-th order terms, we obtain the Hamiltonian of the nonlinear term as 
\begin{equation}\label{equ: H_static of NEMS}
        \hat{\mathcal{H}}_\text{nonlin} = \sum_{n\geq3} g_n^\text{static} \left(\hat{a}^\dagger + \hat{a} \right)^n,
\end{equation}
with nonlinear interaction coefficients $g_n^\text{static}= \frac{1}{n!} E_\text{J} c_{n}^\text{static} \varphi_{\mathrm{zpf}}^n$. Note that the nonlinear terms $ g_n \left(\hat{a}^\dagger + \hat{a} \right)^n $ may also contribute to the frequency shift of the oscillator~\cite{Hillmann2022Designing}, therefore, $ \omega_\text{static}$ in Eq.~\eqref{equ: H_lin of NEMS oscillator} denotes the ``bare" or ``plasma" frequency when no nonlinear terms~($n>2$) are included, instead of the experimentally detectable frequency.

For the dynamic part of the potential due to the drive, it can be approximated as  
\begin{equation}\label{equ: U_driven of NEMS general}
\frac{U_\text{driven}(\varphi,t)}{E_\text{L}} = \sum_{i} r_i \sin\left( \frac{\varphi + \overline{\phi_{ei}}}{n_{i}} \right) \delta \phi_{ei}(t),
\end{equation}
by treating the time-dependent modulations as a first-order perturbation on $U_\text{static}$ [Eq.~(\ref{equ: U_static of NEMS general})]. The effects due to high-order terms of flux modulations are discussed in Appendix Sect.~\ref{subsec: Magnetic driven Hamiltonian in NEMS}. Assuming a driving field with an amplitude of $ \epsilon(t)$ is applied to the NEMS, it can generate a magnetic flux threading the entire device. Consequently, the flux variations in each loop are synchronized and can be expressed as 
\begin{equation}\label{equ: r_phi_ei relative magnetic drive amplitude }
\delta \phi_{ei}(t) = r_{\phi_{ei}}\epsilon(t),
\end{equation}
with parameters $ r_{\phi_{ei}} $ satisfying the normalization condition $ \sum_{i} \abs{ r_{\phi_{ei}} } = 1 $. The corresponding driving Hamiltonian can be written as:
\begin{equation}\label{equ: H_driven of NEMS}
            \hat{\mathcal{H}}_\text{driven} = \sum_{n\geq1} g_n^\text{driven} \epsilon(t) \left(\hat{a}^\dagger + \hat{a} \right)^n,
\end{equation}
where the driven term coefficients are $ g_n^\text{driven}= \frac{1}{n!} E_\text{L}  c_{n}^\text{driven} \varphi_{\mathrm{zpf}}^n$ with $c_{n}^\text{driven}$ being determined by the Taylor expansion { $U_\text{driven}( \varphi, t )/E_\text{L} = \epsilon(t) \sum_{n} \frac{1}{n!} c_{n}^\text{driven} \left(\varphi-\varphi^*\right)^n$ }.

By combining all the Hamiltonians, the system dynamics is governed by the total Hamiltonian
\begin{equation}\label{totalHamiltonian}
\hat{\mathcal{H}} =  \hat{\mathcal{H}}_\text{lin}+ \hat{\mathcal{H}}_\text{nonlin}+\hat{\mathcal{H}}_\text{driven}.
\end{equation}
Consequently, the corresponding nonlinear interaction strengths at a given order $n$ and the nonlinear driven term at a given order $m$ are controllable through tuning the parameters $c_{n}^\text{static}$ and $c_{n}^\text{driven}$, which are determined by the NEMS geometry design parameters $r_i$, $n_{i}$, $\overline{\phi_{ei}}$, and $r_{\phi_{ei}}$.

\subsection{Weak anharmonicity approximation}

For a simplified representation of the system by Eq.~\eqref{totalHamiltonian}, it is essential to ensure that the perturbation approximation for the nonlinearity and the drive remains valid in our device. This leads to a general requirement for designing the NEMS devices: the potential should exhibit a deep single-well profile, ensuring that the system's low-energy behavior closely resembles that of a harmonic oscillator. Such a requirement limits the choice of parameters for the junction size $ E_{\text{J}_i} $ and the external flux $ \phi_{ei}$. Deriving general parameter limitations for complex multi-loop and multi-JJ structures is nontrivial, therefore we provide two general suggestions a priori: First, the Josephson energy of each small JJ should be smaller than the inductive energy, i.e., $ E_{\text{J}_i} < E_\text{L} $, to avoid the formation of a double-well potential. Second, the external flux associated with a multi-JJ branch should be kept close to zero, i.e., $ \abs{\phi_{ei}} < \pi/2 $ for $ n_{i} >1$, to prevent unwanted phase slips. Further details and examples of NEMS parameter selection can be found in Appendix Sect.~\ref{sec: limitation on flux bias}. 

\subsection{Engineering universality}

To fully exploit NEMS devices for practical applications, we must carefully engineer the static and driven nonlinear terms in the Hamiltonian shown in Eq.~\eqref{totalHamiltonian}. {Specifically, the nonlinear device needs to satisfy}
two key requirements: First, in the absence of time-dependent drive fields, i.e., $\vec{\phi}_{e} =  \overline{ \vec{\phi}_{e} }$ or $\epsilon(t)=0$, the potential should closely approximate a harmonic potential with tunable anharmonicity. Second, under dynamic drive fields, the system should generate the desired driven potential for quantum control. Mathematically, these requirements translate to the independent adjusting of the static and driven nonlinearity coefficients, $ g_n^\text{static} $ and $ g_m^\text{driven}$, for arbitrary integers $n$ and $m$, by tuning the design parameters $ r_i $, $ n_{i} $, $ \overline{\phi_{ei}} $, and $ r_{\phi_{ei}} $. 

\begin{figure}
\includegraphics{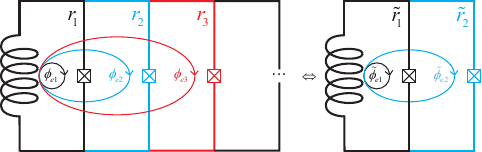}
\caption{Circuit diagram of NEMS device consisting of several single-JJ branches in parallel with an inductor branch. Similar to ATS, the driven potential under a general magnetic drive takes the form of $ \cos(\varphi) $ or $ \sin(\varphi) $,  implying that extra single-JJ branches do not enhance the capability in  Hamiltonian engineering. To independently engineer the coefficients $ c_{n}^\text{driven}$ of each order individually, deploying Josephson branches with multiple JJs is necessary.}
\label{fig: Only_Single_JJ_v1}
\end{figure}

Before the demonstration of arbitrary adjustments of the coefficients by the general framework of NEMS, we first prove the necessity to employ Josephson branches containing multiple junctions. Consider a simple scenario, as depicted in Fig.~\ref{fig: Only_Single_JJ_v1}, where the system comprises many Josephson branches, each containing only a single JJ. Under a general magnetic drive, the driven potential can be expressed as:
\begin{equation}\label{equ: U driven of single junction part}
    \begin{split}
        U_\text{driven}^{n_{i}=1}/E_\text{L} &= \sum_i r_i \sin\left(\varphi + \overline{\phi _{ei}} \right) \delta \phi_{ei} \\
        &= \cos\left(\varphi \right) \epsilon(t) \sum_i r_i \sin \left( \overline{\phi _{ei}} \right) r_{\phi_{ei}}   \\
        &\quad+ \sin\left(\varphi \right) \epsilon(t) \sum_i r_i \cos \left( \overline{\phi_{ei}} \right) r_{\phi_{ei}}. 
    \end{split}
\end{equation}
These driven potentials take the form of $ \cos(\varphi)\epsilon(t) $ or $ \sin(\varphi)\epsilon(t) $, with relative coefficients determined by the parameters $ r_i $, $ \overline{\phi_{ei}} $, and $ r_{\phi_{ei}} $. Through parameter engineering, we can selectively generate odd or even driving terms, similar to the results achieved in ATS. However, the ratios between driving terms of the same parity remain fixed, implying that the addition of {single-JJ} branches does not enhance our Hamiltonian engineering capabilities.

Similarly, for a system containing only {single-JJ} branches, the static potential can be explicitly written as:
\begin{equation}\label{equ: U static of single junction part}
    \begin{split}
        U_\text{static}^{n_{i}=1}/E_\text{L} &= \varphi^2/2  - \sum_i r_i \cos\left(\varphi + \overline{\phi _{ei}} \right) \\
        &= \varphi^2/2 - \cos\left(\varphi \right) \sum_i r_i \cos \left( \overline{\phi _{ei}} \right) \\
        &\quad + \sin\left(\varphi \right) \sum_i r_i \sin \left( \overline{\phi_{ei}} \right).
    \end{split}
\end{equation}
These static potentials take the form of $ \cos(\varphi) $ or $ \sin(\varphi) $ with relative coefficients depending on the parameters $r_i $ and $ \overline{\phi_{ei}} $. To completely cancel the static nonlinearity at a selected order, at least two {single-JJ} branches are required. While we can control the relative strength between odd and even static terms, the ratios between static terms of the same parity remain fixed. We also note that this conclusion also holds for multi-JJ branches if the numbers of identical JJs are all the same in different branches. 

Therefore, to realize the universal control of both the static and driven nonlinearity of a single bosonic mode, it is necessary to introduce multiple branches consisting of different numbers of junctions. Given a NEMS system with $N$ Josephson branches, each containing $n_{i}$ junctions $(i = 1, 2, ..., N)$, we have $4N$ individual control parameters: $\{   r_i,n_{i}, \overline{\phi_{ei}}, r_{\phi_{ei}} \}$. According to Eq.~\eqref{equ: U_driven of NEMS general}, the coefficients $g_{n}^{\text{driven}}$ includes contributions from all branches with the contribution in each branch proportional to $\frac{1}{n!}r_{i}f_{i}$, where $f_{i}$ takes $\mathrm{cosine}$ or $\mathrm{sine}$ function depending on the parity of the nonlinearity order $n$. For odd-order driven nonlinearity, we have
\begin{equation}
\overrightarrow{g}_{\text{driven}}	=	\mathbf{A}\cdot\overrightarrow{x},
\label{eq:universal}
\end{equation}
where$\overrightarrow{g}_{\mathrm{driven}}=\{g_{1},g_{3},g_{5},...\}$ is the nonlinearity coefficient vector, and $\overrightarrow{x}=\{r_{1}f_{1},r_{3}f_{3},r_{5}f_{5},...\}$ is the control parameter vector. The range of values for $x_{i}$ is free with its amplitude proportional to $r_{i}$ and its sign determined by the flux $\overline{\phi_{ei}}$. Proper choice of $\overline{\phi_{ei}}$ ensures a single-well potential profile, which will be further discussed in the following sections. The matrix $\mathbf{A}$ is a Vandermonde matrix
\begin{equation}
\mathbf{A}	=	\left(\begin{array}{cccc}
1 & 1 & 1 & ...\\
\frac{1}{n_{1}^{2}} & \frac{1}{n_{2}^{2}} & \frac{1}{n_{3}^{2}}\\
\frac{1}{n_{1}^{4}} & \frac{1}{n_{2}^{4}} & \frac{1}{n_{3}^{4}}\\
...
\end{array}\right)
\label{Bmatirx}
\end{equation}
with determinant $\mathrm{Det}[\mathbf{A}]=\Pi (\frac{1}{n_{j}^{2}}-\frac{1}{n_{i}^{2}})$. Under WAO approximation applied earlier, the coefficients for the orders larger than the desired nonlinearity order $m$ are suppressed due to the decays of coefficients in the Taylor series. Therefore, the dimension of $\overrightarrow{g}_{\mathrm{driven}}$ and $\mathbf{A}$ can be truncated to a certain order by neglecting higher order effects. An arbitrary driven nonlinearity, with the strengths represented by $\overrightarrow{g}_{\mathrm{driven}}$, can be realized for $\mathrm{Det}[\mathbf{A}]\neq 0$ under the condition that $n_{i}\neq n_{j}$ for all branches ($i$ and $j$). It is important to note that $\mathbf{A}$ may not be a square matrix if the number of branches is larger than the length of $\overrightarrow{g}_{\mathrm{driven}}$. In this case, the universality is allowed when the number of branches with different $n_{i}$ values is greater than the length of $\overrightarrow{g}_{\mathrm{driven}}$.

The above analysis can be separately applied to engineer even-order driven nonlinearity and even-order static nonlinearity. For odd-order static nonlinearity, it actually plays the role of odd-order driving terms with driving frequencies being integer multiples of the mode frequency, {and thus can be substituted by a well-designed odd-order driving term with proper frequency. On the other hand, odd-order static nonlinearities are often used together with electrical drives to achieve lower-order nonlinear processes, which can be directly achieved by lower-order magnetic driving nonlinearities in NEMS devices.}
Therefore, by paralleling circuits of different functions, we can in principle achieve the universal control of the Hamiltonian in a single superconducting LC oscillator or bosonic mode.

In the following sections, we will provide examples demonstrating the design of specialized NEMS device for selected orders of nonlinear driven terms, i.e, pure $m$-th order drive nonlinearity. Specifically, we will present three examples for engineering $ g_3^\text{driven} $, $ g_4^\text{driven} $, and $ g_5^\text{driven} $. The suppression of lower-order terms is achieved through the coherent cancellation of the contributions from multiple Josephson loops under magnetic driving. This unified framework for designing nonlinearity in NEMS can also be generalized to create specialized devices for given distributions of nonlinearity coefficients at various orders.

\subsection{Advantage of Magnetic drive}

The previous analysis focuses solely on the magnetic drive, i.e., evaluating the potential change under a small alternating magnetic field. In contrast, the electric drive applying to the capacitor $C_\text{S}$ is intrinsically linear but can induce nonlinear effects when there are non-zero static nonlinearities $g_n^\text{static}$. Unlike magnetic drive, which directly activates nonlinear operators, electric drive on a WAO is essentially a linear drive that leads to a displacement operation $\hat{a} \rightarrow \hat{a} + \xi(t)$ on the cavity field. Nonlinear phenomenons appear when the displacement is dressed with static nonlinearities~\cite{Blais2021Circuit}: 
\begin{equation}\label{equ: H_driven of electric drive}
        \hat{\mathcal{H}}_\text{ele} = \sum_{n\geq3} g_n^\text{static} \left( \hat{a} + \xi(t) + h.c. \right)^n.\\
\end{equation}
Consequently, the nonlinear effect of electric drive is limited by the static nonlinearity. One requires a large static nonlinear coefficient ($g_n^\text{static}$) to generate strong nonlinear effect. However, this requirement conflicts with our design target of independently controlling the static and driven nonlinearities. Therefore, magnetic drive is preferred in nonlinearity engineering. Despite the difference between the electric and magnetic drive mechanisms, it is possible to combine them and realize more powerful control. For example, the linear electric drive can be employed to compensate for the first-order terms in the magnetic-driven Hamiltonian.

\section{Engineering Odd-Order Driven Nonlinearities}

In this section, we discuss the procedure of engineering a device that only generates odd order $ g_\text{driven}^{n} $, with small static nonlinearities $ g_\text{static}^m = 0 $ for $ m < n $. {There are a variety of designs to meet these requirements following the analysis in Eqs.~\eqref{eq:universal} and \eqref{Bmatirx}. Here, we give the simplest design with a minimal number of branches.}  It is obvious that the odd-order driving terms can be generated simply by sine-like driven potential, which can be obtained by setting $ \phi_{ei} $ to 0 or $ n_{i}\pi $. Moreover, to ensure that $ U_\text{tot} $ has a single well, the external flux bias of multi-JJ branches is set to 0 to avoid phase slips, see Appendix Sect.~\ref{sec: limitation on flux bias}.

According to Eqs.~(\ref{equ: U_static of NEMS general}) and (\ref{equ: U_driven of NEMS general}), the sine-like driving potential requires the static potential to be:
\begin{equation}\label{equ: U_static odd order engineering}
\begin{split}
    \frac{U_\text{static}}{E_\text{L}} = & \frac{\varphi^2}{2} - \sum_{i} s_i n_{i} r_i \cos\left( \frac{\varphi}{n_{i}} \right),
\end{split}
\end{equation}
with $ s_i $ being a sign variable that is assigned to $ \pm 1 $ according to the following rule:
\begin{equation}\label{equ: rule for sign(i)}
    \begin{split}
        s_i &= +1 \text{ if } n_{i} \neq 1 \text{ or } n_{i} = 1~\, \mathrm{and} \,~\phi_{ei} = 0, \\
        s_i &= -1 \text{ if } n_{i} = 1~\,\mathrm{and}\,~ \phi_{ei} = \pi.
    \end{split}
\end{equation}
It is shown that the static potential contains only even-order terms, {such that $\varphi^*=0$.} As long as the static potential remains a single well, the minimum point $ \varphi^* $ will always be 0. Then, the corresponding $ c_{n}^\text{static} $ can be derived explicitly as	
{
\begin{equation}\label{equ: c_static odd order engineering}
    \begin{split}
        c_{n}^\text{static} = & \eval{\frac{\mathrm{d}^n}{\mathrm{d} \varphi^n } \frac{\varphi^2}{2} }_{\varphi = 0} \\
        - & \sum_{i} \frac{r_i }{n_{i}^{n-1}}  s_i \eval{\frac{\mathrm{d}^n}{\mathrm{d} \varphi^n } \cos( \varphi )}_{\varphi = 0}.
    \end{split}
\end{equation}
}
It is easy to see that $ c_{n}^\text{static} = 0 $ when $ n $ is odd, i.e., the potential has no odd-order static nonlinearity. 

Given that $ \varphi^*=0 $, we can simplify the driven potential to
\begin{equation}\label{equ: U_driven odd order engineering}
    \begin{split}
        \frac{U_\text{driven}}{E_\text{L} \epsilon(t)} = & \sum_{i} s_i r_i \sin( \frac{\varphi}{n_{i}} ) r_{\phi_{ei}},
    \end{split}
\end{equation}
with the corresponding coefficients
\begin{equation}\label{equ: c_driven odd order engineering}
    \begin{split}
        c_{n}^\text{driven} = & \sum_{i} \frac{r_i r_{\phi_{ei}} }{n_{i}^n}  s_i \eval{\frac{\mathrm{d}^n}{\mathrm{d} \varphi^n } \sin( \varphi )}_{\varphi = 0}.
    \end{split}
\end{equation}
The latter part $ s_i \eval{\frac{\mathrm{d}^n}{\mathrm{d} \varphi^n } \sin( \varphi )}_{\varphi = 0} $ is just a sign that can be predetermined once the numbers of JJs are fixed. To engineer the coefficient $ c_{n}^\text{driven} $, we only need to find a proper parameter set for the former part $ r_i r_{\phi_{ei}} / n_{i}^n $. We also notice that $ c_{n}^\text{driven} = 0 $ when $ n $ is even, i.e., the potential has no even-order driven nonlinearity.

\subsection{Pure $ g_{3}^\text{driven} $ nonlinearity }

\begin{figure}
\includegraphics{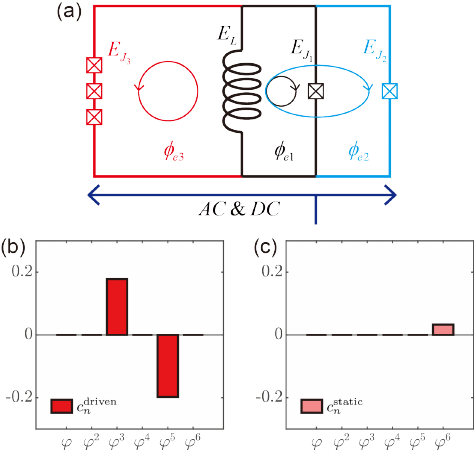}
\caption{(a) Circuit diagram of NEMS-3. NEMS-3 consists of two single-JJ branches and one triple-JJ branch,  producing $ \varphi^3 $ and higher odd-order driving terms under flux modulation. One possible construction of the magnetic driving line is depicted in blue, with the alternating flux amplitude in each loop labeled distinctly. (b) and (c) The coefficients $ c_{n}^\text{driven}$ of driven terms and $ c_{n}^\text{static}$ of static terms, respectively. 
}
\label{fig: NEMS-3 structure and nonlinearity}
\end{figure}

Here we present a simple example of odd-order engineering: the NEMS-3 device with only $ g_{3}^\text{driven} $ nonlinearity. For third-order driven term engineering, we aim for $ c_{1}^\text{driven} = 0 $ and $ c_{3}^\text{driven} \neq 0 $, while neglecting all $ n \geq 5 $ terms because the nonlinear interaction strength scales with order as $ \varphi_{\mathrm{zpf}}^n / n! $. This condition can be satisfied with {two single-JJ branches} and one multi-JJ branch, as shown in Fig.~\ref{fig: NEMS-3 structure and nonlinearity}. The NEMS-3 device is designed to generate only $ g_{3}^\text{driven} $ term under magnetic drive, with minimal static nonlinearity. According to Eqs.~(\ref{equ: c_static odd order engineering}) and (\ref{equ: c_driven odd order engineering}), the corresponding design conditions for a NEMS device with three Josephson branches are given as
\begin{equation}\label{equ: c_driven target of NEMS-3}
    \begin{split}
        c_{1}^\text{driven} = + s_1 r_1 r_{\phi_{e1}} + s_2\frac{r_2 r_{\phi_{e2}}}{n_{2}} + \frac{r_3 r_{\phi_{e3}}}{n_{3}} &= 0, \\
        c_{3}^\text{driven} = - s_1 r_1 r_{\phi_{e1}} - s_2\frac{r_2 r_{\phi_{e2}}}{n_{2}^3} - \frac{r_3 r_{\phi_{e3}}}{n_{3}^3} &\neq 0.
    \end{split}
\end{equation}
For the static nonlinearity, the lowest order coefficient that needs consideration is the $ c_{4}^\text{static} $, because the odd-order static nonlinearity is 0 and $ c_{2}^\text{static} $ is responsible for the harmonic potential. Given that there are plenty of free design parameters, we can further set the constrain
\begin{equation}\label{equ: c_static target of NEMS-3}
   c_4^\text{static} = - r_1 - s_2 \frac{r_2}{n_{2}^3} - \frac{r_3}{n_{3}^3} = 0. 
\end{equation}

Although we present the equations in the form of three branches, we note that the conditions can also be met with only two Josephson branches. For example, the $ r_2 $ branch can be eliminated and then $ r_1 $ and $ r_3 $ can be adjusted to meet the conditions. However, without $ r_2 $, the condition for canceling Kerr interaction may lead to an extreme JJ size ratio of $ r_1 = r_3/n_{3}^3 $, which is challenging for experimental realization. The single-JJ branch $ E_{J_1} $ on the right and the triple-JJ branch $ E_{J_3} $ on the left provide the desired third-order drive, with the second single-JJ branch $ E_{J_1} $ balancing the static nonlinearity and making no contribution to the driven potential. The JJ size requirements are listed below:
\begin{equation}\label{equ: rs of NEMS-3}
\begin{split}
r\equiv r_1 = r_3,&\quad r_2 = \frac{28}{27} r, \\
n_{1} = n_{2} = 1,&\quad n_{3} = 3.
\end{split}
\end{equation}
We design the device to operate at the following flux bias point, with the flux modulation in each loop satisfying a specific ratio:
\begin{equation}\label{equ: phi_e of NEMS-3}
\begin{split}
\overline{\phi_{e1}} = & 0, \quad \delta\phi_{e1} = 1/5 \cdot \epsilon(t), \\ 
\overline{\phi_{e2}} = & \pi, \quad \delta\phi_{e2} = 0, \\ 
\overline{\phi_{e3}} = & 0, \quad \delta\phi_{e3} = -3/5 \cdot \epsilon(t).
\end{split}
\end{equation}
Note that $ \delta\phi_{e2} = 0 $ means that the flux modulation does not influence the $ r_2 $ branch. Careful design of the AC line geometry is necessary to ensure precise distribution of flux fluctuations. {In Fig.~\ref{fig: NEMS-3 structure and nonlinearity}(a), we present one possible construction of the magnetic driving line (depicted in blue). The driving current is evenly divided into two parts, ensuring that the net flux in loop $ \phi_{e2} $ remains constant under the magnetic drive. Furthermore, we relocate the $ r_3 $ branch to the left of the inductor branch so the magnetic flux in the $ \phi_{e3} $ loop is opposite to that of the $ \phi_{e1} $ loop. The relocation does not change the potential function. }Two additional DC flux lines [not depicted in Fig.~\ref{fig: NEMS-3 structure and nonlinearity}(a)] are required to satisfy the static flux bias requirements.

For given $ r_i $ and $ n_{i} $ in a specific device, the total potential can be simplified as 
\begin{equation}\label{equ: U_tot of NEMS-3}
\begin{split}
  \frac{U_\text{tot}}{E_\text{L}} = &  \frac{\varphi^2}{2} - r \cos (\varphi + \phi_{e1} ) - \frac{28}{27} r \cos (\varphi + \phi_{e2} ) \\
        & - 3 r\cos\left( \frac{\varphi + \phi_{e3}}{3} \right).
\end{split}
\end{equation}
With this total potential, the system behavior at different flux bias points and the device spectrum can be thoroughly analyzed, as detailed in Appendix Sect.~\ref{subsec: NEMS-3 spectrum}. At the flux bias point required in Eq.~(\ref{equ: phi_e of NEMS-3}), the static potential energy is
\begin{equation}\label{equ: U_static of NEMS-3}
    \begin{split}
        \frac{U_\text{static}}{E_\text{L}} = & \frac{\varphi^2}{2} + \frac{ r }{ 27 }  \cos\left( \varphi \right) - 3 r \cos\left( \frac{\varphi}{3} \right) \\
        = & \frac{\varphi^2}{2} + \frac{8}{27}r\varphi^2 + O\left( \varphi^6 \right),  
    \end{split}
\end{equation}
and the driving potential is
\begin{equation}\label{equ: U_driven of NEMS-3}
\begin{split}
        \frac{U_\text{driven}}{\epsilon(t)E_\text{L}} = & + \frac{1}{5} r \sin(\varphi)  - \frac{3}{5} r\sin\left( \frac{\varphi }{3}\right)  \\
        = & - \frac{8}{45} \cdot r \cdot \frac{\varphi^3}{3!} + O\left( \varphi^5 \right). 
\end{split}
\end{equation}
It is shown that the first-order term $ \varphi $ disappears but the third-order term $ \varphi^3 $ persists. {The nonlinear coefficients $ c_n^\text{static} $ and $ c_n^\text{driven} $ of different orders are depicted in Figs.~\ref{fig: NEMS-3 structure and nonlinearity}(b) and \ref{fig: NEMS-3 structure and nonlinearity}(c), respectively.}
    
Next, we straightforwardly derive the system Hamiltonian under the specific flux bias point according to Eq.~(\ref{equ: H_static of NEMS}) and Eq.~(\ref{equ: H_driven of NEMS}). The static and driven Hamiltonian can be written as
\begin{eqnarray}
\hat{\mathcal{H}}_\text{static} & \approx & \omega \hat{a}^\dagger \hat{a} + g_6^\text{static} ( \hat{a}^\dagger + \hat{a} )^6 + \cdots,\\
    \label{equ: H_static of NEMS-3}
    \hat{\mathcal{H}}_\text{driven} & \approx & \epsilon(t) g_3^\text{driven} \left( \hat{a}^\dagger + \hat{a} \right) ^3 + \cdots.
    \label{equ: H_driven of NEMS-3}
\end{eqnarray}
Although the device does not have $ g_4^\text{static} $ nonlinear term, there may still be some non-zero Kerr nonlinearity coming from $ g_6^\text{static} $ or higher-order nonlinear terms. The driven Hamiltonian contains only three-photon interaction and no single-photon excitation, and $ g_3^\text{driven} = 8/45 \cdot r/3! \cdot E_\text{L}\varphi_{\mathrm{zpf}}^3 $. Note that we only keep the first-order expansion of the magnetic drive $ \epsilon(t) $, since the driving amplitude is typically small. Under a strong magnetic drive, the high-order components such as $ \epsilon(t)^2 $ will affect the frequency and Kerr nonlinearity of the NEMS-3 oscillator, as discussed in Appendix Sect.~\ref{subsec: Magnetic driven Hamiltonian in NEMS}.

\subsection{ Pure $ g_{5}^\text{driven} $ nonlinearity }

\begin{figure}
    \includegraphics{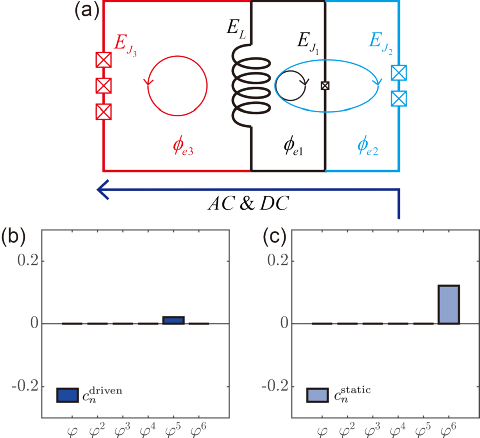}
    \caption{ (a) Circuit diagram of NEMS-5. NEMS-5 has one single-JJ branch, one double-JJ branch, and one triple-JJ branch with different junction sizes, producing $ \varphi^5 $ and higher odd-order driving terms under a specific magnetic drive. The magnetic driving lines are depicted in blue, with the alternating flux amplitude in each loop labeled distinctly.  The coefficients $ c_{n}^\text{driven}$ of driven terms and $ c_{n}^\text{static}$ of static terms are shown in (b) and (c), respectively. Canceling lower-order terms comes with the price of smaller higher-order terms.}
    \label{fig: NEMS-5 structure and nonlinearity}
\end{figure}

Similar to NEMS-3, fifth-order driven term engineering can be achieved by a NEMS-5 device, as shown in Fig.~\ref{fig: NEMS-5 structure and nonlinearity}(a). NEMS-5 is designed for $c_{1}^\text{driven} = c_{3}^\text{driven} = 0$ and $c_{5}^\text{driven} \neq 0$, while minimizing $c_{4}^\text{static}$. The device consists of three Josephson branches with single-, double-, and triple-JJs, with a magnetic drive satisfying $\abs{\delta\phi_{ei}(t)} \propto n_i \epsilon(t)$. The solution (see Appendix Sect.~\ref{subsec: Numerical results of stabilizing four-cat states with NEMS-5} for details) yields specific ratios for junction sizes as
\begin{equation}\label{equ: rs of NEMS-5}
\begin{split}
        &r_1 = \frac{5}{32}r,\quad r_2 = r,\quad r_3 = \frac{27}{32} r, \\
        &n_{1} = 1,\quad n_{2} = 2,\quad n_{3} = 3,
\end{split}
\end{equation}
and the magnetic control parameters as
\begin{equation}\label{equ: phi_e of NEMS-5}
\begin{split}
        \overline{\phi_{e1}} = \pi, \quad & \delta\phi_{e1}(t) = 1/5 \cdot \epsilon(t), \\ 
        \overline{\phi_{e2}} = 0, \quad & \delta\phi_{e2}(t) = 2/5 \cdot \epsilon(t), \\ 
        \overline{\phi_{e3}} = 0, \quad & \delta\phi_{e3}(t) = -3/5 \cdot \epsilon(t).
\end{split}
\end{equation}
Here, $ r \lesssim 1 $ is a parameter limiting the size of small JJs. The flux bias point is chosen such that the {single-JJ branch} provides $ +\cos{ \varphi } $ potential, while the {multi-JJ branches} provide $ -\cos{ \left( \varphi/n_{i} \right) } $ potential.  {In Fig.~\ref{fig: NEMS-5 structure and nonlinearity}(a), we present one possible construction of the magnetic driving line (depicted in blue). We relocate the $ r_3 $ branch to the left of the inductor branch so the magnetic flux in the $ \phi_{e3} $ loop is opposite to that of the $ \phi_{e1} $ and $ \phi_{e2} $ loops. The relocation does not change the potential function. } Two additional DC flux lines [not depicted in Fig.~\ref{fig: NEMS-5 structure and nonlinearity}(a)] are required to satisfy the static flux bias requirements.
{The nonlinear coefficients $ c_n^\text{static} $ and $ c_n^\text{driven} $ of different orders are depicted in Figs.~\ref{fig: NEMS-5 structure and nonlinearity}(b) and \ref{fig: NEMS-5 structure and nonlinearity}(c), respectively.}

The resulting static and driven Hamiltonians are
\begin{eqnarray}\label{equ: H_static of NEMS-5}
    \hat{\mathcal{H}}_\text{static} &\approx& \omega \hat{a}^\dagger \hat{a} + g_6 ( \hat{a}^\dagger + \hat{a} )^6 + \cdots,\\
    \hat{\mathcal{H}}_\text{driven} &\approx& \epsilon(t) g_5^\text{driven} \left( \hat{a}^\dagger + \hat{a} \right) ^5 + \cdots.
\end{eqnarray}
The nonlinear coefficient is $g_5^\text{driven} = 1/48 \cdot r / 5! \cdot E_\text{L}\varphi_{\mathrm{zpf}}^5$, limited by the decay of Taylor expansion coefficients for high orders and also an additional factor of $1/48$ due to the coefficients canceling between different branches. The strength of the fifth-order nonlinear drive could potentially be increased by relaxing the elimination conditions for lower-order terms or optimizing flux driving ratios.

\section{Engineering Even-Order Driven Nonlinearities}

The engineering of even-order driving terms is more challenging than that of odd-order terms, because the construction of even-order (cosine-like) driving terms normally results in odd-order (sine-like) static terms, according to Eq.~(\ref{equ: U_static of NEMS general}) and Eq.~(\ref{equ: U_driven of NEMS general}). {The odd-order static terms may change the minimum point $ \varphi^* $ of the static potential, increasing the difficulty in engineering driving terms.} 

Here we present the \textit{symmetric-double-branch} method that utilizes the circuit symmetry to cancel the static odd-order nonlinearity. This construction requires two identical Josephson branches with opposite flux bias to construct one cosine-like driving term, with the corresponding static terms being still cosine-like.  For example, we assign two identical multi-JJ branches with opposite flux bias, and then the corresponding potential of \textit{double multi-JJ (DMJ) branches} $ U_\text{dmJ} $ is    
\begin{equation}\label{equ: U_dmJ total }
    \begin{split}
        \frac{ U_\text{dmJ} }{ E_\text{L} } &= - n_{i} r_{i} \cos\left( \frac{\varphi + \phi_{ei} }{ n_{i} } \right) - n_{i} r_{i} \cos\left( \frac{\varphi - \phi_{ei} }{ n_{i} } \right)  \\ 
        & = - 2 n_{i} r_{i} \cos\left( \frac{\varphi}{n_{i}} \right) \cos( \frac{\phi_{ei}}{n_{i}} ). 
    \end{split}
\end{equation}
Unlike the single-branch potential in Eq.~(\ref{equ: U_static of NEMS general}) and Eq.~(\ref{equ: U_driven of NEMS general}) which only provides static and driven terms with different parity, the double-branch potential provides even terms for both static and driven components:
\begin{equation}\label{equ: U_dmJ static and driven }
    \begin{split}
        \frac{ U_\text{dmJ}^\text{static} }{ E_\text{L} } &= - 2 n_{i} r_{i} \cos\left( \frac{\varphi}{n_{i}} \right) \cos( \frac{ \overline{\phi_{ei}} }{ n_{i} } ),  \\
        \frac{ U_\text{dmJ}^\text{driven} }{ E_\text{L} } &= 2 r_{i} \cos\left( \frac{\varphi}{n_{i}} \right) \sin( \frac{ \overline{\phi_{ei}} }{ n_{i} }  ) \delta\phi_{ei}. 
    \end{split}
\end{equation}
The flux bias of the DMJ branches is still limited by the phase slip dynamics, meaning that the effective flux bias $ \phi_{ei} $ is limited within the range of $ \left( -\pi,\pi \right) $. Therefore, for $ n_{i} \geq2 $, the static potential of the DMJ branches will always have a minus sign, i.e., the static potential will resemble $ -\cos(\varphi) $. 
    
Similarly, we can design \textit{double single-JJ (DSJ) branches} to provide static and driven terms with different signs for Hamiltonian engineering:
\begin{equation}\label{equ: U_smJ total }
    \begin{split}
        \frac{ U_\text{dsJ} }{ E_\text{L} } &= - r_{i} \cos\left( \varphi + \phi_{ei} \right) - r_{i} \cos\left( \varphi - \phi_{ei} \right) \\ 
        & = - 2 r_{i} \cos\left( \varphi  \right) \cos(\phi_{ei} ).  
    \end{split}
\end{equation}
The corresponding static and driven potentials can be written as:
\begin{equation}\label{equ: U_smJ static and driven }
    \begin{split}
        \frac{ U_\text{dsJ}^\text{static} }{ E_\text{L} } &= - 2 r_{i} \cos\left( \varphi \right) \cos( \overline{\phi_{ei}} ),  \\
        \frac{ U_\text{dsJ}^\text{driven} }{ E_\text{L} } &= 2 r_{i} \cos\left( \varphi \right) \sin( \overline{\phi_{ei}} ) \delta\phi_{ei}.  
    \end{split}
\end{equation}
The flux bias of the single-JJ branch is not limited by the phase slip dynamics, so their flux bias can be assigned to near $ \pi $ to provide negative static terms. Therefore, the static potential of the DSJ branches could have a plus or minus sign, i.e., the static potential could be $ \pm\cos(\varphi) $. {The \textit{symmetric-double-branch} method guarantees that $\varphi^*$ will always be 0, simplifying the engineering of nonlinear terms.}

Now we can engineer even-order terms in the same way as odd-order terms. First, we assign several DMJ branches to a specific flux bias point $ \overline{\phi_{ei}}= n_{i}\overline{\phi_{e0}} < \pi $, so that their static and driven potentials have the same coefficient $ \cos(\overline{\phi_{e0}}) $ or $ \sin(\overline{\phi_{e0}}) $. Then, we assign several DSJ branches to the flux bias point such that $ \overline{\phi_{ei}} = \phi_{e0} $ or $ \abs{\overline{\phi_{ei}}} = \phi_{e0} \pm \pi $, so they can provide $ \pm\cos(\varphi) $ terms for Hamiltonian engineering. Finally, we collect all the terms and derive static and driven nonlinear coefficients as follows: 
\begin{equation}\label{equ: c_static even order engineering}
    \begin{split}
        c_{n}^\text{static} = & \eval{\frac{\mathrm{d}^n}{\mathrm{d} \varphi_L^n } \frac{\varphi_L^2}{2} }_{\varphi = 0} \\
        -& 2 \cos( \overline{\phi_{e0}} ) \sum_{i} \frac{ r_i }{n_{i}^{n-1}} s_i \eval{\frac{\mathrm{d}^n}{\mathrm{d} \varphi^n } \cos( \varphi )}_{\varphi = 0},
    \end{split}
\end{equation}
\begin{equation}\label{equ: c_driven even order engineering}
    \begin{split}
        c_{n}^\text{driven} = & 2 \sin( \overline{\phi_{e0}} ) \sum_{i} \frac{ r_i r_{\phi_{ei}} }{n_{i}^n} s_i \eval{\frac{\mathrm{d}^n}{\mathrm{d} \varphi^n } \cos( \varphi )}_{\varphi = 0}.
    \end{split}
\end{equation}
The sign $ s_i $ still follows the rule of Eq.~(\ref{equ: rule for sign(i)}). The latter part $ s_i \eval{ \mathrm{d}^n \left( \cos( \varphi ) \right) / \mathrm{d} \varphi^n }_{\varphi = 0} $ is just a sign that can be predetermined once the numbers of JJs are fixed. The former part $  r_i r_{\phi_{ei}} / n_{i}^n $ can be carefully designed to engineer the coefficient $ c_{n}^\text{driven} $. It is easy to see that $ c_{n}^\text{driven} = 0 $ when $ n $ is odd, i.e., the potential has no odd-order driven nonlinearity.

The construction process indicates that engineering even-order terms is harder than engineering odd-order terms. To avoid phase slip in each loop, the flux bias coefficient $ \overline{\phi_{e0}} $ is limited within the range of $ \left( -\pi/n_{i},\pi/n_{i} \right) $ for the largest $ n_{i} $. So the coefficient of driven nonlinearity decreases as $ \sin{\left( \overline{\phi_{e0}} \right)} \propto 1 / n_{i} $ . Consequently, the even-order terms diminish $ 1 / n_{i} $ faster than the odd-order terms. When $ n_{i} $ is large, the multi-JJ branch eventually behaves like a linear inductor, with all the nonlinear terms vanishing.

\subsection{Pure $ g_4^\text{driven} $ nonlinearity }

\begin{figure}
    \includegraphics{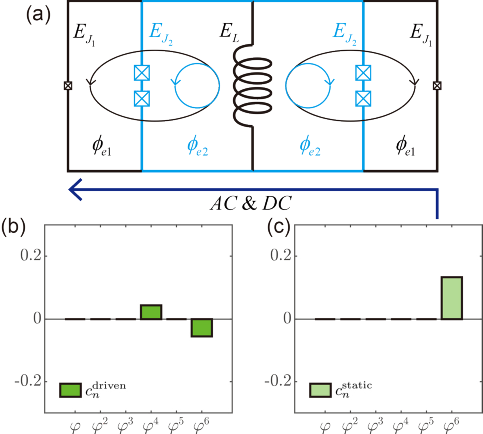}
    \caption{(a) Circuit diagram of NEMS-4. NEMS-4 has two double-JJ branches and two single-JJ branches, producing $ \varphi^4 $ and higher even-order driving terms under a specific magnetic drive. {The coefficients $ c_{n}^\text{driven}$ of driven terms and $ c_{n}^\text{static}$ of static terms are shown in (b) and (c), respectively.} 
    Canceling lower-order terms comes with the price of smaller higher-order terms.}
    \label{fig: NEMS-4 structure and nonlinearity}
\end{figure}

As an example of even-order nonlinearity engineering, we design the NEMS-4 device for selectively generating $ g_4^\text{driven}$ nonlinearity. Our goal is to construct a target potential with even-order driving terms while simultaneously eliminating the second-order driving term and minimizing the static nonlinearity. These conditions are met using two single-JJ branches biased at $ \pm ( \phi_{e0} + \pi ) $ and two multi-JJ branches biased at $ \pm n_{2}\phi_{e0} $. The structure of the NEMS-4 is illustrated in Fig.~\ref{fig: NEMS-4 structure and nonlinearity}(a). The corresponding design requirements for driving terms are
\begin{equation}\label{equ: c_driven target of NEMS-4}
    \begin{split}
        \frac{c_{2}^\text{driven}}{ 2\sin(\overline{\phi_{e0}}) } = - r_1 r_{\phi_{e1}} + \frac{r_2 r_{\phi_{e2}}}{n_{2}^2} &= 0, \\
        \frac{c_{4}^\text{driven}}{ 2\sin(\overline{\phi_{e0}}) } = + r_1 r_{\phi_{e1}} - \frac{r_2 r_{\phi_{e2}}}{n_{2}^4} &\neq 0,
    \end{split}
\end{equation}
and the requirement for suppressing static term is
\begin{equation}\label{equ: c_static target of NEMS-4}
        \frac{c_4^\text{static}}{ 2\cos( \overline{\phi_{e0}} ) } = + r_1 - \frac{r_2}{n_{2}^3} = 0. 
\end{equation}
Note that each $ r_i $ now corresponds to two identical branches with opposite flux bias. A convenient approach to satisfy these requirement is to set $ n_2 = 2$ and $\phi_{e0} = \pi/4 $, with the solution:   
\begin{equation}\label{equ: rs of NEMS-4}
\begin{split}
&r_1 = \frac{1}{8}r,\quad r_2 = r,\\
&n_{1} = 1,\quad n_{2} = 2, \\ 
&r_{ \phi_{e1} } = 2 r_{ \phi_{e2} },
\end{split}
\end{equation}
where $ r \lesssim 1 $ is a parameter limiting the size of small JJs. The corresponding control magnetic fluxes are taken as 
\begin{equation}\label{equ: phi_e of NEMS-4th}
    \begin{split}
        \overline{\phi_{e1}} = \frac{5\pi}{4}, \quad & \delta\phi_{e1} = \frac{ 1 }{ 2 } \epsilon(t), \\ 
        \overline{\phi_{e2}} = \frac{\pi}{2}, \quad & \delta\phi_{e2} = \frac{ 1 }{ 4 } \epsilon(t). 
    \end{split}
\end{equation}
The magnetic drive strength has been normalized to ensure that the total magnetic flux change has an amplitude of $ \epsilon(t) $. {In Fig.~\ref{fig: NEMS-4 structure and nonlinearity}(a), we present one possible construction of the magnetic driving line (depicted in blue). The symmetric double branches are positioned symmetrically on both sides of the inductor branch so the magnetic fluxes in the symmetric loops remain opposite. The positioning does not change the potential function. Typically, three additional DC flux lines [not depicted in Fig.~\ref{fig: NEMS-4 structure and nonlinearity}(a)] are required to satisfy the static flux bias requirements.}

With all parameter restrictions applied, the total potential of the NEMS-4 device is given as
\begin{eqnarray}\label{equ: U_tot of NEMS-4th}
        \frac{U_\text{tot}}{E_\text{L}} & = & \frac{\varphi^2}{2} - \frac{1}{4} r \cos\left( \varphi \right) \cos\left( \phi_{e1} \right)\nonumber\\
        & &- 4 r \cos\left( \frac{\varphi}{2} \right) \cos\left( \frac{\phi_{e2}}{2} \right). 
\end{eqnarray}
The corresponding static potential is
\begin{equation}\label{equ: U_static of NEMS-4th}
    \begin{split}
        \frac{U_\text{static}}{E_\text{L}} = & \frac{\varphi^2}{2} + \frac{\sqrt{2}}{8} r \cos \left( \varphi \right) - 2\sqrt{2} r \cos\left( \frac{ \varphi }{2} \right)\\
        = & \frac{\varphi^2}{2} + \frac{3\sqrt{2}}{16} r \varphi^2 + O\left( \varphi^6 \right),
    \end{split}
\end{equation}
and the driven potential takes the form of
\begin{equation}\label{equ: U_driven of NEMS-4th}
    \begin{split}
        \frac{ U_\text{driven} }{ \epsilon(t) E_\text{L} } & = - \frac{\sqrt{2}}{16} r \cos\left( \varphi \right)+ \frac{\sqrt{2}}{4} r \cos\left( \frac{\varphi}{2} \right) \\
        & = - \frac{ 3\sqrt{2} }{64} \cdot r \cdot \frac{ \varphi^4}{4!} + O\left(\varphi^6\right).
    \end{split}
\end{equation}
Here, since the static potential still only contains even-order terms, the minimum position is still $ \varphi^*=0 $. {The nonlinear coefficients $ c_n^\text{static} $ and $ c_n^\text{driven} $ of different orders are depicted in Figs.~\ref{fig: NEMS-4 structure and nonlinearity}(b) and \ref{fig: NEMS-4 structure and nonlinearity}(c), respectively.}

Finally, we arrive at the device Hamiltonian with 
\begin{equation}\label{equ: H_static of NEMS-4th}
        \hat{\mathcal{H}}_\text{static} \approx \omega \hat{a}^\dagger \hat{a} + g_6 ( \hat{a}^\dagger + \hat{a} )^6 + \cdots,
\end{equation}
and
\begin{equation}\label{equ: H_driven of NEMS-4th}
    \hat{\mathcal{H}}_\text{driven} \approx \epsilon(t) g_4^\text{driven} \left( \hat{a}^\dagger + \hat{a} \right) ^4 + \cdots.
\end{equation}
The device is dominated by a fourth-order driven Hamiltonian, with the nonlinear coefficient given by $  g_4^\text{driven} = 3\sqrt{2}/64 \cdot r / 4! \cdot E_\text{L}\varphi_{\mathrm{zpf}}^4 $.

\section{Applications}

{The NEMS devices have the potential to realize selected orders of nonlinear Hamiltonian, which can be used in applications involving multi-photon interactions. In the following, we present two possible applications of NEMS devices. The first is to implement a Kerr-cat BPCNOT gate with a NEMS-3 device, which involves constructing a three-photon interaction. The second is to stabilize 4-cat states with a NEMS-5 device, which requires constructing a five-photon interaction.}

\subsection{Manipulate Kerr-cat qubit}

\begin{figure}
    \includegraphics{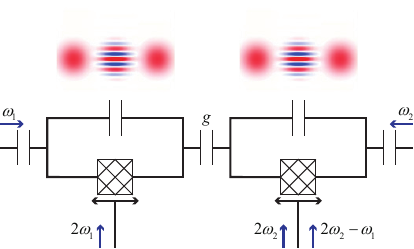}
    \caption{ Circuit diagram of linearly coupled two Kerr-cat qubits based on NEMS-3. A slight deviation from the working point is required for the NEMS-3 device to stabilize the cat states, followed by a double-frequency magnetic drive through a magnetic drive line. Single-qubit control is achieved by applying a resonant drive on the electric drive line. The two-qubit BPCNOT gate is realized by applying a magnetic drive at $ 2\omega_2 - \omega_1 $ on one of the cat-qubits. }
    \label{fig: NEMS 2Cat}
\end{figure}

The Kerr-cat encoding scheme stabilizes the 2-cat state $ \ket{\mathcal{C}^{\pm}_{\alpha}} \equiv \mathcal{N}_{\pm} \left( \ket{\alpha} \pm \ket{-\alpha} \right)$ through the combination of Kerr nonlinearity and two-photon driving, with $ \mathcal{N}_{\pm} $ defined as the normalization factor. In the code space, the 2-cat qubit exhibits the biased-noise property~\cite{Lescanne2020Exponential,Grimm2020Stabilization,Berdou2023OneHundred,Frattini2024Observation,Reglade2024NatureDCat10s}, which is useful in concatenated encoding schemes like surface-cat codes and repetition-cat codes. However, demonstrating the full potential of the Kerr-cat qubit requires implementing a BPCNOT gate that preserves the biased-noise property.
Realizing the BPCNOT gate requires implementing the following Hamiltonian~\cite{Puri2020Bias-preserving}
\begin{equation}\label{equ: H of Kerr-cat BPCX middle}
    \hat{\mathcal{H}}_2 = - K \left( \hat{a}_2^{\dagger2} - \frac{\alpha_2^2}{\alpha_1} \hat{a}_1^\dagger \right) \left( \hat{a}_2^2 - \frac{\alpha_2^2}{\alpha_1} \hat{a}_1 \right),
\end{equation}
where $\hat{a}_{1,2}$ are the annhilation operators of the two WAOs, $\alpha_{1,2}$ are the sizes of the two Kerr-cat qubits, and $K$ is the Kerr nonlinearity of the target Kerr-cat qubit.

The BPCNOT Hamiltonian requires simultaneous engineering of Kerr nonlinearity $K \hat{a}_2^{\dagger2} \hat{a}_2^2 $ and a three-photon interaction $ K \alpha_2^2/\alpha_1 \cdot \hat{a}_2^{\dagger2} \hat{a}_1 $, which is challenging to realize experimentally. For example, the widely used SNAIL nonlinear element in Kerr-cat experiments operates in a regime of strong $ g_3^\text{static} $ and small $ g_4^\text{static} $~\cite{Frattini2017SNAIL,Grimm2020Stabilization,Hajr2024TaKerrCat,Frattini2024Observation,Venkatraman2024DetunedKerrCat}, with an electric drive as control. Based on Eq.~(\ref{equ: H_driven of electric drive}), achieving a strong three-photon interaction with an electric drive requires a large driving amplitude while maintaining a small $ g_4^\text{static} $. Consequently, implementing the BPCNOT gate on a SNAIL platform may necessitate strong drives, potentially resulting in extra residual process~\cite{Huang2023Residual,Yaxing2019Engineering,Chapman2023SNAILBeamSplitter,Miano2023Hamiltonian}. An alternative approach is to implement the Kerr-cat BPCNOT gate with ATS, which possesses the desired small Kerr nonlinearity and strong $g_3^\text{driven}$ term. However, as previously discussed in Eq.~(\ref{equ: U driven of single junction part}), ATS can only select odd-order driving terms, and the strong residual first-order terms may cause leakage and adversely affect the gate fidelity.

NEMS-3 can selectively generate a clean $ g_3^\text{driven} $ term while providing controllable even-order static nonlinearity. Therefore, it is straightforward to implement the Kerr-cat BPCNOT gate with NEMS-3. Figure~\ref{fig: NEMS 2Cat} illustrates the circuit diagram of two linearly coupled Kerr-cat qubits, where two cat modes are coupled through a linear capacitor with a coupling strength $g$ typically smaller than their detuning $ \Delta \equiv |\omega_2 - \omega_1| $. Both cat modes exhibit small static nonlinearity, making it appropriate to apply the Bogoliubov transformation for analyzing their interaction, as detailed in the Appendix Sect.~\ref{subsec: Coupled Two Kerr cat qubits}. The desired three-photon interaction Hamiltonian can be realized by applying a magnetic drive on the target mode $ \hat{a}_2 $ at the frequency $ \omega_\text{d} = 2\omega_2 - \omega_1 $. By applying the rotating wave approximation, the magnetic-driven Hamiltonian for BPCNOT gate is given by 
\begin{eqnarray}\label{equ: H of Magnetic BPCX drive term}
        \hat{\mathcal{H}}_\text{BPCNOT} & = & g_3^\text{driven}  \left( \frac{1}{2} \epsilon_\text{d} \mathrm{e}^{-\mathrm{i} \omega_\text{d} t } + c.c. \right)\nonumber\\
        & & \quad \times \left( \hat{a}_2 \mathrm{e}^{-\mathrm{i} \omega_2 t } +  \frac{g}{\Delta} \hat{a}_1 \mathrm{e}^{-\mathrm{i} \omega_1 t } + h.c. \right)^3 \nonumber \\
        &  \approx & \frac{3}{2} \epsilon_\text{d} g_3^\text{driven} \frac{g}{\Delta} \hat{a}_2^{\dagger2} \hat{a}_1 + h.c.
\end{eqnarray}
Numerical results show that NEMS-3 enables direct implementation of Kerr-cat stabilization and BPCNOT gate, with a moderate driving strength and minimal residual terms, as detailed in Appendix Sect.~\ref{subsec: Numerical Simulation}. 
    
\subsection{Stabilizing 4-cat state}

\begin{figure}
    \includegraphics{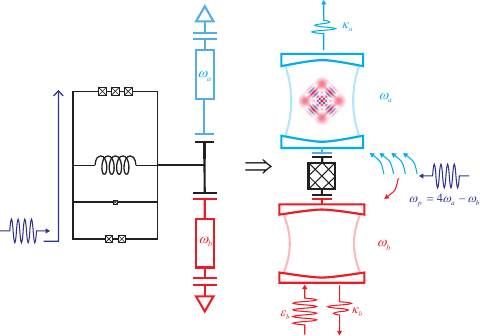}
    \caption{(Left) Circuit diagram of stabilizing a 4-cat state with a NEMS-5 device. A crucial requirement is to engineer a 4-to-1 coupling $ g_\text{c} \hat{a}^{\dagger 4} \hat{b} + h.c. $ between the storage mode $ \hat{a} $ (shown in cyan) and the buffer mode $ \hat{b} $ (shown in red) through a nonlinear coupler (shown in black). Through adiabatic elimination of the buffer mode, a single-photon driven dissipation process in the buffer mode can be converted to a four-photon driven dissipation process in the storage mode, thereby stabilizing a 4-cat state. (Right) Physical device for stabilizing a 4-cat state with NEMS-5, which linearly couples to both the storage and buffer modes. Under flux modulation, the NEMS-5 generates the desired five-photon interaction, thereby inducing an active 4-to-1 coupling between the two modes when pumped at the appropriate frequency. 
    }
    \label{fig: NEMS_4Cat}
\end{figure}    
In this section, we propose that NEMS-5 devices have the potential to realize the stabilization of a 4-cat code~\cite{Mirrahimi2014Dynamically}. Specifically, We aim to construct a four-photon driven dissipation process, which can be described by a master equation in the form as~\cite{Mirrahimi2014Dynamically}:
\begin{equation}\label{equ: 4-ph DD process for 4cat stabilization}
    \begin{split}
        \dot{\rho} &= \left[\varepsilon_\text{4ph}\hat{a}^{\dagger 4} + \varepsilon_\text{4ph}^*\hat{a}^{4}, \rho \right] +\kappa_\text{4ph} \mathcal{D}\left[\hat{a}^{4}\right] \rho  \\
        & = \kappa_\text{4ph} \mathcal{D}\left[\hat{a}^{4} - \alpha^4 \right] \rho,
    \end{split}
\end{equation}
with $\alpha = \left( {2\varepsilon_\text{4ph} }/{\kappa_\text{4ph}} \right) ^ {1/4}$. To construct such a four-photon driven dissipation process on a target long-lifetime storage mode, it is necessary to nonlinearly couple it with a rapidly dissipating buffer mode. In stabilizing the 4-cat state, we design a nonlinear interaction that converts four photons from the target mode ($ \hat{a} $) to one photon in the buffer mode ($ \hat{b} $), as shown in Fig.~\ref{fig: NEMS_4Cat}, and vice versa. The 4-to-1 coupling Hamiltonian is given by:
\begin{equation}
\hat{\mathcal{H}}_\text{c} = g_\text{c} \hat{a}^{\dagger 4} \hat{b} + h.c.
\end{equation}
When the decay rate of the buffer mode ($ \kappa_b$) is significantly larger than the coupling strength, an effective nonlinear dissipation of the storage mode can be achieved by adiabatic eliminating the buffer mode. This system evolution transformation also applies to the linear drive $ \varepsilon_b\hat{b}^\dagger + \varepsilon_b^*\hat{b} $ on the buffer mode. Consequently, the resulting four-photon driving and dissipation strengths in the storage mode are as follows:
\begin{equation}
    \varepsilon_\text{4ph} = -\frac{2 i g_\text{c} \varepsilon_b}{\kappa_b}, \, \kappa_\text{4ph} = \frac{2 g_\text{c}^2 }{\kappa_b}.
\end{equation}

The NEMS-5 device provides an ideal candidate for directly generating the 4-to-1 five-photon interaction between the storage mode and the buffer mode. The circuit and device configurations are illustrated in Fig.~\ref{fig: NEMS_4Cat}. The NEMS-5 under magnetic drive (flux pump) can be described by the Hamiltonian:
\begin{eqnarray}
    \hat{\mathcal{H}}_\text{c} & = & g_5^\text{{driven}} \left( \frac{1}{2} \epsilon_\text{d} \mathrm{e}^{-\mathrm{i} \omega_\text{d} t } + c.c. \right) \nonumber \\
    & & \quad \times \left( p_b \hat{b} \mathrm{e}^{-\mathrm{i} \omega_b t } + p_a\hat{a} \mathrm{e}^{-\mathrm{i} \omega_a t }  + h.c. \right)^5 \nonumber \\
    & \approx & \frac{5}{2} \epsilon_\text{d} g_5^\text{driven} p_a^4 p_b a^{\dagger4}b + h.c..
\end{eqnarray}
where $ \omega_\text{d} $ is the frequency of the magnetic drive, $ p_a $ and $ p_b $ represent the participation ratios~\cite{Minev2021pyEPR} of modes $ \hat{a} $ and $ \hat{b} $ in the  device, respectively.The corresponding frequency matching condition for the 4-to-1 interaction is $ \omega_\text{d} = 4 \omega_a - \omega_b $. The circuit configuration can be interpreted as simultaneously coupling two resonators to a nonlinear coupler and then activating the nonlinear process between the two resonators through an external drive. 

\section{Conclusion and Discussion}

In conclusion, we propose a novel set of inductive elements, termed NEMS, which exhibit universal controllability in shaping static nonlinearity and generating selective nonlinear drives on demand. Inspired by ATS, NEMS incorporates one inductor branch alongside several Josephson branches with varying junction sizes and numbers, providing a sufficient parameter space for engineering high-order interactions. We discuss different approaches for engineering odd- and even-order terms, highlighting the challenges associated with generating even-order driving terms compared to odd-order terms. To address this, we propose the \textit{symmetric-double-branch} method for engineering general even-order terms. We present three configurations: NEMS-3, NEMS-4, and NEMS-5, each capable of generating nonlinear interaction Hamiltonians with target orders. We then demonstrate the feasibility of implementing Kerr-cat qubits with NEMS-3 and stabilizing the 4-cat state with NEMS-5. 
Our results suggest that NEMS devices are promising candidates for complex high-order Hamiltonian engineering. 
    
In practical experiments, the alternating magnetic field may cause residual electric drive on the nonlinear device, which should be properly treated to avoid parasitic nonlinear effects. The residual electric drives typically arise in two ways: First, the large shunting capacitor is influenced by the alternating magnetic field~\cite{Riwar2022Circuit}. Since both ATS and NEMS have a large shunting capacitor comparable in size to the microwave wavelength, it is crucial to carefully design the spatial distribution of the magnetic field to avoid additional electric drives. Using lumped capacitors with smaller footprints can be beneficial in minimizing residual electric drive. Second, the alternating magnetic field generates an electromotive force on the JJ capacitors~\cite{You2019Circuit,Bryon2023FluxDriveFluxonium}.
We discuss the electromotive force caused by JJ capacitors in the Appendix Sect.~\ref{subsec: Lagrangian under fluctuating magnetic fields} and Appendix Sect.~\ref{subsec: ATS Residual Electric Drive}, demonstrating that this electromotive force is zero for ATS and NEMS-3 device. However, for other nonlinear devices, careful consideration of the JJ capacitors is necessary to minimize the residual electric drives.

\begin{acknowledgments}
We thank Mazyar Mirrahimi and Zaki Leghtas for helpful discussions.
This work is funded by the National Natural Science Foundation of China (Grants No. 11925404, 92165209, 92365301, 92265210, 11890704, 92365206),  Innovation Program for Quantum Science and Technology (Grant No.~2021ZD0300203, 2021ZD0301800, and 2021ZD0300201), and the National Key R\&D Program (2017YFA0304303). M.L. and C.-L.Z. are also supported by the Fundamental Research Funds for the Central Universities and USTC Research Funds of the Double First-Class Initiative. 

\end{acknowledgments}

\appendix
    
\section{Single-well condition and limitation on flux bias}
\label{sec: limitation on flux bias}

A comprehensive calibration of NEMS devices requires a general knowledge of the spectrum. However, it is hard to calculate the spectrum at certain flux bias points, due to the complex multi-junction multi-loop structure. For a deep potential of a single minimum, {the system behaves as a WAO. Thus,} we can directly give the frequency and nonlinear coefficients of each order by expanding the phase variable to different orders. In certain parameter regions, the potential may contain multiple minima, and then the spectrum can only be given through numerical diagonalization. Identifying {the WAO region} not only simplifies the spectrum calculation but also provides insight into the working regime of NEMS.

The number of local minima of a single-loop Josephson circuit has been studied thoroughly by Ref.~\cite{Miano2023Hamiltonian}. Multiple minima may appear when the external flux in one loop is close to $ \pi $. For NEMS devices, two cases may cause multiple potential minima. First, the single-JJ branches provide $ - E_{\text{J}_i} \cos(\varphi) $ potential when $ \phi_{ei} \sim \pm\pi $. When {the Josephson energy exceeds the energy from the inductor}, i.e., $ E_{\text{J}_i} > E_\text{L}$, the system may work as a fluxonium~\cite{Manucharyan2009Fluxonium,Pop2014Fluxonium}. Second, the external flux bias of the multi-JJ branch is close to $ \pm\pi $. The flux may change by $ 2\pi $ due to phase slip~\cite{Matveev2002Persistent,Astafiev2012Coherent,Manucharyan2012Evidence}, causing extra dephasing due to charge noise.

\subsection{Multiple minima associated with a single-JJ branch}

\begin{figure}
    \includegraphics{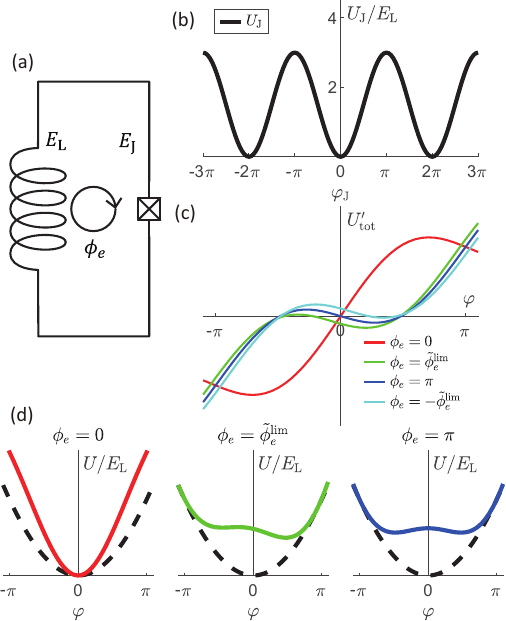}
    \caption{ (a) Circuit diagram of a single-JJ branch ($ E_\text{J} / E_\text{L} = 1.5$) and an inductor branch connected in parallel. (b) The potential energy of the single-JJ branch, which is single-valued and periodic. (c) The derivative of the total potential energy $ U_\text{tot}^\prime $ under different magnetic flux bias $ \phi_e $. Within the {WAO region} where $ \abs{\tilde{\phi}_e} < \tilde{\phi}_e^\text{lim} $, the derivative curve crosses the horizontal axis only once, corresponding to only one minimum. Outside the WAO region, the system may have multiple minima. (d) The total potential energy $ U_\text{tot} $ under different magnetic flux bias $ \phi_e $. The solid lines and the dashed lines correspond to the total potential energy $ U_\text{tot} $ and the inductor energy $ E_\text{L} \varphi^2/2 $, respectively. At the critical point where $ \phi_e = \tilde{\phi}_e^\text{lim} $, the depth of the single well is significantly smaller than in the case of $ \phi_e = 0 $, implying that the system can accommodate fewer photons.}
    \label{fig: Single junction branch and inductor}
\end{figure}

To understand the multiple minima associated with a single-JJ branch, we analyze a simple case when an inductor is connected to a single-JJ branch, as shown in Fig.~\ref{fig: Single junction branch and inductor}(a). The potential energy $ U_\text{J} $ associated with a single JJ is a single-valued function of the flux variable $ \varphi_\text{J} $, as shown in Fig.~\ref{fig: Single junction branch and inductor}(b):
\begin{equation}
    U_\text{J} = -E_\text{J} \cos(\varphi_\text{J})
\end{equation}
where  $ E_\text{J} $ is the Josephson energy of the JJ. 
{Here we define the phase difference across the inductor branch as the generalized variable $\varphi$, such that the phase difference across the single-JJ branch is expressed as $\varphi_\text{J} = \varphi - \phi_{e} $.}
Therefore, the total energy is given by:	
\begin{equation}\label{equ: U of single junction branch and inductor}
    U_\text{tot}(\varphi) = \frac{E_\text{L} \varphi^2 }{2} - E_\text{J}\cos( \varphi - \phi_{e}),
\end{equation}
where $ E_\text{L} $ is the energy of the inductor. {If the inductor is formed by $ n_\text{L}$ large JJs with equal junction size $ E_{\text{J}_\text{L}} $, the effective inductor energy is defined as $ E_\text{L} \equiv E_{\text{J}_\text{L}} / n_\text{L} $}. The minimum total energy requires $ U_\text{tot}'(\varphi) = 0 $, meaning 	
\begin{equation}\label{equ: U of single junction branch and inductor min condition1}
    E_\text{L} \varphi + E_\text{J}\sin( \varphi - \phi_{e}) = 0.
\end{equation} 
When $ E_\text{J} < E_\text{L} $, it is easy to see that the system can have only one minimum regardless of the value of $ \phi_{e} $. When $ E_\text{J} \gtrsim E_\text{L} $, we give an approximate single-minimum condition as:
\begin{equation}\label{equ: phi limit for single-JJ branch}
    \abs{\tilde{\phi}_e} < \tilde{\phi}_e^\text{lim} \equiv \pi - \left( E_\text{J} / E_\text{L} - 1 \right),
\end{equation}
where $ \tilde{\phi}_{e} $ is defined as a truncated external flux:
\begin{equation}\label{equ: tilde phi_e definition}
    \tilde{\phi}_{e} \equiv \phi_{e} - \left\lfloor\frac{\phi_{e} + \pi }{2\pi}\right\rfloor \cdot 2\pi \in \left(-\pi,\pi\right).
\end{equation} 

In Fig.~\ref{fig: Single junction branch and inductor}(c), we plot $ U_\text{tot}'(\varphi) $ for different external flux $ \phi_{e} $. {Here we set $ E_\text{J} / E_\text{L} = 1.5 $ to reveal the possible multiple-minima behavior under the case that $ E_\text{J} \gtrsim E_\text{L} $.}
When $ \phi_e = \pi $, the derivative curve crosses the horizontal axis three times, indicating that the system has two local minima and one local maximum. When $\phi_e < \tilde{\phi}_e^\text{lim} $, the derivative curve crosses the horizontal axis only once, so there will be only one minimum. The total potential under different flux biases is shown in Fig.~\ref{fig: Single junction branch and inductor}(d). At the critical point $ \phi_e = \tilde{\phi}_e^\text{lim} $, the system still has only one minimum, but the depth of the single well is significantly smaller than the case of $ \phi_e = 0 $. A shallower potential well implies that the system can accommodate fewer photons, so it is not recommended to operate near the critical point. 

Generally, all the small JJs in NEMS are required to satisfy conditions $ E_\text{J} \lesssim E_\text{L} $ to avoid multiple minima. However, the combination of several single-JJ branches may cause multiple minima. Recall that the static potential of several single-JJ branches is the same as a single JJ:
\begin{equation}
    \begin{split}
        U_\text{J}^\text{eff} &= - \sum_{i} E_{\text{J}_i} \cos(\varphi - \phi_{ei}) \\
                &= - E_\text{J}^\text{eff} \cos( \varphi - \phi_{e}^\text{eff} ). 
    \end{split}
\end{equation}
Therefore, $ E_\text{J}^\text{eff} $ and $ \phi_{e}^\text{eff} $ should also satisfy Eq.~(\ref{equ: phi limit for single-JJ branch}) to ensure that the NEMS device works as a WAO. At the working point of NEMS-3 in Eq.~(\ref{equ: phi_e of NEMS-3}) where {$ \phi_{e2}=0 $}, the effective SQUID energy $ E_\text{J}^\text{eff} $ is close to 0, therefore the potential will not be affected by the single-JJ branches. 

\subsection{Multiple minima associated with multi-JJ branches}

\begin{figure}
    \includegraphics{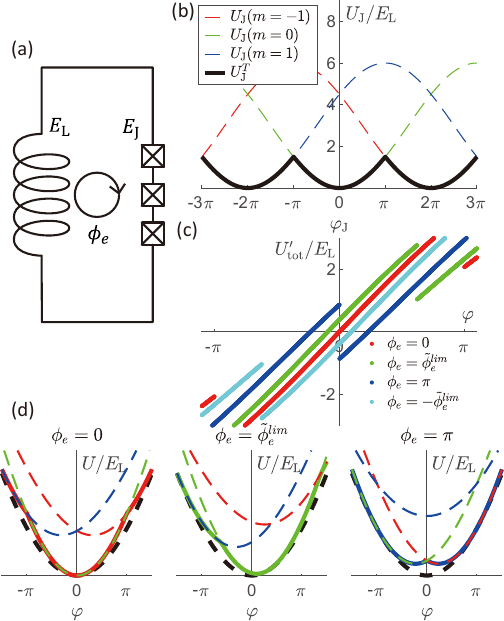}
    \caption{ (a) Circuit diagram of a multi-JJ branch ($ n_\text{J} = 3, E_\text{J} / E_\text{L} = 1$) and an inductor branch connected in parallel. (b) The potential energy of the multi-JJ branch. The colored dashed lines correspond to the potential with different phase-slip numbers $ m $, which are $ 2 n_\text{J}\pi $ periodic individually. The black solid lines denote the minimum potential energy, which is $ 2\pi $ periodic.  (c) The derivative of the total potential energy $ U_\text{tot}^\prime $ under different magnetic flux bias $ \phi_e $. Within the {WAO region} where $ \abs{\tilde{\phi}_e} < \tilde{\phi}_e^\text{lim} $, the derivative curve crosses the horizontal axis only once, corresponding to only one minimum. Outside the WAO region, the system may have multiple minima. (d) The total potential energy $ U_\text{tot} $ under different magnetic flux bias $ \phi_e $. The colored dashed lines correspond to the potential with different phase-slip numbers. The solid lines and the black dashed lines correspond to the total potential energy $ U_\text{tot} $ and the inductor energy $ E_\text{L} \varphi^2/2 $, respectively. Within the {WAO region} where jumping between different $ m $ is unlikely to happen, the system can be approximated as a nonlinear oscillator and we can calculate the spectrum directly. In the extreme case when $ \phi_{e} = \pm \pi $, the system has two degenerate ground states located in two potential wells with different $ m $. }
    \label{fig: Multi-JJ branch and inductor}
\end{figure}

To understand the multiple minima associated with the multi-JJ branches, we analyze a simple case when an inductor is connected to a multi-JJ branch, as shown in Fig.~\ref{fig: Multi-JJ branch and inductor}(a). 

We first analyze the energy-phase relation of a multi-JJ branch. In steady state, the currents $ I_\text{J} $ flowing through multiple serial junctions are the same when neglecting current shunting across capacitors. {For {the $j^\text{th}$ junction inside a multi-JJ branch,} the relationship between the overall current $ I_\text{J} $ and the phase difference $ \varphi_{\text{J}_j} $ across the junction is nonlinear and multi-valued}:
\begin{equation}
    I_\text{J} = I_{\text{c}_j} \sin(\varphi_{\text{J}_j}), \quad I_{\text{c}_j} = \frac{2eE_{\text{J}_j}}{\hbar},
\end{equation}
{where $I_{\text{c}_j}$ are the critical currents.} For a given current $ I_\text{J} $, the possible value of $ \varphi_{\text{J}_j} $ is given by: 
\begin{equation}
    \varphi_{\text{J}_j} = z \pi + (-1)^z\arcsin(\frac{I_\text{J}}{I_{\text{c}_j}}).
\end{equation}
To minimize the energy $ U_{\text{J}_j} = - E_{\text{J}_j} \cos(\varphi_{\text{J}_j}) $, $ z $ should be chosen as an even number to guarantee $ \cos(\varphi_{\text{J}_j}) > 0 $. Therefore, having the same current ensures that the phase difference across each junction differs by $ 2\pi $. 

The phase difference $ \varphi_\text{J} $ across the multi-JJ branch is the sum of the phase differences $ \varphi_{\text{J}_j} $ across each junction. Since a change of $ 2\pi $ in $ \varphi_{\text{J}_j} $ does not affect the current or energy, the total phase difference $ \varphi_\text{J} $ can differ by multiple $ 2\pi $. When the junctions are of equal size, the total phase difference in the multi-JJ branch is expressed as:
\begin{equation}
    \begin{split}
          \varphi_\text{J} & = \sum_{j}\varphi_{\text{J}_j} + 2\pi m\\ 
        & = n_\text{J} \arcsin(\frac{I_\text{J}}{I_{\text{c}_j}}) + 2\pi m. 
    \end{split}
\end{equation}
{Here we simplify the expression by considering that the multi-JJ branch normally contains $n_\text{J}$ JJs with the same Josephson energy $ E_\text{J} \equiv E_{J_i}$. }Therefore, the relation between energy and phase difference of the multi-JJ branch is given by:
\begin{equation}
    U_\text{J}\left(\varphi_\text{J},m\right) = n_\text{J} E_\text{J} \cos( \frac{ \varphi_\text{J} - 2\pi m }{n_\text{J}} ), 
\end{equation}
where $ m $ is referred to as the phase-slip number. In the multi-JJ branches of the NEMS device, we take no more than three small JJs in series, {so a transition of $ m $ results in a dramatic change of energy}. In a steady state, $ m $ should take the value that minimizes the total energy. At given $ \varphi_\text{J} $, the minimum potential energy of the multi-JJ branch can be simplified to a periodic form:
\begin{equation}\label{equ: U of Multi-JJ branch3}
    U_\text{J}^T(\varphi_\text{J}) = - n_\text{J} E_\text{J} \cos\left( \frac{\tilde{\varphi}_\text{J}}{n_\text{J}} \right), 
\end{equation}
with $ \tilde{\varphi}_\text{J} \equiv \varphi_\text{J} - \left\lfloor(\varphi_\text{J} + \pi) /2\pi\right\rfloor \cdot 2\pi $ being the truncated flux variable. 

Figure~\ref{fig: Multi-JJ branch and inductor}(b) plots the potential of the multi-JJ branch for the case $ n_\text{J} = 3 $. The dashed lines with different colors correspond to the potential with phase-slip number $ m\in\left\lbrace 0,\pm1\right\rbrace $, which are $ 2 n_\text{J}\pi $ periodic individually. The blue solid line denotes the minimum potential energy $ U_\text{J}^T(\varphi_\text{J}) $, which is $ 2\pi $-periodic. Note that at the boundary where $ \tilde{\varphi}_\text{J} = \pm\pi $, the phase-slip number $ m $ undergoes a sudden change by $ 1 $, which means that the phase difference $ \varphi_{\text{J}_i} $ across one of the JJs changes by $ 2\pi $. 

Now, we analyze the total potential energy, taking the inductor into account. The flux variable of the inductor $ \varphi $ and multi-JJ branch $ \varphi_\text{J} $ are related by the constraint $ \varphi - \varphi_\text{J} = \phi_{e} $. Therefore, the total energy is given by:	
\begin{equation}\label{equ: U of Multi-JJ branch and inductor}
    U_\text{tot}(\varphi) = \frac{E_\text{L} \varphi^2 }{2}  + U_\text{J}^T( \varphi - \phi_{e}).
\end{equation}
The minimum total energy requires $ U_\text{tot}'(\varphi) = 0 $, which means 	
\begin{equation}\label{equ: U of Multi-JJ branch and inductor min condition1}
     E_\text{L} \varphi + E_\text{J}\sin( \frac{\varphi - \phi_{e} - 2\pi m }{n_\text{J}}) = 0,
\end{equation} 
where $m$ takes the value that $ \varphi - \phi_{e3} - 2\pi m \in \left(-\pi,\pi\right)$.
Therefore, it is easy to verify that the minimum only appears when $ \abs{\varphi} < \varphi^\text{lim} \equiv E_\text{J}/E_\text{L} \cdot \sin( \pi/n_\text{J} ) $. To further ensure the system has only one minimum, we need to make sure no phase-slip happens within this region. Note that the phase-slip happens when $ \varphi - \phi_{e} = 2\pi m $, and thus the condition transforms to 
\begin{equation}\label{equ: phi limit for multi-JJ branch}
    \left|\tilde{\phi}_{e}\right| < \tilde{\phi}_{e}^\text{lim} \equiv \pi - \varphi^\text{lim} = \pi - \frac{E_\text{J}}{E_\text{L}} \sin(\frac{\pi}{n_\text{J}}).
\end{equation} 

The $ U_\text{tot}^\prime\left(\varphi_L\right) $ for different external flux bias are plotted in Fig.~\ref{fig: Multi-JJ branch and inductor}(c). We set the parameters to $ n_\text{J} = 3$, and {$ E_\text{J} / E_\text{L} = 1 $.}
Similar to the single-JJ case, the derivative curve crosses the horizontal line only once when - $\tilde{\phi}_{e}^\text{lim} < \phi_{e} < \tilde{\phi}_{e}^\text{lim} $, but three times when  $ \phi_{e} = \pi $. The total potential is plotted in Fig.~\ref{fig: Multi-JJ branch and inductor}(d). The dashed lines in blue, red, and green correspond to the total potential with different phase-slip numbers, and the dark dashed line denotes the energy of the inductor. The minimal total energy is depicted in the blue solid line, which shows discontinuities when $ \varphi - \phi_{e} = \pm \pi $. In the extreme case when $ \phi_{e} = \pm \pi $, the system will have two degenerate ground states located in two potential wells with different phase-slip numbers $ m $. Even at $ \phi_e = 0 $, where the potential well is the deepest, jumping between different $ m $ still limits the maximum number of photons that can be tolerated in the device. The jumping probability will be discussed in the next subsection.

Within the {WAO region} where the phase-slip number is nearly constant, $ m $ can be determined simply by the external flux $ \phi_e $. Therefore, the potential of the multi-JJ branch can be replaced with a simpler form: 
\begin{equation}\label{equ: U of Multi-JJ branch0}
    U_\text{J} = - n_\text{J} E_\text{J} \cos\left(\frac{\varphi - \tilde{\phi}_{e} }{n_\text{J}} \right).
\end{equation}
  The periodic spectrum in Fig.~\ref{fig: NEMS-3_spectrum_plot2Dand3D} is determined in this way.

When the magnetic drive is included, the system may jump between different periods of the periodic function {$ U_{\text{J}}^T $.} 
To be safe, it requires that the external flux does not exceed the {WAO region} even at the peak of the magnetic drive, so the driving amplitude $ \delta\phi_{e3} $ is limited by:
\begin{equation}\label{equ: phi epsilon limit for multi-JJ branch}
    |\tilde{\phi}_{e} \pm \delta\phi_{e}| < \pi - \frac{E_\text{J}}{E_\text{L}}  \sin(\frac{\pi}{n_\text{J}}).
\end{equation}

\subsection{Phase-slip dynamics and charge Noise }

A complete discussion of the dynamics of phase-slip number $ m $ requires considering the intermediate nodes in the multi-junction branch, which involves introducing additional flux variables and becomes more complex. Since we are discussing small oscillations around the equilibrium state, with vibration frequencies lower than the plasma frequency of the junctions (typically 10$ \sim $40~GHz), we expect the transition probability of different $ m $  to be relatively small. In this case, we can introduce the changes in $ m $ as a transition matrix into the Hamiltonian, i.e.:
\begin{equation}
    \begin{split}
        \hat{\mathcal{H}}_\text{tot} & = \hat{\mathcal{H}}_\text{tot}^{(m)} \ket{m}\bra{m} + \sum_{m} \left(E_\text{S} \ket{m}\bra{m+1}+h.c.\right),\\
        \hat{\mathcal{H}}_\text{tot}^{(m)} & = 4E_\text{C} \hat{n}^2 + U_\text{L}\left(\varphi\right) + U_{sJ}\left(\varphi_{sJ}\right)+U_\text{J}\left(\varphi_\text{J},m\right). 
    \end{split}
\end{equation}
Because a change in $ m $ may correspond to a $ 2\pi $ phase jump occurring on different junctions, the total transition energy is the sum of different transition paths, i.e., $ E_\text{S} = \sum_{j}E_{\text{S}_j} \exp(i\theta_{\text{S}_j}) $. When the phase slip occurs on the $ j^{\text{th}} $-junction, the corresponding transition energy $ E_{\text{S}_j} $ is given by~\cite{Matveev2002Persistent}:
\begin{equation}
    \left| E_{\text{S}_j} \right| = \sqrt{ \frac{2}{\pi} } \left( 8^3 E_{\text{J}_j}^3 E_{\text{C}_j} \right)^{\frac{1}{4}} \exp(-\sqrt{\frac{8 E_{\text{J}_j} }{E_{\text{C}_j}}}).
\end{equation}
The phase $ \exp(i\theta_{\text{S}_j}) $ of the transition matrix element $ E_{\text{S}_j} $ is determined by the charge on adjacent nodes, which is sensitive to charge noise. 

For the NEMS devices discussed in the main text, the impact of charge noise is limited. On one hand, the transition matrix elements themselves are relatively small. For parameters in Table~\ref{tab: Nonlinearity comparison between NEMS ATS and SNAIL}, the sizes of the smaller junctions in the loop exceed those of typical transmons, resulting in transition energies on the order of $ \left| E_{\text{S}_j} \right| \sim 100\text{-}1000$ kHz, with the exponential factor $ \exp(-\sqrt{8 E_{\text{J}_j} / E_{\text{C}_j} } ) \sim 10^{-5} $. If one further increases the junction size and reduces $ E_\text{C} $ to 100 MHz, the transition energy will reduce to 10-100 Hz, while $ \exp(-\sqrt{8 E_{\text{J}_j} / E_{\text{C}_j} } ) \sim 10^{-10} $. On the other hand, the number of junctions in the multi-junction branch does not exceed $ n_r = 3 $, so when $ m $ changes, there is a significant change in the overall energy of the system, as shown in Fig.~\ref{fig: Multi-JJ branch and inductor}(d). The variation in equilibrium energy for different $ m $ values can exceed $ ~100 $ GHz, significantly suppressing transitions of $ m $.	

\subsection{ NEMS-3 spectrum }
\label{subsec: NEMS-3 spectrum}

\begin{figure}
    \includegraphics{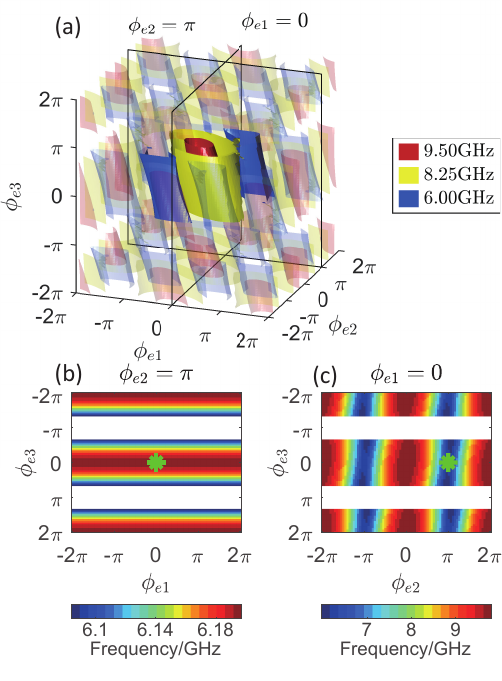}
    \caption{(a) The primary region of NEMS-3's three-dimensional spectrum, where the potential has a single well with limited asymmetry. The solid part represents a single period, with the red, green, and blue colors indicating distinct frequency isosurfaces. The transparent part represents the remaining periods. (b) A two-dimensional cross-section of the spectrum at $ \phi_{e2} = \pi $. (c) A two-dimensional cross-section of the spectrum at $ \phi_{e1} = 0 $. In both cross-sectional figures (b) and (c), the green star indicates the operational point, which is a frequency saddle point. }
    \label{fig: NEMS-3_spectrum_plot2Dand3D}
\end{figure}

Figure \ref{fig: NEMS-3_spectrum_plot2Dand3D}(a) depicts the spectrum of NEMS-3 { within the WAO region}. The spectral analysis of NEMS-3 can be conducted straightforwardly as a nonlinear oscillator, under the condition that the potential is confined to a single well with limited asymmetry. Fortunately, most of the spectrum falls within the single-well region, including the NEMS-3 working point. The spectrum exhibits a $ 2\pi $ periodicity for all three external flux variables $ \phi_{ei}$, which is useful in experimental calibration. The solid part in Fig.~\ref{fig: NEMS-3_spectrum_plot2Dand3D}(a) represents a single period, with the red, green, and blue colors indicating distinct frequency isosurfaces. The transparent part represents the remaining periods. The blank regions generally do not meet the {WAO} condition, indicating that the potential energy may be approaching a double-well, resulting in a complex spectrum. By setting $ \phi_{e2} = \pi $ and $ \phi_{e1} = 0 $, the two-dimensional cross-section of the spectrum is depicted in Figs.~\ref{fig: NEMS-3_spectrum_plot2Dand3D}(b-c). The green star denotes the NEMS-3 working point, which is a saddle point where flux noise is suppressed.

\section{CQED analysis of NEMS }

\subsection{ Lagrangian under fluctuating magnetic fields}
\label{subsec: Lagrangian under fluctuating magnetic fields}

The Lagrangian in the main text only considers the large shunting capacitor. However, the capacitor of the JJs may also affect the system behavior, especially under fluctuating magnetic fields. Specifically, a residual electric drive may come from an alternating magnetic field due to voltage fluctuations across the capacitors caused by electromotive force (EMF) ~\cite{You2019Circuit,Riwar2022Circuit,Yao2023Highfidelity,Bryon2023FluxDriveFluxonium}. {Here we define the phase difference across the inductor branch as the generalized variable $\varphi$} and rewrite the Lagrangian with each flux variable of each loop:

\begin{equation}\label{equ: L total driven of NEMS with CJ}
    \begin{split}
        \mathcal{L}_\text{NEMS} & = \frac{1}{2}C_\text{S} \dot{\varphi}^2 + \sum_{i = 1,2,3}\frac{1}{2}C_{i} \dot{\varphi}_{i}^2 \\
        & \quad - U_\text{NEMS}(\varphi_L, \varphi_{i}), \\ 
    \end{split}
\end{equation}
where $ \varphi_{i} $ is defined as the flux variable associated with {the $i^\text{th}$} branch:
\begin{equation}\label{equ: phi_j for each branch of NEMS with CJ}
    \begin{split}
        \varphi_i & = \varphi + \phi_{ei},  \\
        \dot{\varphi}_{i} &= \dot{\varphi} + \dot{\phi}_{ei} = \dot{\varphi} + r_{\phi_{ei}} \dot{\epsilon}(t).
    \end{split}
\end{equation}
For JJ capacitors, the residual drive can be calculated based on junction size. Typically, the capacitance of {a} sandwich-shaped JJ is proportional to the junction energy{, while the total capacitance of a Josephson branch is inversely proportional to the number of JJs, i.e., $ C_{i} \propto E_{\text{J}_i} / n_{i} = r_i E_{\text{L}} / n_{i} $. By defining a reference capacitor $C_r$, we observe that the capacitors in different branches are related by the parameters $r_i$ and $n_i$ : }
\begin{equation}\label{equ: C_j for each branch of NEMS with CJ}
    \begin{split}
        C_{i} \equiv r_i C_r / n_{i}. \\
    \end{split}
\end{equation}
{It is found that the reference capacitor $ C_{r} \propto E_{\text{L}} $ with the same coefficient, indicating that $ C_{r} $ happens to be equal to the capacitor of a JJ with a Josephson energy equal to $ E_{\text{L}} $. }
So the total Lagrangian is reduced to:
\begin{equation}\label{equ: L total driven of NEMS with CJ2}
    \begin{split}
        \mathcal{L}_\text{NEMS} & = \frac{1}{2} C_\text{tot} \dot{\varphi}^2 + \sum_{i} \frac{r_i r_{\phi_{ei}} }{n_i} C_r\dot{\varphi}\dot{\epsilon}(t) - U_\text{NEMS}(\varphi), \\ 
    \end{split}
\end{equation}
where $ C_\text{tot} \equiv \left(C_\text{S} + \sum_{i} C_i \right) $ and we drop the term $ \propto \dot{\epsilon}(t)^2 $. When $ \sum_i\frac{r_i r_{\phi_{ei}} }{n_i} = 0 $, the above Lagrangian reduces back to {Eq.~(\ref{equ: L_total of NEMS}) {in the main text} with $ C_\text{tot} $ replaced by $ C_\text{S} $}. {It can be found directly that the even-order NEMS devices with symmetric double branches already satisfy the EMF canceling condition because there are always pairs of branches with the same junction size and numbers but opposite magnetic drive amplitude. In the case of NEMS-3, we verify that the EMF canceling condition is also met, as long as $ r_1 - r_3 = 0 $. Unfortunately, the  setup does not satisfy the EMF canceling condition.}     

\subsection{Magnetic driven Hamiltonian in NEMS}
\label{subsec: Magnetic driven Hamiltonian in NEMS}
When analyzing the driven potential in the main text, only the first-order expansion of the magnetic drive is kept as Eq.~(\ref{equ: U_driven of NEMS general}). A thorough analysis of the high-order expansion is required to see the residual nonlinear effect caused by a strong magnetic drive. We first rewrite the driven potential as the following unexpanded form:
\begin{equation}\label{equ: U_driven of NEMS general complete}
    \begin{split}
        & U_\text{driven}/E_\text{L} = \\
        & - \sum_{i} n_i r_i \cos\left( \frac{\varphi + \overline{\phi_{ei}}}{n_i} \right) \left( \cos\left(\frac{\delta \phi_{ei}}{n_i}\right) - 1 \right) \\
        & + \sum_{i} n_i r_i \sin\left( \frac{\varphi + \overline{\phi_{ei}}}{n_i} \right) \sin\left(\frac{\delta \phi_{ei}}{n_i}\right). 
    \end{split}
\end{equation}
For the applications we propose in the cat code experiment, the magnetic drive normally has a form of $ \epsilon(t) = \epsilon_\text{d} \cos(\omega_\text{d} t) $, with frequency $ \omega_\text{d} $ around GHz and an amplitude around 1. Therefore, the driving terms can be expanded with Bessel functions. Define $ \epsilon_\text{d}^i \equiv r_{\phi_{ei}} \epsilon_\text{d}/n_i $ as the relative driving amplitude of each loop, the Bessel expansion follows:
\begin{equation}\label{equ: Bessel function of flux drive expansion}
    \begin{split}
        \cos \frac{\delta \phi_{ei}}{n_i} &= J_0 \left( \epsilon_\text{d}^i \right) - 2J_2 \left( \epsilon_\text{d}^i \right) \cos(2\omega_\text{d} t) +\cdots,\\
        \sin \frac{\delta \phi_{ei}}{n_i} &= 2J_1 \left( \epsilon_\text{d}^i \right) \cos(\omega_\text{d} t) - 2J_3 \left( \epsilon_\text{d}^i \right) \cos(3\omega_\text{d} t) +\cdots,
    \end{split}
\end{equation}
in which $ J_n(x) $ is the n\textsuperscript{th}-order Bessel function. When the magnetic driving amplitude is relatively small, the high-order terms in $ \epsilon_\text{d}^i $ can be neglected, and then the driven potential returns to the form of Eq.~(\ref{equ: U_driven of NEMS general}). For not too large amplitude $ ( \epsilon_\text{d}^i < 0.2 ) $, we keep the expansion terms up to third order, as shown below:
\begin{equation}\label{equ: Bessel function up to 3rd of flux drive expansion}
    \begin{split}
        J_0(\epsilon_\text{d}^i) - 1 &= - \left(\epsilon_\text{d}^i/2\right)^2 + O\left(\left(\epsilon_\text{d}^i/2\right)^4\right), \\
        J_1(\epsilon_\text{d}^i) &= \epsilon_\text{d}^i/2 - \left(\epsilon_\text{d}^i/2\right)^3 / 2 + O\left(\left(\epsilon_\text{d}^i/2\right)^5\right), \\
        J_2(\epsilon_\text{d}^i) &= \left(\epsilon_\text{d}^i/2\right)^2 / 2 + O\left(\left(\epsilon_\text{d}^i/2\right)^4\right), \\
        J_3(\epsilon_\text{d}^i) &= \left(\epsilon_\text{d}^i/2\right)^3 / 6 + O\left(\left(\epsilon_\text{d}^i/2\right)^5\right).
    \end{split}
\end{equation}
Then we get the expanded potential by collecting terms with equal frequency:
\begin{equation}\label{equ: U_driven wp of NEMS}
    \begin{split}
        U_{\text{driven}}^{DC}/E_\text{L} \approx & \sum_{i} n_i r_i \cos\left( \frac{\varphi + \overline{\phi_{ei}}}{n_i} \right)\left(\frac{\epsilon_\text{d}^i}{2}\right)^2,  \\
        U_{\text{driven}}^{\omega_\text{d}} /E_\text{L} \approx & \sum_{i} n_i r_i \sin\left( \frac{\varphi + \overline{\phi_{ei}}}{n_i} \right) \left( \epsilon_\text{d}^i - \left(\frac{\epsilon_\text{d}^i}{2}\right)^3 \right) \cos(\omega_\text{d} t), \\
        U_{\text{driven}}^{2\omega_\text{d}}/E_\text{L} \approx & -\sum_{i} n_i r_i \cos\left( \frac{\varphi + \overline{\phi_{ei}}}{n_i} \right)  \left(\frac{\epsilon_\text{d}^i}{2}\right)^2 \cos(2\omega_\text{d} t), \\
        U_{\text{driven}}^{3\omega_\text{d}} /E_\text{L}\approx & \sum_{i} n_i r_i \sin\left( \frac{\varphi + \overline{\phi_{ei}}}{n_i} \right) \left(\frac{\epsilon_\text{d}}{2}\right)^3 \cos(3\omega_\text{d} t).
    \end{split}
\end{equation}

The non-zero $ U_{\text{driven}}^{DC} $ indicates that a strong magnetic drive can influence the static potential, and the influence is proportional to the square of the driving amplitude. For example, we can estimate the frequency and Kerr shift of the NEMS-3 device when a strong magnetic drive is applied:
\begin{align}
    \delta\omega \approx & \frac{r_1 + r_3 / n_{3}}{1 + r_3 / n_{3}} \frac{ \omega_\text{static} }{2}\left( - \left(0.1\epsilon_\text{d}\right)^2 \right),  \label{equ: delta omega of NEMS} \\
    \delta K \approx & \frac{r_1 + r_3 / n_{3}^3}{r_3 / n_{3}^3} K_\text{static} \left( - \left(0.1\epsilon_\text{d}\right)^2 \right). 	\label{equ: delta Kerr of NEMS-3}
\end{align}
Considering the required driving amplitude for the Kerr-cat BPCNOT gate in Table~\ref{tab: Nonlinearity comparison between NEMS ATS and SNAIL}, it is necessary to compensate for the frequency shift during gate operation.

Since $ U_\text{driven}^{\omega_\text{d}} $ depends on the static flux bias $ \overline{\phi_{ei}} $, it is easy to add extra driving terms with a small deviation from the working point. For example, a small static flux bias and doubled-frequency drive on a NEMS-3 device can be applied to generate two-photon driving Hamiltonian $ \hat{a}^2 $ for a Kerr-cat. Specifically, $ \overline{\phi_{e1}} $ can be shifted from 0 to a small angle $ \overline{\Delta\phi_{e1}} $, while keeping $ \overline{\phi_{e2}} = \pi $ and $ \overline{\phi_{e3}} = 0 $ unchanged. When the deviation is small, the static potential remains nearly unchanged, while driven potential changes to
\begin{equation}\label{equ: U_driven of NEMS-3 2ph_drive}
    \begin{split}
        \frac{U_\text{driven}}{\epsilon(t)E_J} = & r \sin(\varphi + \overline{\Delta\phi_{e1}} ) \cdot \frac{1}{5}  - 3 r\sin\left( \frac{\varphi }{3}\right) \cdot \frac{3}{5} \cdot \frac{1}{3}  \\
        \approx & - \frac{\overline{\Delta\phi_{e1}} }{5} \cdot r \cdot \frac{\varphi^2}{2!}  - \frac{8}{45} \cdot r \cdot \frac{\varphi^3}{3!} + \cdots .
    \end{split}
\end{equation}
With the extra $ \varphi^2 $ term, the deformed driven Hamiltonian follows:
\begin{equation}\label{equ: H deformed driven of NEMS-3 2ph_drive}
    \begin{split}
        \hat{\mathcal{H}}_\text{driven}^\text{deformed} & \approx \epsilon(t) g_2^\text{driven} \left( \hat{a}^\dagger + \hat{a} \right) ^2 \\
        &\quad + \epsilon(t) g_3^\text{driven} \left( \hat{a}^\dagger + \hat{a} \right) ^3 + \cdots
    \end{split}
\end{equation} 
with $ g_2^\text{driven} \propto r \overline{\Delta \phi_{e1}} $. The deformed driven Hamiltonian contains the desired two-photon terms for implementing a Kerr-cat qubit.

\subsection{Environmental coupling associated with flux drive}
\label{subsec: Environmental coupling associated with flux drive}
A major difference between flux nonlinear driving in NEMS and traditional electric driving is that the magnetic field coupling to the NEMS are inherently nonlinear {while the traditional electric driving is intrinsically linear.}
Such coupling also implies nonlinear NEMS-environment interactions, which potentially give rise to irreversible nonlinear dissipation processes in contrary to the single-photon dissipation in electric drivings.

{The NEMS-environment coupling Hamiltonian is modeled by replacing the classical magnetic driving amplitude $\epsilon_\text{d}$ of the magnetic driving Hamiltonian with the vacuum fluctuation operator $\hat{c}_{\vec{k}}$ : }
\begin{equation}\label{equ: H of Magnetic drive environment coupling}
    \begin{split}
        \hat{\mathcal{H}}_\text{I} &= \int_{ \vec{k} } \mathrm{d}\vec{k} g_{\vec{k},n}^\text{{driven}}  \left( \epsilon_{\mathrm{zpf}}(\vec{k}) \hat{c}_{\vec{k}}^{\dagger} \mathrm{e}^{\mathrm{i} \omega_c(\vec{k}) t } + h.c. \right)\\
        & \qquad\qquad\qquad  \cdot \left( \hat{a}^\dagger \mathrm{e}^{\mathrm{i} \omega_a t } + h.c. \right)^n. 
    \end{split}
\end{equation}
{Here, $\vec{k}$ represents the momentum of environment modes, i.e., the continuum modes in the magnetic driving line, and $g_{\vec{k},n}^\text{{driven}} $ is the corresponding nonlinear coupling coefficient of the NEMS device. $\epsilon_{\mathrm{zpf}}(\vec{k})$ is the amplitude of the flux throughout the whole Josephson loop area caused by the vacuum fluctuation of continuum mode $\vec{k}$, i.e., the effective magnetic driving amplitude caused by the environment. Different modes generally have different magnetic field distributions, resulting in different flux fluctuation amplitude in each loop. Consequently, $r_{\phi_{ei}}$ should change to $r_{\phi_{ei}}(\vec{k})$ for different modes, resulting in different $g_{\vec{k},n}^\text{{driven}} $.}

{The dissipation can be described by master equations with environment modes being traced out. In general, we obtain the nonlinear dissipation operator $ \kappa_n[\hat{a}^{\dagger n_1}\hat{a}^{n_2}] $ of the NEMS, with $n_1 + n_2 = n$. By engineering the mode density of the environment, we omit the contribution of modes with frequency mismatch and only keep the environment modes satisfying $\omega_c(\vec{k}) \approx n \omega_a$, resulting in the multi-photon dissipation $ \kappa_n[\hat{a}^{n}] $. For a well-designed magnetic driving line, we expect that the major modes affect each Josephson loop in the same way as we design such that $g_{\vec{k},n}^\text{{driven}} \approx g_n^\text{{driven}} $. Consequently, the dissipation strength is proportional to the squares of both the magnetic driving coefficient and the vacuum fluctuation amplitude:}
\begin{equation}\label{equ: kappa of Magnetic drive environment coupling}
    \begin{split}
        \kappa_n &\propto\abs{g_n^\text{driven}\epsilon_{\mathrm{zpf}} \left(\vec{k},\omega_c(\vec{k})=n\omega_a\right)}^2.\\
    \end{split}
\end{equation}
{Therefore, one can engineer the nonlinear dissipation by either designing the Josephson loop to change $g_n^\text{driven}$, or engineering the microwave response of the magnetic driving line to change $\epsilon_{\mathrm{zpf}}(\vec{k})$.}

\section{ATS as a special case of NEMS}
\label{sec: ATS as a special case of NEMS}

Asymmetrically Threaded SQUID (ATS) consists of only two loops, with the bias flux in each loop at the operating point being 0 and $ \pi $, respectively. The static nonlinearity of ATS at the operating point is small, and it can generate only odd-order driving Hamiltonians under magnetic flux driving. By setting the bias flux to be symmetric, with both loops at $ \pi/2 $, we obtain the Symmetrically Threaded SQUID (STS)~\cite{Kwon2022Autonomous,Bhandari2024STS}. STS is still a weakly nonlinear resonator but can generate only even-order driving Hamiltonians under magnetic flux drive. Furthermore, by setting the bias flux between ATS and STS, we can control the relative strength of odd-order and even-order magnetic drivings, which is used to implement the {small-bias ATS} for realizing a Kerr-cat qubit. Next, we will analyze the structure of ATS and how it achieves different magnetic driving Hamiltonians.

\subsection{Magnetic drive Hamiltonian}
\label{subsec: ATS Magnetic Drive Hamiltonian}

\begin{figure}
    \includegraphics{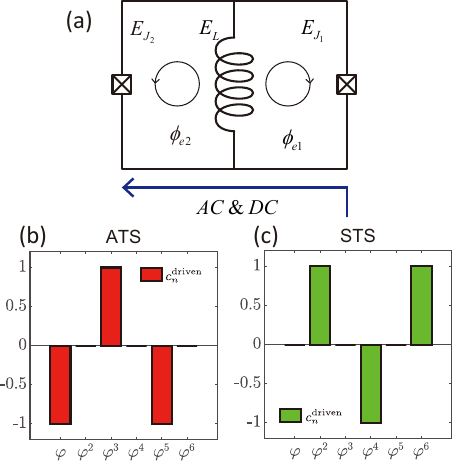}
    \caption{ (a) Circuit structure of ATS and STS. The circuit contains one inductor and {two symmetric JJ branches}, while the flux of the two loops can be biased at different working points, resulting in even or odd driving terms. The coefficients $ c_{n}^\text{driven}$ of driven terms for the ATS and STS working points, [Eq.~(\ref{equ: U static of ATS}) and  Eq.~(\ref{equ: U driven of STS})], are shown in (b) and (c), respectively. In ideal cases, the static nonlinearity vanishes for both ATS and STS working points.}
    \label{fig: ATS_circuit}
\end{figure}

The inductive part of ATS contains one large inductor and two small JJs connected in parallel. The large inductor is formed by several large JJs in series and possesses a small static nonlinearity. The inductive energy of ATS is written as:
\begin{equation}\label{equ: U of ATS}
    \begin{split}
        \frac{U_\text{ATS}}{E_\text{L}}  = & \frac{1}{2} \varphi^2 - r_1 \cos (\varphi + \phi_{e1} ) - r_2 \cos (\varphi + \phi_{e2} ), \\
    \end{split}
\end{equation}
where $ E_\text{L} $ is the inductor energy and $ r_{1,2} $ is the ratio between the small JJ energy and the inductor energy. 

To control the nonlinearity of ATS, we generally set the sizes of the two small JJs to be the same, i.e., $ r_1 = r_2 $. For the static potential energy, it is easy to see that if the static magnetic flux biases of the two single-JJ branches differ by $ \pi $, i.e., $ \abs{\overline{\phi_{e1}}-\overline{\phi_{e2}}} = \pi $, the contribution of the two small JJs cancel each other:
\begin{equation}\label{equ: U static of ATS}
    \begin{split}
        U_\text{ATS}^\text{static} / E_\text{L} = \frac{1}{2} \varphi^2.
    \end{split}
\end{equation}	
So the oscillator will have only a relatively weak static Kerr nonlinearity due to the finite number of junctions in the inductive branch. For non-ideal cases, the condition $ r_1 = r_2 $ may not be satisfied due to fabrication errors. Then the static potential will be like SNAIL with an effective small JJ part $ -\tilde{E}_{sJ}\cos( \varphi + \tilde{\phi}_{E_\text{J}} ) $, where $ \tilde{E}_{sJ} $ and $ \tilde{\phi}_{E_\text{J}} $ are the effective junction size and external flux bias. Although the junction size is subject to fabrication uncertainties, the flux bias can be adjusted as needed after sample preparation. For applications requiring a small Kerr, the residual single-JJ part can be used to compensate for the nonlinearity from the inductor branch to fully eliminate static Kerr. For applications requiring three-wave mixing, the residual single-JJ part can be used to generate the $ g_3 $ nonlinearity on demand.  

When applying alternating magnetic drive, the driven potential can be expressed as:
\begin{equation}\label{equ: U total driven of ATS}
    \begin{split}
        &U_\text{ATS}^\text{driven}/E_\text{L} = \\
        & \quad - \sum_{i = 1,2} r_i \cos (\varphi + \overline{\phi_{ei}} ) \left( \cos( \delta\phi_{ei} ) - 1 \right) \\
        & \quad + \sum_{i = 1,2} r_i \sin (\varphi + \overline{\phi_{ei}} ) \sin( \delta\phi_{ei} ).\\
    \end{split}
\end{equation}

When $ \overline{\phi_{e1}} = 0 $, the device generates odd-order terms of $ \varphi $ under the magnetic drive $ \epsilon(t) $, which is the normal ATS working point:
\begin{equation}\label{equ: Bias and drive of ATS}
    \begin{split}
        \overline{\phi_{e1}} = & 0, \quad \delta\phi_{e1} = \epsilon(t), \\ 
        \overline{\phi_{e2}} = & \pi, \quad \delta\phi_{e2} = - \epsilon(t). \\
    \end{split}
\end{equation}
The static potential is the same as Eq.~(\ref{equ: U static of ATS}), while the driven potential is given by:
\begin{equation}\label{equ: U driven of ATS}
    \begin{split}
        U_\text{ATS}^\text{driven}/E_\text{L} = 2r \sin(\varphi)\epsilon(t).
    \end{split}
\end{equation}
The driven potential contains only odd-order terms. The ATS working point can be treated as a simple case of odd-order engineering.

When $ \overline{\phi_{e1}} = \pi/2 $, the device generates even-order terms of $ \varphi $ under the magnetic drive $ \epsilon(t) $. Here we refer to this working point as the symmetric threaded SQUID (STS) point because the flux bias is symmetric:
\begin{equation}\label{equ: Bias and drive of STS}
    \begin{split}
        \overline{\phi_{e1}} = & \pi/2, \quad \delta\phi_{e1} = \epsilon(t), \\ 
        \overline{\phi_{e2}} = & -\pi/2, \quad \delta\phi_{e2} = -\epsilon(t). \\
    \end{split}
\end{equation}
The static potential is the same as Eq.~(\ref{equ: U static of ATS}), while the driven potential is given by:
\begin{equation}\label{equ: U driven of STS}
    \begin{split}
        U_\text{STS}^\text{driven} & /E_\text{L} =  2r \cos (\varphi ) \epsilon(t).
    \end{split}
\end{equation}
The driven potential contains only even-order terms that can be used to realize a two-photon or four-photon drive for two-cat or four-cat stabilization. Notice that the two single-JJ branches have the same JJ size and opposite flux bias, {therefore, we can treat the STS point as a simple case of the symmetric double branches for even-order engineering.}

When the flux bias $ \overline{\phi_{ei}} $ is between $ 0 $ and $ \pi/2 $, the device can generate both even and odd terms in $ \varphi $ simultaneously. For Kerr-cat implementation, we need a strong $ \varphi^3 $ driven term for the BPCNOT gate and a not-too-small $ \varphi^2 $ driven term for the two-photon drive. A small deviation $ \overline{\Delta\phi_{e1}} = 0.05\,\pi $ from the ATS working point is sufficient, with a driven potential
\begin{equation}\label{equ: U driven of ATS 2 Cat}
    \begin{split}
        U_\text{ATS}^\text{driven} & /E_\text{L} =  2r \epsilon(t) \left( \sin ( \varphi ) + \cos( \varphi ) \overline{\Delta\phi_{e1}} \right).
    \end{split}
\end{equation}
Note that $ \overline{\phi_{e2}} $ should be changed to $ 1.05\,\pi $ to satisfy the condition of canceling static nonlinearity. The small Kerr term required by the Kerr-cat qubit can be obtained from the residual nonlinearity in the inductor branch. Therefore, ATS under a slightly varied flux bias point meets the requirement of Kerr-cat stabilization and BPCNOT gate implementation, except for the problem of detrimental first-order residual terms.

\subsection{Residual electromotive force}
\label{subsec: ATS Residual Electric Drive}

A residual electric drive may come from an alternating magnetic field due to voltage fluctuations across the capacitors caused by EMF ~\cite{You2019Circuit,Riwar2022Circuit,Yao2023Highfidelity,Bryon2023FluxDriveFluxonium}. To minimize the residual electric drive, consideration must be given to both geometric and JJ capacitors. The residual drive on geometric capacitors can be mitigated by optimizing geometry using electromagnetic simulation methods. For JJ capacitors, the residual drive can be calculated based on junction size. Typically, the capacitance of sandwich-shaped JJs is proportional to the junction energy, providing the opportunity to compute the electromagnetic force on the JJs capacitor.

Now we analyze the EMF on the JJs in the ATS device. When the capacitors of JJs are included, we can write the Lagrangian of capacitively shunted ATS as:
\begin{equation}\label{equ: L total driven of ATS with CJ}
    \begin{split}
        \mathcal{L}_\text{ATS} & =\frac{1}{2}C_\text{S} \dot{\varphi}^2 + \sum_{i = 1,2}\frac{1}{2}C_i \dot{\varphi_i}^2 - U_\text{ATS}(\varphi), \\ 
    \end{split}
\end{equation}
where {$C_i$ and $\varphi_i $ are the capacitance and the flux variable associated with the JJ branch, respectively,} and $ \varphi $ {is the generalized phase variable across the inductor branch}. The relation between $ \varphi $ and $\varphi_i $ depends on the external flux:
\begin{equation}\label{equ: phi_j for each branch of ATS with CJ}
    \begin{split}
        \varphi_1 & = \varphi + \phi_{e1}, \dot{\varphi_1} = \dot{\varphi} + \dot{\epsilon}(t), \\
        \varphi_2 & = \varphi + \phi_{e2}, \dot{\varphi_2} = \dot{\varphi} - \dot{\epsilon}(t). \\
    \end{split}
\end{equation}
Here we already consider the flux operating scheme described in Eqs.~(\ref{equ: Bias and drive of ATS}-\ref{equ: Bias and drive of STS}). {The capacitance of each Josephson branch is proportional to the junction energy. By defining $ C_r $ as the capacitance reference, the capacitance of each branch can be calculated correspondingly:}
\begin{equation}\label{equ: C_j for each branch of ATS with CJ}
    \begin{split}
        C_i = r_i C_r. \\
    \end{split}
\end{equation}
{It is shown that the reference capacitance $ C_{r} \propto E_{\text{L}} $ with the same coefficient, meaning that $ C_{r} $ happens to be equal to the capacitance of a JJ with a Josephson energy equal to $ E_{\text{L}} $.} Then the total Lagrangian is reduced to:
\begin{equation}\label{equ: L total driven of ATS with CJ2}
    \begin{split}
        \mathcal{L}_\text{ATS} & = \frac{1}{2} C_\text{tot} \dot{\varphi}^2 + \left( r_1 - r_2 \right)C_r\dot{\varphi}\dot{\epsilon}(t) - U_\text{ATS}(\varphi), \\ 
    \end{split}
\end{equation}
where $ C_\text{tot} \equiv \left(C_\text{S} + \left(r_1 + r_2 \right) C_r \right) $ and we drop the term $ \propto \dot{\epsilon}(t)^2 $. When $ r_1 = r_2 $, the Lagrangian {contains no extra $\dot{\varphi}$ terms}. {So with the current JJ size and magnetic drive scheme, the ATS device has no extra {EMF} associated with the JJ capacitors.}

\section{Implementing Kerr-cat BPCNOT gate with NEMS-3 }
\label{sec: Implementing Kerr-cat BPCNOT gate with NEMS-3 }

The Kerr-cat encoding scheme stabilizes the cat state through the combined effects of Kerr nonlinearity and two-photon driving, as described by the Hamiltonian:
\begin{equation}\label{equ: H of Kerr-Cat}
    \begin{split}
        \hat{\mathcal{H}}_{KC} &= K \alpha ^ 2  \hat{a}^{\dagger 2} - K \alpha^{*2}  \hat{a}^2 + K \hat{a}^{\dagger 2} \hat{a}^2 \\
        &= - K \left( \hat{a}^{\dagger 2} - \alpha^{*2} \right) \left( \hat{a}^2 - \alpha^2 \right)+ K \alpha^{*2} \alpha^2.
    \end{split}
\end{equation}
The two degenerate ground states of the system are given by $ \ket{\mathcal{C}^{\pm}_{\alpha}} = \mathcal{N}_{\pm} \left( \ket{\alpha} \pm \ket{-\alpha} \right)$, with $ \mathcal{N}_{\pm} $ defined as the normalization factor. The two-cat qubit is encoded on two basis states $ \ket{0} \equiv \frac{ \ket{\mathcal{C}^{+}_{\alpha}} + \ket{\mathcal{C}^{-}_{\alpha}} }{\sqrt{2}} $ and $ \ket{1} \equiv \frac{ \ket{\mathcal{C}^{+}_{\alpha}} - \ket{\mathcal{C}^{-}_{\alpha}} }{\sqrt{2}} $. 

The two-qubit interaction Hamiltonian $\hat{\mathcal{H}}=\hat{\mathcal{H}}_1+\hat{\mathcal{H}}_2$ for the BPCNOT gate is realized by a controlled rotation on the target cat qubit depending on the state of the control cat qubit with
\begin{equation}\label{equ: H of Kerr-cat BPCX}
    \begin{split}
        \hat{\mathcal{H}}_1 = &- K \left( \hat{a}_1^{\dagger2} - \alpha_1^2 \right) \left( \hat{a}_1^2 - \alpha_1^2 \right), \\
        \hat{\mathcal{H}}_2 = &- K \left( \hat{a}_2^{\dagger2} - \frac{\alpha_2^2}{2\alpha_1} \left( \alpha_1 + \hat{a}_1^\dagger \right) - \frac{ \alpha_2^2 e^{2i\phi(t)} }{2\alpha_1} \left( \alpha_1 - \hat{a}_1^\dagger \right) \right) \\
        &\times \left( \hat{a}_2^2 - \frac{\alpha_2^2}{2\alpha_1} \left( \alpha_1 + \hat{a}_1 \right) - \frac{ \alpha_2^2 e^{-2i\phi(t)} }{2\alpha_1} \left( \alpha_1 - \hat{a}_1 \right) \right).
    \end{split}
\end{equation}
Here $ \phi(t) $ adiabatically changes from 0 to $ \pi $ during the gate operation.

In this section, we illustrate the feasibility of implementing Kerr-cat qubits with NEMS-3, supported by numerical analysis. We first present the technique for creating individual Kerr-cat qubits using magnetic drives and static flux bias. Then, we examine the implementation of the BPCNOT gate between two Kerr-cat qubits. By conducting a numerical comparison with ATS and SNAIL, we illustrate that NEMS-3 can easily implement the Kerr-cat BPCNOT, mitigating the impact of residual terms or large pumping amplitude.

\subsection{Single Kerr-cat qubit with deformed NEMS-3 and ATS } 
\label{subsec: Single Kerr-cat Qubit with deformed NEMS-3 and ATS }  

For NEMS-3, the two-photon drive $ \hat{a}^2 + \hat{a}^{\dagger^{2}} $ is not directly realizable using magnetic drives, because the driven Hamiltonian Eq.~(\ref{equ: H_driven of NEMS-3}) lacks even-photon interactions. To circumvent this, we deform the driven Hamiltonian with a static flux bias to induce even-order terms, as described in Eqs.~(\ref{equ: U_driven of NEMS-3 2ph_drive}-\ref{equ: H deformed driven of NEMS-3 2ph_drive}). The deformation results in a magnetic-driven Hamiltonian with even-photon interactions:
\begin{equation}\label{equ: H of Magnetic 2ph_drive}
    \begin{split}
        \hat{\mathcal{H}}_{KC}^\text{mag} & = \left(\frac{1}{2}\epsilon_\text{d} \mathrm{e}^{-\mathrm{i} \omega_\text{d} t }  + c.c. \right) g_2^\text{driven} \left( \hat{a} \mathrm{e}^{-\mathrm{i} \omega_a t } +h.c. \right) ^2 +\cdots\\
        & = \epsilon_\text{d} g_2^\text{driven} \hat{a}^{\dagger2} + h.c.
    \end{split}
\end{equation} 
{Here we have already considered the frequency matching condition that $ \omega_\text{d} = 2 \omega_a $,} with $\omega_a$ being the mode frequency. 

Alternatively, we can use an electric drive, wherein the three-photon interaction comes from the static fourth-order nonlinearity, yielding an electric-driven Hamiltonian:
\begin{equation}\label{equ: H of Electric 2ph_drive}
    \begin{split}
        \hat{\mathcal{H}}_{KC}^\text{ele} &= g_3^\text{static} \left( \hat{a} \mathrm{e}^{-\mathrm{i} \omega_a t } + \xi_p\mathrm{e}^{-\mathrm{i} \omega_\text{d} t } + h.c. \right)^3 \\
        &\approx 3 \xi_p g_3 \hat{a}^{\dagger2} + h.c.
    \end{split}
\end{equation}
The static third-order nonlinearity may come from a deformed static potential. {An odd-order static nonlinearity will break the symmetry of the static potential, resulting in a shift of potential minimum position, which is hard to estimate analytically. Consequently, in this case, we can only obtain the nonlinear coefficient $g_3$ through numerical simulations.}

{ To obtain a finite Kerr nonlinearity for Kerr-cat implementation, we release the Kerr-cancelling condition Eq.~(\ref{equ: c_static target of NEMS-3}) of a NEMS-3 device to allow a finite fourth-order static nonlinearity. Meanwhile, the inductor branch may also provide residual fourth-order static nonlinearity coming from the finite number $ n_\text{L} $ of large JJs. The deviation of flux bias point also results in a small third-order nonlinearity. Consequently, the NEMS-3 device for Kerr-cat implementation is designed to have small extra third- and fourth-order static Hamiltonians:
\begin{equation}\label{equ: H_Kerr of NEMS-3 cat}
    \begin{split}
        H_\text{static} \approx g_3^\text{static} \left( \hat{a}^\dagger + \hat{a} \right)^3 + g_4^\text{static} \left( \hat{a}^\dagger + \hat{a} \right)^4.
    \end{split}
\end{equation}
Recall that the nonlinear interaction coefficients are defined as $g_n^\text{static}= \frac{1}{n!} E_\text{J} c_{n}^\text{static} \varphi_{\mathrm{zpf}}^n$. Here $ c_{3,4}$ has small but nonzero values. According to Ref.~\cite{Hillmann2022Designing}, both the third- and fourth-order static nonlinear terms contribute to the effective Kerr nonlinearity:
\begin{equation}\label{equ: nonlinearity_harmonic of NEMS}
    \begin{split}
        K_{\text{static}} = - 6 g_4^\text{static} + 30\frac{ \left(g_3^\text{static}\right)^2}{\omega}.
    \end{split}
\end{equation}
Here $ K_{\text{static}} $ is defined as the coefficient of operator $ \hat{a}^\dagger \hat{a}^\dagger \hat{a} \hat{a}$. The small residual nonlinearity should be carefully addressed in numerical simulations.}

\subsection{Coupled two Kerr-cat qubits}
\label{subsec: Coupled Two Kerr cat qubits}

Constructing a two-qubit gate requires a coupling between the two bosonic modes. Here we present a simple case of direct capacitive coupling.
The Hamiltonian of the coupled system can be written as the sum of a linear part and a nonlinear part:
\begin{equation}\label{equ: H of coupled NEMS}
    \begin{split}
        \hat{\mathcal{H}}_\text{L} &= \omega_{1} \hat{a}_1^\dagger \hat{a}_1 + \omega_{2} \hat{a}_2^\dagger \hat{a}_2 + g_{12} \left( \hat{a}_1^\dagger \hat{a}_2 + \hat{a}_2^\dagger \hat{a}_1 \right), \\
        \hat{\mathcal{H}}_{NL} &= K_{1} \hat{a}_1^\dagger \hat{a}_1^\dagger \hat{a}_1 \hat{a}_1 + K_{2} \hat{a}_2^\dagger \hat{a}_2^\dagger \hat{a}_2 \hat{a}_2. \\
    \end{split}
\end{equation}
Here, we only consider the static Hamiltonian Eq.~(\ref{equ: H_static of NEMS-3}) with Kerr nonlinearity. For linear capacitive coupling, $ g_{12} $ is independent of the magnetic drive. The linear part can be diagonalized with the Bogoliubov transformation:
\begin{equation}\label{equ: Bogoliubov transformation}
    \hat{U}_\text{BT} \equiv \exp(\Lambda\left( \hat{a}_1^\dagger \hat{a}_2 - \hat{a}_2^\dagger \hat{a}_1 \right)),
\end{equation}
which transfers the oscillator ladder operator as:
\begin{equation}\label{equ: Bogoliubov transformation2}
    \begin{split}
        \hat{U}_\text{BT}^\dagger \hat{a}_1 \hat{U}_\text{BT} &= \cos(\Lambda ) \hat{a}_1 + \sin(\Lambda) \hat{a}_2, \\
        \hat{U}_\text{BT}^\dagger \hat{a}_2 \hat{U}_\text{BT} &= \cos(\Lambda ) \hat{a}_2 - \sin(\Lambda) \hat{a}_1. 
    \end{split}	
\end{equation}
If we choose $ \Lambda = \frac{1}{2}\arctan(2\abs{g_{12}/\Delta}) $ with $ \Delta \equiv \omega_{1} - \omega_{2} $, then the Bogoliubov transformation diagnalizes the linear Hamiltonian $ \hat{\mathcal{H}}_\text{L} $:
\begin{equation}\label{equ: H_BT linear}
    \hat{U}_\text{BT}^\dagger \hat{\mathcal{H}}_\text{L} \hat{U}_\text{BT} = \tilde{\omega}_{1} \hat{a}_1^\dagger \hat{a}_1 + \tilde{\omega}_{2} \hat{a}_2^\dagger \hat{a}_2,
\end{equation}
with the dressed frequencies $ \tilde{\omega}_{1,2} = \frac{1}{2} ( \omega_{1} + \omega_{2} \pm \sqrt{4g_{12}^2 + \Delta^2 } ) $. 

For the nonlinear part, we apply the same transformation to the nonlinear Hamiltonian $ \hat{\mathcal{H}}_{NL} $:
\begin{equation}\label{equ: H_BT nonlinear}
    \begin{split}
        \hat{U}_\text{BT}^\dagger \hat{\mathcal{H}}_{NL} \hat{U}_\text{BT} &\approx \tilde{K}_{1} \hat{a}_1^\dagger \hat{a}_1^\dagger \hat{a}_1 \hat{a}_1 + \tilde{K}_{2} \hat{a}_2^\dagger \hat{a}_2^\dagger \hat{a}_2 \hat{a}_2 \\
        &\qquad + \chi_{12} \hat{a}_1^\dagger \hat{a}_1 \hat{a}_2^\dagger \hat{a}_2.
    \end{split}
\end{equation}

The Kerr and cross-Kerr coefficients are given by:
\begin{equation}\label{equ: K_BT and chi_BT nonlinear}
    \begin{split}
        \tilde{K}_1 &\equiv K_1 \cos(\Lambda)^4 + K_2 \sin(\Lambda)^4, \\
        \tilde{K}_2 &\equiv K_2 \cos(\Lambda)^4 + K_1 \sin(\Lambda)^4, \\
        \chi_2 &\equiv 2\left( K_1 + K_2 \right) \cos(\Lambda)^2 \sin(\Lambda)^2.\\
    \end{split}
\end{equation}

For the driving Hamiltonian, applying the same transformation gives us the coupled two-mode drive. For example, the magnetic-driven Hamiltonian Eq.~(\ref{equ: H deformed driven of NEMS-3 2ph_drive}) in $\hat{a}_2$ mode transfers to a simultaneous magnetic drive on two modes:
\begin{equation}
    \label{equ: H_BT nonlinear magnetic drive}
    \begin{split}
        \hat{U}_\text{BT}^\dagger &\hat{\mathcal{H}}_\text{driven} \hat{U}_\text{BT} \\
        \approx & \sum_{n} g_n^\text{driven} \left(\frac{1}{2}\epsilon_\text{d}\mathrm{e}^{-\mathrm{i} \omega_\text{d} t } + c.c. \right)  \\
        &\quad \times \left( \cos(\Lambda)\hat{a}_2 - \sin(\Lambda)\hat{a}_1+ h.c. \right)^n.
    \end{split}
\end{equation}
In case of realizing a BPCNOT gate, it requires the drive to satisfy the frequency matching condition $ \omega_\text{d} = 2\omega_2 - \omega_1 $, {with $\hat{a}_2$ as the target cat qubit and $\hat{a}_1$ as the control qubit}. The driven third-order terms mainly contribute in the magnetic BPCNOT gate. Under RWA, we neglect fast rotating terms and obtain the effective BPCNOT Hamiltonian with magnetic drive:
\begin{equation}\label{equ: H_BPCNOT with magnetic drive}
    \begin{split}
        \hat{\mathcal{H}}_\text{BPCNOT}^\text{mag} \approx \frac{3}{2}g_3^\text{driven} \epsilon_\text{d} \cos(\Lambda)^2\sin(\Lambda) \hat{a}_2^{\dagger2} \hat{a}_1 + h.c.
    \end{split}
\end{equation}
Similarly, the electric-driven Hamiltonian in one mode transfers to a simultaneous electric drive on two modes:
\begin{equation}\label{equ: H_BT nonlinear electric drive}
    \begin{split}
        &\hat{U}_\text{BT}^\dagger \hat{\mathcal{H}}_\text{ele} \hat{U}_\text{BT} \\
        &\approx \sum_n g_n^\text{static}\left( \cos(\Lambda)\hat{a}_2 - \sin(\Lambda)\hat{a}_1+ \xi_p\mathrm{e}^{-\mathrm{i} \omega_\text{d} t } + h.c. \right)^n.
    \end{split}
\end{equation}	
The frequency matching condition for a BPCNOT gate is the same as the magnetic drive. The static fourth-order terms mainly contribute in the electric BPCNOT gate. Under RWA, we neglect fast rotating terms and obtain the effective BPCNOT Hamiltonian with electric drive:
\begin{equation}\label{equ: H_BPCNOT with electric drive}
    \begin{split}
        \hat{\mathcal{H}}_\text{BPCNOT}^\text{ele} \approx 12 g_4^\text{static} \xi_d \cos(\Lambda)^2\sin(\Lambda) \hat{a}_2^{\dagger2} \hat{a}_1 + h.c.
    \end{split}
\end{equation}
The systems designed for cat-qubit experiments generally have small fourth-order static nonlinearity, making it challenging to realize a three-photon interaction with an electric drive.

\subsection{ Numerical Simulation }
\label{subsec: Numerical Simulation}

\begin{table*}
    \renewcommand{\arraystretch}{1.2}
    \begin{tabular}{c|c|c|c} 
    \hline
                & NEMS-3	& ATS	& SNAIL
    \\
    \hline
    \multicolumn{4}{c}{ Circuit Parameter }
    \\
    \hline
    Inductor Branch $ E_{\text{J}_\text{L}}, n_\text{L} $ 
        & $ E_{\text{J}_\text{L}} = 90~\text{GHz}, n_\text{L} = 5 $		
        & $ E_{\text{J}_\text{L}} = 120~\text{GHz}, n_\text{L} = 5 $	
        & $ E_{\text{J}_\text{L}} = 114~\text{GHz}, n_\text{L} = 3 $	
    \\ 
    Linear Capacitor $ E_\text{C} $				& 200~MHz	& 200~MHz	& 200~MHz	
    \\ 
    Josephson Branches $ E_{\text{J}_i},n_i $		
        & \makecell{$ E_{\text{J}_1} = 18.0~\text{GHz}, n_1 = 1 $ \\ $ E_{\text{J}_2} = 18.9~\text{GHz}, n_2 = 1 $ \\ $ E_{\text{J}_3} = 18.0~\text{GHz}, n_3 = 3 $}		
        & \makecell{$ E_{\text{J}_1} = 24.0~\text{GHz} , n_1 = 1 $ \\ $ E_{\text{J}_2} = 24.0~\text{GHz}, n_2 = 1  $ }		
        & $ E_{\text{J}_1} = 11.4~\text{GHz} , n_1 = 1 $		
    \\ 
      Flux Control			
        & \makecell{$ \phi_{e1} = 0.05\pi + 0.2\epsilon(t) $ \\ $ \phi_{e2} = 1.05\pi $ \\ $ \phi_{e3} = 0 + 0.6\epsilon(t) $}  
        & \makecell{$ \phi_{e1} = 0.05\pi + 0.5\epsilon(t) $ \\ $ \phi_{e2} = 1.05\pi - 0.5\epsilon(t) $ } 
        & $ \phi_{e} = 0.78\pi $ 
    \\ 		
    \hline
    \multicolumn{4}{c}{ Static Hamiltonian } 
    \\
    \hline
    Frequency			& 6.08~GHz	& 6.19~GHz	& 6.05~GHz 
    \\  
    $ \varphi_{\mathrm{zpf}} $	& 0.36		& 0.36		& 0.36 
    \\  
    $ g_3 (\hat{a}^\dagger + \hat{a})^3$ & 1.09~MHz			& 0			& -48~MHz 
    \\  
    $ g_4 (\hat{a}^\dagger + \hat{a})^4$ & -0.35~MHz	& -0.66~MHz	& 1.38~MHz 
    \\ 
    \hline
    \multicolumn{4}{c}{ Flux Driven Hamiltonian } 
    \\ 
    \hline
    $ g_1^\text{driven} \epsilon_\text{d} (\hat{a}^\dagger + \hat{a})$ & 17.3~MHz $ \cdot \epsilon_\text{d} $			& 8.5~GHz $ \cdot \epsilon_\text{d} $		&  
    \\    
      $ g_2^\text{driven} \epsilon_\text{d} (\hat{a}^\dagger + \hat{a})^2$ & 38.3~MHz $ \cdot \epsilon_\text{d} $			& 242~MHz $ \cdot \epsilon_\text{d} $		&  
    \\  
      $ g_3^\text{driven} \epsilon_\text{d} (\hat{a}^\dagger + \hat{a})^3$ & 25.0~MHz $ \cdot \epsilon_\text{d} $			& 183~MHz $ \cdot \epsilon_\text{d} $		&  
    \\ 
    \hline 
    \multicolumn{4}{c}{ Kerr-cat   Parameter   $ \overline{n} = 4 $  $ g/\Delta = 0.1 $ }
    \\
    \hline
      Kerr				& -2.2~MHz	& -4.0~MHz	& -3.1~MHz 
    \\  
      Two-photon Drive 			& $ \epsilon_\text{d} = 0.45 $ & $ \epsilon_\text{d} = 0.13 $ & $ \xi = 0.09 $ 
    \\
      BPCNOT Drive 		& $ \epsilon_\text{d} = 1.15 $ & $ \epsilon_\text{d} = 0.29 $ & $ \xi = 3.7 $
    \\
      Residual 1ph Drive $ \Omega_1^\mathrm{res}\hat{a} + h.c. $  & 10~MHz & 1.2~GHz & 
        \\
    Residual 2ph Drive $ \Omega_2^\mathrm{res}\hat{a}^2 + h.c. $ & 22~MHz & 35~MHz & 
    \\
    \hline
    \end{tabular}
    \caption{A numerical comparison of NEMS, ATS, and SNAIL in implementing Kerr-cat qubits. We choose similar $ \varphi_{\mathrm{zpf}} $ to maintain consistent scaling ratios for nonlinear coefficients. The coupling strength between the cat modes is set to $ g/\Delta \approx 0.1 $. Implementing the BPCNOT gate with SNAIL demands a large driving strength, which is difficult to achieve. The ATS implementation introduces a strong first-order residual term, resulting in additional errors and dissipation. The NEMS-3 demands a moderate driving strength while effectively suppressing the residual terms, therefore suitable for BPCNOT gate implementation.}
    \label{tab: Nonlinearity comparison between NEMS ATS and SNAIL}
\end{table*}

Here, we demonstrate the advantages of NEMS-3 as a Kerr-cat implementation by numerical comparison with ATS and SNAIL. We estimate the necessary driving strengths and low-order residual terms in Kerr-cat stabilization and operations across different scenarios. Here we use a magnetic drive for ATS and NEMS-3 and an electric drive for SNAIL.     
To maintain a consistent scaling ratio for nonlinear coefficients of each order, We set similar $ \varphi_{\mathrm{zpf}} $ across different scenarios. Additionally, the coupling strength is fixed at $ g_{12}/ \Delta \approx 0.1 $. 

The numerical estimation results are summarized in Table~\ref{tab: Nonlinearity comparison between NEMS ATS and SNAIL}. SNAIL demands a relatively large driving strength ($ \xi>1 $) for the BPCNOT gate, which is challenging in experimental realization. Moreover, the emergence of complex higher-order nonlinear terms when $ \xi>1 $ may significantly impact the system~\cite{Yaxing2019Engineering,Chapman2023SNAILBeamSplitter}. For ATS, a strong residual single-photon drive emerges during BPCNOT operations, leading to additional excitation and dissipation. 

\begin{figure}
    \centering
    \includegraphics{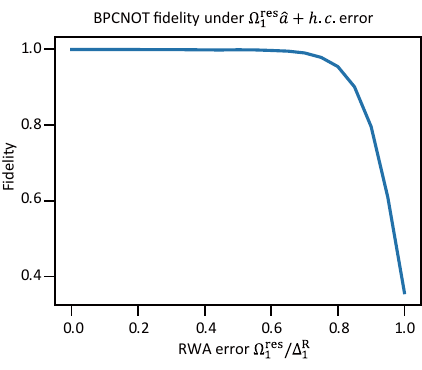}
    \caption{The BPCNOT gate fidelity with respect to the residual first-order term $ \Omega_1^{\text{res}}\hat{a} + h.c.$ The frequency difference $ \Delta_1^\text{R} $ is set to 500~MHz, a moderate value in experiments. The gate fidelity decreases rapidly as the residual first-order term $ \Omega_1^{res} $ approaches the detuning $ \Delta_1^\text{R} $, indicating a breakdown of the RWA condition. To guarantee the gate fidelity, the upper-bound of the residual first-order term is approximately 300~MHz, which is unreachable for ATS. }
    \label{fig: BPCNOT_fidelity_under_relative_g1_terms}
\end{figure}

The strong residual drive in ATS implementation may induce undesired excitation, leading to additional errors. Figure \ref{fig: BPCNOT_fidelity_under_relative_g1_terms} illustrates the impact of the residual term on the BPCNOT gate fidelity under numerical simulation. The simulation uses the data from Table~\ref{tab: Nonlinearity comparison between NEMS ATS and SNAIL}, with the detuning set to $ \Delta = $ 500~MHz. {The frequency of the residual first-order driving term in the rotating frame is:}
\begin{equation}
    \Delta_1^{\text{R}} = \omega_2 - \omega_\text{d} = \omega_1 - \omega_2 = \Delta.
\end{equation}
It can be inferred from Fig.~\ref{fig: BPCNOT_fidelity_under_relative_g1_terms}(a) that the gate fidelity decreases rapidly as the residual first-order term $ \Omega_1^{res} $ approaches the detuning $ \Delta_1^{\text{R}} $, indicating a breakdown of the RWA condition. Consequently, for the ATS implementation scheme, the detuning between adjacent qubits must be larger than the amplitude of the residual first-order term, which is hard to achieve in the multi-qubit situation. Theoretically, it is possible to balance the residual terms by applying driving terms with opposite phases, such as applying a single-photon electric drive through another port to balance the residual first-order term. However, the compensation requires large pulse amplitude and precise timing, potentially leading to additional problems like heating and crosstalk. In contrast, NEMS offers the advantage of controlling the residual term amplitude through nonlinearity engineering, thereby enhancing the gate fidelity and alleviating the frequency congestion problem.

The strong residual driving terms also cause additional dissipation through the magnetic driving line, as discussed in Appendix Sect.~\ref{subsec: Environmental coupling associated with flux drive}. Moreover, the nonlinearity of the magnetic driving terms implies that the environmental coupling can exhibit nonlinear behavior, resulting in a nonlinear dissipation. Applying reverse pulses cannot suppress the additional dissipation, since the coupling is always on and the environmental fluctuations occur randomly. Dissipation through a magnetic driving line can only be suppressed by designing the field distribution $ \epsilon_{\mathrm{zpf}} $ and the coupling coefficient $ g_n^\text{driven} $. The NEMS-3 device has smaller residual driving terms, effectively suppressing environmental dissipation.

\section{ Stablizing four-cat states }

\subsection{ Numerical results of stabilizing four-cat states with NEMS-5 }
\label{subsec: Numerical results of stabilizing four-cat states with NEMS-5}

\begin{table*}
    \renewcommand{\arraystretch}{1.2}
    \begin{tabular}{c|c|c|c}
        \hline
        &  NEMS-5  		 &    		NEMS-3 &    		ATS 
        \\ 
        \hline
        \multicolumn{4}{c}{ Circuit Parameter }
        \\
        \hline
        Inductor Branch $ E_{\text{J}_\text{L}}, n_\text{L} $  &
        $ E_{\text{J}_\text{L}} = 180~\text{GHz}, n_\text{L} = 10 $ &
        $ E_{\text{J}_\text{L}} = 180~\text{GHz}, n_\text{L} = 10 $ &
        $ E_{\text{J}_\text{L}} = 180~\text{GHz}, n_\text{L} = 10 $ 
        \\
        Capacitor $ E_\text{C} $ &
        246~MHz &
        194~MHz &
        151~MHz 
        \\
        \begin{tabular}[c]{@{}c@{}}Josephson Branch\\ $ E_{\text{J}_i}, n_i$\end{tabular} &
        \begin{tabular}[c]{@{}c@{}}$ E_{\text{J}_1} = 2.88~\text{GHz}, n_1 = 1 $\\ $ E_{\text{J}_2} = 18.0~\text{GHz}, n_2 = 2 $\\ $ E_{\text{J}_3} = 15.12~\text{GHz}, n_3 = 3 $\end{tabular} &
        \begin{tabular}[c]{@{}c@{}}$ E_{\text{J}_1} = 18.0~\text{GHz}, n_1 = 1 $\\ $ E_{\text{J}_2} = 18.9~\text{GHz}, n_2 = 1 $\\ $ E_{\text{J}_3} = 18.0~\text{GHz}, n_3 = 3 $\end{tabular} &
        \begin{tabular}[c]{@{}c@{}}$ E_{\text{J}_1} = 18.0~\text{GHz}, n_1 = 1 $\\ $ E_{\text{J}_2} = 18.0~\text{GHz}, n_2 = 1 $\end{tabular} 
        \\
        \begin{tabular}[c]{@{}c@{}}Flux Control\\ Scheme\end{tabular}  &
        \begin{tabular}[c]{@{}c@{}}$ \phi_{e1} = \pi + 0.2\epsilon(t) $\\ $ \phi_{e2} = 0 + 0.4\epsilon(t) $\\ $ \phi_{e3} = 0 + 0.6\epsilon(t) $\end{tabular} &
        \begin{tabular}[c]{@{}c@{}}$ \phi_{e1} = 0 + 0.2\epsilon(t) $\\ $ \phi_{e2} = \pi$\\ $ \phi_{e3} = 0 + 0.6\epsilon(t)$\end{tabular} &
        \begin{tabular}[c]{@{}c@{}}$ \phi_{e1} = 0 + 0.5\epsilon(t) $\\ $ \phi_{e2} = \pi - 0.5\epsilon(t)$\end{tabular} 
        \\ 
        \hline
        \multicolumn{4}{c}{ Static Hamiltonian }
        \\
        \hline
        Frequency &
        7.58~GHz &
        5.99~GHz &
        4.67~GHz 
        \\
        $ \varphi_{\mathrm{zpf}} $ &
        0.36 &
        0.36 &
        0.36 
        \\
        $ g_3 (\hat{a}^\dagger + \hat{a})^3$ &
        0 &
        0 &
        0 
        \\
        $ g_4 (\hat{a}^\dagger + \hat{a})^4$ &
        -0.12~MHz &
        0.04~MHz &
        -0.12~MHz 
        \\ 
        \hline
        \multicolumn{4}{c}{ Flux Driven Hamiltonian }
        \\
        \hline
        $ g_1^\text{driven} \epsilon_\text{d} (\hat{a}^\dagger + \hat{a})$ &
        0 &
        0 &
        6.48~GHz $ \cdot \epsilon_\text{d} $ 
        \\
        $ g_2^\text{driven} \epsilon_\text{d} (\hat{a}^\dagger + \hat{a})^2$ &
        0 &
        0 &
        0 
        \\
        $ g_3^\text{driven} \epsilon_\text{d} (\hat{a}^\dagger + \hat{a})^3$ &
        0 &
        24.8~MHz $ \cdot \epsilon_\text{d} $ &
        -140~MHz $ \cdot \epsilon_\text{d} $ 
        \\
        $ g_4^\text{driven} \epsilon_\text{d} (\hat{a}^\dagger + \hat{a})^4$ &
        0 &
        0 &
        0 
        \\
        $ g_5^\text{driven} \epsilon_\text{d} (\hat{a}^\dagger + \hat{a})^5$ &
        18.9~kHz $ \cdot \epsilon_\text{d} $ &
        179~kHz $ \cdot \epsilon_\text{d} $ &
        907~kHz $ \cdot \epsilon_\text{d} $ 
        \\ 
        \hline
    \end{tabular}
    \caption{A numerical comparison between NEMS-5, NEMS-3, and ATS in implementing higher-order driven Hamiltonians. We choose similar $ \varphi_{\mathrm{zpf}} $ to maintain consistent scaling ratios for nonlinear coefficients. NEMS-5 completely eliminates the first- and third-order terms, but reduces the fifth-order term to $ 1/50 $ compared to ATS. For experimental convenience, an alternative option is to use NEMS-3 instead of NEMS-5 in generating a five-photon interaction. NEMS-3 fully eliminates the most detrimental first-order term, while maintaining a moderate fifth-order term at $ 1/5 $ of that of ATS.}
    \label{tab: Nonlinearity comparison between NEMS-3 NEMS-5 and ATS}
\end{table*}

{We first complete the explicit expression of the potential function of NEMS-5.}
For fifth-order driven term engineering, we want $ c_{1}^\text{driven} = 0 $, $ c_{3}^\text{driven} = 0 $, and $ c_{5}^\text{driven} \neq 0 $, and neglect all $ n \geq 7 $ terms because the nonlinear Hamiltonian scales with $ \varphi_{\mathrm{zpf}}^n $. The condition can be satisfied with one single-JJ branch biased at $ \pi $ and two multi-JJ branches biased at $ 0 $. The corresponding design requirements are given as 
\begin{equation}\label{equ: c_driven target of NEMS-5}
    \begin{split}
        c_{1}^\text{driven} = - r_1 r_{\phi_{e1}} + \frac{r_2 r_{\phi_{e2}}}{n_2} + \frac{r_3 r_{\phi_{e3}}}{n_3} &= 0, \\
        c_{3}^\text{driven} = + r_1 r_{\phi_{e1}} - \frac{r_2 r_{\phi_{e2}}}{n_2^3} - \frac{r_3 r_{\phi_{e3}}}{n_3^3} &= 0, \\
        c_{5}^\text{driven} = - r_1 r_{\phi_{e1}} + \frac{r_2 r_{\phi_{e2}}}{n_2^5} + \frac{r_3 r_{\phi_{e3}}}{n_3^5} &\neq 0.
    \end{split}
\end{equation}
To make $ g_{5}^\text{driven} $ term dominate the nonlinear behavior, the device is required to have minimal $ c_{4}^\text{static} $ static nonlinearity. The corresponding design requirement is given as 
\begin{equation}\label{equ: c_static target of NEMS-5}
    \begin{split}
        c_4^\text{static} = + r_1 - \frac{r_2}{n_2^3} - \frac{r_3}{n_3^3} &= 0.
    \end{split}
\end{equation} 
{Since NEMS-5 has more design requirements than NEMS-3, it is hard to find a design with nearly uniform junction sizes and magnetic driving field distribution. Therefore, we only require a nearly uniform magnetic field and can tolerate large variations in junction sizes. Here, we present one NEMS-5 design containing three Josephson branches with \mbox{single-}, double-, and triple-JJs, with a magnetic drive satisfying $ \abs{\delta\phi_{ei}(t)} \propto n_i \epsilon(t) $, as shown in Fig.~\ref{fig: NEMS-5 structure and nonlinearity}(a). Consequently, the solution of Eqs.~(\ref{equ: c_driven target of NEMS-5}-\ref{equ: c_static target of NEMS-5}) yields: }
\begin{equation}\label{equ: rs of NEMS-5 appendix}
\begin{split}
        &r_1 = \frac{5}{32}r,\quad r_2 = r,\quad r_3 = \frac{27}{32} r, \\
        &n_{1} = 1,\quad n_{2} = 2,\quad n_{3} = 3,
\end{split}
\end{equation}
with the magnetic control parameters as
\begin{equation}\label{equ: phi_e of NEMS-5 appendix}
\begin{split}
        \overline{\phi_{e1}} = \pi, \quad & \delta\phi_{e1}(t) = 1/5 \cdot \epsilon(t), \\ 
        \overline{\phi_{e2}} = 0, \quad & \delta\phi_{e2}(t) = 2/5 \cdot \epsilon(t), \\ 
        \overline{\phi_{e3}} = 0, \quad & \delta\phi_{e3}(t) = -3/5 \cdot \epsilon(t).
\end{split}
\end{equation}
Here, $ r \lesssim 1 $ is a parameter limiting the size of small JJs. 

For given $ r_i $ and $ n_i $ in a specific device, we simplify the total potential of the NEMS-5 to
\begin{equation}\label{equ: U_tot of NEMS-5}
    \begin{split}
        \frac{U_\text{tot}}{E_\text{L}} = & \frac{\varphi^2}{2} - \frac{5}{32} r \cos\left( \varphi + \phi_{e1} \right) \\ &- 2 r  \cos\left( \frac{\varphi + \phi_{e2}}{2} \right) - \frac{81}{32} r  \cos\left( \frac{\varphi + \phi_{e3}}{3} \right).
    \end{split}
\end{equation} 
Without flux modulation, the static potential will be:
\begin{equation}\label{equ: U_static of NEMS-5}
    \begin{split}
        \frac{U_\text{static}}{E_\text{L}} = & \frac{\varphi^2}{2} + \frac{5}{32} r \cos\left( \varphi \right) \\ &- 2 r  \cos\left( \frac{\varphi}{2} \right) - \frac{81}{32} r  \cos\left( \frac{\varphi}{3} \right)\\
        =& \frac{ \varphi^2 }{ 2 }  + \frac{5}{16} r\varphi^2 + O(\varphi^6).
    \end{split}
\end{equation} 	
The chosen junction sizes precisely satisfy the conditions for eliminating the $ \varphi^4 $ term. However, this elimination condition does not apply to the linear part $ \varphi^2 $ or nonlinearities beyond $ \varphi^6 $. As a result, the device still exhibits nonzero $ \varphi^6 $ and higher-order nonlinearities.

{When the magnetic drive is applied as required, the NEMS-5 device only generates a $ \varphi^5 $ driving term:
\begin{equation}\label{equ: U_driven of NEMS-5}
    \begin{split}
        \frac{ U_\text{driven} }{ \epsilon(t)E_\text{J} } = & - \frac{1}{32} \sin\left( \varphi \right) + \frac{2}{5} r \sin\left( \frac{\varphi}{2}\right) - \frac{81}{160} r \sin\left( \frac{\varphi}{3} \right)\\
        = & -\frac{1}{48} \cdot r \cdot \frac{\varphi^5 }{5!}  + O(\varphi^7).
    \end{split}
\end{equation}
Obtaining a pure fifth-order term requires precise parameter control and comes at the cost of reducing the strength of the driving term by an order of magnitude.}

We compare the NEMS-5 with ATS and NEMS-3 for generating odd-order interactions. Similar $ \varphi_{\mathrm{zpf}} $ are chosen for both comparisons to maintain consistent scaling ratios for nonlinear coefficients. The numerical results are given in Table~\ref{tab: Nonlinearity comparison between NEMS-3 NEMS-5 and ATS}. The numerical comparison of the odd-order interactions shows the advantages and limitations of the NEMS-5 device. For the advantage, NEMS-5 can eliminate driving terms below the fifth order. Besides, the static Kerr nonlinearity in NEMS-5 is similar to those in NEMS-3 and ATS. The trade-off is that the desired fifth-order term produced by NEMS-5 is  $ \sim1/10 $ of that of NEMS-3 and $ \sim1/50 $ of that of ATS, under the same magnetic driving conditions. For experimental convenience, an alternative option is to use NEMS-3 instead of NEMS-5 in generating a five-photon interaction. NEMS-3 eliminates the most detrimental first-order term while maintaining a moderate fifth-order term at $ 1/5 $ of that of ATS. In the experiment, the {participation factors $ p_a $ and $ p_b $}~\cite{Minev2021pyEPR} are limited by the linear coupling strength between the modes and the nonlinear device. Utilizing a shared inductor possibly increases the participation factors, offering a way to enhance the coupling strength.

\subsection{ Stabilizing four-cat states with NEMS-4 }
\label{subsec:  Stabilizing four-cat states with NEMS-4}

\begin{table*}
    \renewcommand{\arraystretch}{1.2}
    \begin{tabular}{c|c|c}
        \hline
        &
        NEMS-4 &
        STS  
        \\ 
        \hline
        \multicolumn{3}{c}{ Circuit Parameter }
        \\
        \hline
        Inductor Branch $ E_{\text{J}_\text{L}}, n_\text{L} $ &
        $ E_{\text{J}_\text{L}} = 180~\text{GHz}, n_\text{L} = 10 $ &
        $ E_{\text{J}_\text{L}} = 180~\text{GHz}, n_\text{L} = 10 $ 
        \\
        Capacitor $ E_\text{C} $ &
        231~MHz &
        151~MHz 
        \\
        Josephson Branch $ E_{\text{J}_i}, n_i$  &
        \begin{tabular}[c]{@{}c@{}}$ E_{\text{J}_1} = 4.5~\text{GHz}, n_1 = 1 $\\ $ E_{\text{J}_2} = 18.0~\text{GHz}, n_2 = 2 $\end{tabular} &
        \begin{tabular}[c]{@{}c@{}}$ E_{\text{J}_1} = 18.0~\text{GHz}, n_1 = 1 $\\ $ E_{\text{J}_2} = 18.0~\text{GHz}, n_2 = 1 $\end{tabular}  
        \\
        \begin{tabular}[c]{@{}c@{}}Flux Control\\ Scheme\end{tabular} &
        \begin{tabular}[c]{@{}c@{}}$ \phi_{e1} = 1.25\pi + 0.5\epsilon(t) $\\ $ \phi_{e2} = 0.5\pi + 0.25\epsilon(t) $\end{tabular} &
        \begin{tabular}[c]{@{}c@{}}$ \phi_{e1} = 0.5\pi + 0.5\epsilon(t) $\\ $ \phi_{e2} = 0.5\pi - 0.5\epsilon(t) $\end{tabular}  
        \\ 
        \hline
        \multicolumn{3}{c}{ Static Hamiltonian }
        \\
        \hline
        Frequency &
        7.14~GHz &
        4.67~GHz  
        \\
        $ \varphi_{\mathrm{zpf}} $ &
        0.36 &
        0.36 
        \\
        $ g_3 (\hat{a}^\dagger + \hat{a})^3$ &
        0 &
        0 
        \\
        $ g_4 (\hat{a}^\dagger + \hat{a})^4$ &
        -0.12~MHz &
        -0.12~MHz  
        \\ 
        \hline
        \multicolumn{3}{c}{ Flux Driven Hamiltonian }
        \\
        \hline$ g_1^\text{driven} \epsilon_\text{d} (\hat{a}^\dagger + \hat{a})$ &
        0  &
        0  
        \\
        $ g_2^\text{driven} \epsilon_\text{d} (\hat{a}^\dagger + \hat{a})^2$ &
        0 &
        -1.17~GHz$ \cdot \epsilon_\text{d} $ 
        \\
        $ g_3^\text{driven} \epsilon_\text{d} (\hat{a}^\dagger + \hat{a})^3$ &
        0 &
        0  
        \\
        $ g_4^\text{driven} \epsilon_\text{d} (\hat{a}^\dagger + \hat{a})^4$ &
        -0.84~MHz $ \cdot \epsilon_\text{d} $ &
        12.5~MHz$ \cdot \epsilon_\text{d} $ 
        \\
        $ g_5^\text{driven} \epsilon_\text{d} (\hat{a}^\dagger + \hat{a})^5$ &
        0 &
        0 
        \\ 
        \hline
    \end{tabular}
    \caption{{A numerical comparison between NEMS-4 and STS in implementing higher-order driven Hamiltonians. We choose similar $ \varphi_{\mathrm{zpf}} $ to maintain consistent scaling ratios for nonlinear coefficients. Compared to STS, NEMS-4 fully eliminates the second-order term but reduces the fourth-order term to $ 1/15 $. } }
    \label{tab: Nonlinearity comparison between NEMS-4 and STS}
\end{table*}

\begin{figure}
    \includegraphics{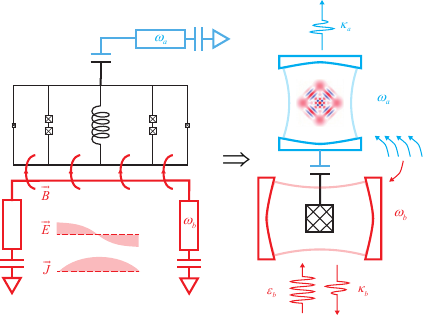}
    \caption{Diagram of stabilizing four-cat states with NEMS-4. A NEMS-4 device linearly couples to a storage mode, with the magnetic driving line replaced by a magnetic driving buffer mode. The magnetic field in the buffer mode induces the fourth-order driving terms in the NEMS-4 mode, effectively establishing a passive 4-to-1 coupling. The circuit is equivalent to placing a nonlinear material inside a buffer cavity to induce magnetic coupling, which is then linearly connected to a storage cavity. }
    \label{fig:NEMS-4_for_4cat}
\end{figure}

Considering the nonlinear coupling to the environment caused by the magnetic driving line, here we propose another method of stabilizing the four-cat state with a NEMS-4 device as the nonlinear coupler. The schematic diagram is shown in Fig.~\ref{fig:NEMS-4_for_4cat}(a). In this setup, an effective buffer mode is constructed by replacing the magnetic driving line with a transmission line resonator, while the storage mode linearly couples to the NEMS-4 device. At the NEMS-4 working point, the alternating magnetic field in the driving line resonator primarily induces fourth-order drive in the NEMS-4 mode. Considering the quantum nature of photons in the driving line resonator, we obtain the nonlinear coupling Hamiltonian between the storage and the buffer modes:
\begin{equation}\label{equ: H_c for NEMS-4 four-cat stabilization}
    \hat{\mathcal{H}}_c = g_4^\text{driven} \left( \epsilon_{\mathrm{zpf}} \hat{b} + \epsilon_{\mathrm{zpf}} \hat{b}^\dagger \right) \left( p_a\hat{a} + p_a\hat{a}^\dagger \right)^4.
\end{equation}
{Compared to Eq.~(\ref{equ: H of Magnetic drive environment coupling}), here we only consider the vacuum fluctuation of one buffer mode.}
$ p_a $ denotes the participation factor of the storage mode $ \hat{a} $ in the NEMS-4 device. 
A similar design has also been employed in stabilizing two-cat states~\cite{Marquet2024Autoparametric}. The coupling Hamiltonian only comes into play when the frequency matching condition $ \omega_b = 4\omega_a $ is satisfied between the two resonators. The circuit can be viewed as embedding a nonlinear device NEMS-4 within a buffer resonator to induce magnetic coupling while connecting it linearly to a storage mode. The electromagnetic field in the buffer resonator and the nonlinear device is coupled through mutual inductance, thus inducing the desired 4-to-1 coupling interaction.

For comparison, we conduct a numerical comparison between the NEMS-4 and STS for generating even-order interactions, as shown in Table~\ref{tab: Nonlinearity comparison between NEMS-4 and STS}. The numerical result of generating even-order interactions reveals the advantages and limitations of the NEMS-4 device. On the positive side, NEMS-4 significantly reduces the undesired second-order driving term, surpassing STS by three orders of magnitude. Therefore NEMS-4 effectively minimizes the impact of low-order residual interactions. On the downside, the desired fourth-order driving term is reduced by one order of magnitude, limiting the intensity of the four-photon driven-dissipation process. Increasing the magnetic flux density by adopting high-inductance connecting wires may enhance the coupling strength.

\bibliographystyle{Zou}
\bibliography{Library_for_Paper_hzy}	

\begin{thebibliography}{75}%
\makeatletter
\providecommand \@ifxundefined [1]{%
 \@ifx{#1\undefined}
}%
\providecommand \@ifnum [1]{%
 \ifnum #1\expandafter \@firstoftwo
 \else \expandafter \@secondoftwo
 \fi
}%
\providecommand \@ifx [1]{%
 \ifx #1\expandafter \@firstoftwo
 \else \expandafter \@secondoftwo
 \fi
}%
\providecommand \natexlab [1]{#1}%
\providecommand \enquote  [1]{``#1''}%
\providecommand \bibnamefont  [1]{#1}%
\providecommand \bibfnamefont [1]{#1}%
\providecommand \citenamefont [1]{#1}%
\providecommand \href@noop [0]{\@secondoftwo}%
\providecommand \href [0]{\begingroup \@sanitize@url \@href}%
\providecommand \@href[1]{\@@startlink{#1}\@@href}%
\providecommand \@@href[1]{\endgroup#1\@@endlink}%
\providecommand \@sanitize@url [0]{\catcode `\\12\catcode `\$12\catcode
  `\&12\catcode `\#12\catcode `\^12\catcode `\_12\catcode `\%12\relax}%
\providecommand \@@startlink[1]{}%
\providecommand \@@endlink[0]{}%
\providecommand \url  [0]{\begingroup\@sanitize@url \@url }%
\providecommand \@url [1]{\endgroup\@href {#1}{\urlprefix }}%
\providecommand \urlprefix  [0]{URL }%
\providecommand \Eprint [0]{\href }%
\providecommand \doibase [0]{http://dx.doi.org/}%
\providecommand \selectlanguage [0]{\@gobble}%
\providecommand \bibinfo  [0]{\@secondoftwo}%
\providecommand \bibfield  [0]{\@secondoftwo}%
\providecommand \translation [1]{[#1]}%
\providecommand \BibitemOpen [0]{}%
\providecommand \bibitemStop [0]{}%
\providecommand \bibitemNoStop [0]{.\EOS\space}%
\providecommand \EOS [0]{\spacefactor3000\relax}%
\providecommand \BibitemShut  [1]{\csname bibitem#1\endcsname}%
\let\auto@bib@innerbib\@empty
\bibitem [{\citenamefont {Joshi}\ \emph {et~al.}(2021)\citenamefont {Joshi},
  \citenamefont {Noh},\ and\ \citenamefont {Gao}}]{Joshi2021bosonicqubits}%
  \BibitemOpen
  \bibfield  {author} {\bibinfo {author} {\bibfnamefont {A.}~\bibnamefont
  {Joshi}}, \bibinfo {author} {\bibfnamefont {K.}~\bibnamefont {Noh}}, \ and\
  \bibinfo {author} {\bibfnamefont {Y.~Y.}\ \bibnamefont {Gao}},\ }\bibfield
  {title} {\enquote {\bibinfo {title} {Quantum information processing with
  bosonic qubits in circuit qed},}\ }\href {\doibase 10.1088/2058-9565/abe989}
  {\bibfield  {journal} {\bibinfo  {journal} {Quantum Sci Technol}\ }\textbf
  {\bibinfo {volume} {6}},\ \bibinfo {pages} {033001} (\bibinfo {year}
  {2021})}\BibitemShut {NoStop}%
\bibitem [{\citenamefont {Cai}\ \emph {et~al.}(2021)\citenamefont {Cai},
  \citenamefont {Ma}, \citenamefont {Wang}, \citenamefont {Zou},\ and\
  \citenamefont {Sun}}]{Cai2021Bosonic}%
  \BibitemOpen
  \bibfield  {author} {\bibinfo {author} {\bibfnamefont {W.}~\bibnamefont
  {Cai}}, \bibinfo {author} {\bibfnamefont {Y.}~\bibnamefont {Ma}}, \bibinfo
  {author} {\bibfnamefont {W.}~\bibnamefont {Wang}}, \bibinfo {author}
  {\bibfnamefont {C.-L.}\ \bibnamefont {Zou}}, \ and\ \bibinfo {author}
  {\bibfnamefont {L.}~\bibnamefont {Sun}},\ }\bibfield  {title} {\enquote
  {\bibinfo {title} {Bosonic quantum error correction codes in superconducting
  quantum circuits},}\ }\href {\doibase
  https://doi.org/10.1016/j.fmre.2020.12.006} {\bibfield  {journal} {\bibinfo
  {journal} {Fundamental Research}\ }\textbf {\bibinfo {volume} {1}},\ \bibinfo
  {pages} {50} (\bibinfo {year} {2021})}\BibitemShut {NoStop}%
\bibitem [{\citenamefont {Copetudo}\ \emph {et~al.}(2024)\citenamefont
  {Copetudo}, \citenamefont {Fontaine}, \citenamefont {Valadares},\ and\
  \citenamefont {Gao}}]{Copetudo2024ReviewBosonicQEC}%
  \BibitemOpen
  \bibfield  {author} {\bibinfo {author} {\bibfnamefont {A.}~\bibnamefont
  {Copetudo}}, \bibinfo {author} {\bibfnamefont {C.~Y.}\ \bibnamefont
  {Fontaine}}, \bibinfo {author} {\bibfnamefont {F.}~\bibnamefont {Valadares}},
  \ and\ \bibinfo {author} {\bibfnamefont {Y.~Y.}\ \bibnamefont {Gao}},\
  }\bibfield  {title} {\enquote {\bibinfo {title} {Shaping photons: Quantum
  information processing with bosonic cqed},}\ }\href {\doibase
  10.1063/5.0183022} {\bibfield  {journal} {\bibinfo  {journal} {Appl. Phys.
  Lett.}\ }\textbf {\bibinfo {volume} {124}},\ \bibinfo {pages} {080502}
  (\bibinfo {year} {2024})}\BibitemShut {NoStop}%
\bibitem [{\citenamefont {Devoret}\ and\ \citenamefont
  {Schoelkopf}(2013)}]{Devoret2013SQEC}%
  \BibitemOpen
  \bibfield  {author} {\bibinfo {author} {\bibfnamefont {M.~H.}\ \bibnamefont
  {Devoret}}\ and\ \bibinfo {author} {\bibfnamefont {R.~J.}\ \bibnamefont
  {Schoelkopf}},\ }\bibfield  {title} {\enquote {\bibinfo {title}
  {Superconducting circuits for quantum information: An outlook},}\ }\href
  {\doibase doi:10.1126/science.1231930} {\bibfield  {journal} {\bibinfo
  {journal} {Science}\ }\textbf {\bibinfo {volume} {339}},\ \bibinfo {pages}
  {1169} (\bibinfo {year} {2013})}\BibitemShut {NoStop}%
\bibitem [{\citenamefont {Grimsmo}\ \emph {et~al.}(2020)\citenamefont
  {Grimsmo}, \citenamefont {Combes},\ and\ \citenamefont
  {Baragiola}}]{Grimsmo2020PRXCatandBinomial}%
  \BibitemOpen
  \bibfield  {author} {\bibinfo {author} {\bibfnamefont {A.~L.}\ \bibnamefont
  {Grimsmo}}, \bibinfo {author} {\bibfnamefont {J.}~\bibnamefont {Combes}}, \
  and\ \bibinfo {author} {\bibfnamefont {B.~Q.}\ \bibnamefont {Baragiola}},\
  }\bibfield  {title} {\enquote {\bibinfo {title} {Quantum computing with
  rotation-symmetric bosonic codes},}\ }\href {\doibase
  10.1103/PhysRevX.10.011058} {\bibfield  {journal} {\bibinfo  {journal} {Phys.
  Rev. X}\ }\textbf {\bibinfo {volume} {10}},\ \bibinfo {pages} {011058}
  (\bibinfo {year} {2020})}\BibitemShut {NoStop}%
\bibitem [{\citenamefont {Gouzien}\ \emph {et~al.}(2023)\citenamefont
  {Gouzien}, \citenamefont {Ruiz}, \citenamefont {Le~R{\'e}gent}, \citenamefont
  {Guillaud},\ and\ \citenamefont {Sangouard}}]{Gouzien2023Performance}%
  \BibitemOpen
  \bibfield  {author} {\bibinfo {author} {\bibfnamefont {{\'E}.}~\bibnamefont
  {Gouzien}}, \bibinfo {author} {\bibfnamefont {D.}~\bibnamefont {Ruiz}},
  \bibinfo {author} {\bibfnamefont {F.-M.}\ \bibnamefont {Le~R{\'e}gent}},
  \bibinfo {author} {\bibfnamefont {J.}~\bibnamefont {Guillaud}}, \ and\
  \bibinfo {author} {\bibfnamefont {N.}~\bibnamefont {Sangouard}},\ }\bibfield
  {title} {\enquote {\bibinfo {title} {Performance analysis of a repetition cat
  code architecture: Computing 256-bit elliptic curve logarithm in 9 hours with
  126 133 cat qubits},}\ }\href {\doibase 10.1103/PhysRevLett.131.040602}
  {\bibfield  {journal} {\bibinfo  {journal} {Phys. Rev. Lett.}\ }\textbf
  {\bibinfo {volume} {131}},\ \bibinfo {pages} {040602} (\bibinfo {year}
  {2023})}\BibitemShut {NoStop}%
\bibitem [{\citenamefont {Liu}\ \emph {et~al.}(2024)\citenamefont {Liu},
  \citenamefont {Singh}, \citenamefont {Smith}, \citenamefont {Crane},
  \citenamefont {Martyn}, \citenamefont {Eickbusch}, \citenamefont {Schuckert},
  \citenamefont {Li}, \citenamefont {Sinanan-Singh}, \citenamefont {Soley}
  \emph {et~al.}}]{Liu2024HybridOscillatorQubit}%
  \BibitemOpen
  \bibfield  {author} {\bibinfo {author} {\bibfnamefont {Y.}~\bibnamefont
  {Liu}}, \bibinfo {author} {\bibfnamefont {S.}~\bibnamefont {Singh}}, \bibinfo
  {author} {\bibfnamefont {K.~C.}\ \bibnamefont {Smith}}, \bibinfo {author}
  {\bibfnamefont {E.}~\bibnamefont {Crane}}, \bibinfo {author} {\bibfnamefont
  {J.~M.}\ \bibnamefont {Martyn}}, \bibinfo {author} {\bibfnamefont
  {A.}~\bibnamefont {Eickbusch}}, \bibinfo {author} {\bibfnamefont
  {A.}~\bibnamefont {Schuckert}}, \bibinfo {author} {\bibfnamefont {R.~D.}\
  \bibnamefont {Li}}, \bibinfo {author} {\bibfnamefont {J.}~\bibnamefont
  {Sinanan-Singh}}, \bibinfo {author} {\bibfnamefont {M.~B.}\ \bibnamefont
  {Soley}},  \emph {et~al.},\ }\bibfield  {title} {\enquote {\bibinfo {title}
  {Hybrid oscillator-qubit quantum processors: Instruction set architectures,
  abstract machine models, and applications},}\ }\href {\doibase
  10.48550/arXiv.2407.10381} {\ ,\ \bibinfo {pages} {arXiv:2407.10381}
  (\bibinfo {year} {2024})}\BibitemShut {NoStop}%
\bibitem [{\citenamefont {Crane}\ \emph {et~al.}(2024)\citenamefont {Crane},
  \citenamefont {Smith}, \citenamefont {Tomesh}, \citenamefont {Eickbusch},
  \citenamefont {Martyn}, \citenamefont {K{\"u}hn}, \citenamefont {Funcke},
  \citenamefont {DeMarco}, \citenamefont {Chuang}, \citenamefont {Wiebe} \emph
  {et~al.}}]{Crane2024HybridOscillatorQubit}%
  \BibitemOpen
  \bibfield  {author} {\bibinfo {author} {\bibfnamefont {E.}~\bibnamefont
  {Crane}}, \bibinfo {author} {\bibfnamefont {K.~C.}\ \bibnamefont {Smith}},
  \bibinfo {author} {\bibfnamefont {T.}~\bibnamefont {Tomesh}}, \bibinfo
  {author} {\bibfnamefont {A.}~\bibnamefont {Eickbusch}}, \bibinfo {author}
  {\bibfnamefont {J.~M.}\ \bibnamefont {Martyn}}, \bibinfo {author}
  {\bibfnamefont {S.}~\bibnamefont {K{\"u}hn}}, \bibinfo {author}
  {\bibfnamefont {L.}~\bibnamefont {Funcke}}, \bibinfo {author} {\bibfnamefont
  {M.~A.}\ \bibnamefont {DeMarco}}, \bibinfo {author} {\bibfnamefont {I.~L.}\
  \bibnamefont {Chuang}}, \bibinfo {author} {\bibfnamefont {N.}~\bibnamefont
  {Wiebe}},  \emph {et~al.},\ }\bibfield  {title} {\enquote {\bibinfo {title}
  {Hybrid oscillator-qubit quantum processors: Simulating fermions, bosons, and
  gauge fields},}\ }\href {\doibase 10.48550/arXiv.2409.03747} {\ ,\ \bibinfo
  {pages} {arXiv:2409.03747} (\bibinfo {year} {2024})}\BibitemShut {NoStop}%
\bibitem [{\citenamefont {Flurin}\ \emph {et~al.}(2017)\citenamefont {Flurin},
  \citenamefont {Ramasesh}, \citenamefont {Hacohen-Gourgy}, \citenamefont
  {Martin}, \citenamefont {Yao},\ and\ \citenamefont
  {Siddiqi}}]{Flurin2017Observing}%
  \BibitemOpen
  \bibfield  {author} {\bibinfo {author} {\bibfnamefont {E.}~\bibnamefont
  {Flurin}}, \bibinfo {author} {\bibfnamefont {V.~â.}\ \bibnamefont
  {Ramasesh}}, \bibinfo {author} {\bibfnamefont {S.}~\bibnamefont
  {Hacohen-Gourgy}}, \bibinfo {author} {\bibfnamefont {L.~â.}\ \bibnamefont
  {Martin}}, \bibinfo {author} {\bibfnamefont {N.~â.}\ \bibnamefont {Yao}}, \
  and\ \bibinfo {author} {\bibfnamefont {I.}~\bibnamefont {Siddiqi}},\
  }\bibfield  {title} {\enquote {\bibinfo {title} {Observing topological
  invariants using quantum walks in superconducting circuits},}\ }\href
  {\doibase 10.1103/PhysRevX.7.031023} {\bibfield  {journal} {\bibinfo
  {journal} {Phys. Rev. X}\ }\textbf {\bibinfo {volume} {7}},\ \bibinfo {pages}
  {031023} (\bibinfo {year} {2017})}\BibitemShut {NoStop}%
\bibitem [{\citenamefont {Hu}\ \emph {et~al.}(2018)\citenamefont {Hu},
  \citenamefont {Ma}, \citenamefont {Xu}, \citenamefont {Wang}, \citenamefont
  {Ma}, \citenamefont {Liu}, \citenamefont {Wang}, \citenamefont {Song},
  \citenamefont {Yung},\ and\ \citenamefont {Sun}}]{Hu2018Simulation}%
  \BibitemOpen
  \bibfield  {author} {\bibinfo {author} {\bibfnamefont {L.}~\bibnamefont
  {Hu}}, \bibinfo {author} {\bibfnamefont {Y.-C.}\ \bibnamefont {Ma}}, \bibinfo
  {author} {\bibfnamefont {Y.}~\bibnamefont {Xu}}, \bibinfo {author}
  {\bibfnamefont {W.-T.}\ \bibnamefont {Wang}}, \bibinfo {author}
  {\bibfnamefont {Y.-W.}\ \bibnamefont {Ma}}, \bibinfo {author} {\bibfnamefont
  {K.}~\bibnamefont {Liu}}, \bibinfo {author} {\bibfnamefont {H.-Y.}\
  \bibnamefont {Wang}}, \bibinfo {author} {\bibfnamefont {Y.-P.}\ \bibnamefont
  {Song}}, \bibinfo {author} {\bibfnamefont {M.-H.}\ \bibnamefont {Yung}}, \
  and\ \bibinfo {author} {\bibfnamefont {L.-Y.}\ \bibnamefont {Sun}},\
  }\bibfield  {title} {\enquote {\bibinfo {title} {Simulation of molecular
  spectroscopy with circuit quantum electrodynamics},}\ }\href {\doibase
  https://doi.org/10.1016/j.scib.2018.02.001} {\bibfield  {journal} {\bibinfo
  {journal} {Science Bulletin}\ }\textbf {\bibinfo {volume} {63}},\ \bibinfo
  {pages} {293} (\bibinfo {year} {2018})}\BibitemShut {NoStop}%
\bibitem [{\citenamefont {Wang}\ \emph {et~al.}(2020)\citenamefont {Wang},
  \citenamefont {Curtis}, \citenamefont {Lester}, \citenamefont {Zhang},
  \citenamefont {Gao}, \citenamefont {Freeze}, \citenamefont {Batista},
  \citenamefont {Vaccaro}, \citenamefont {Chuang}, \citenamefont {Frunzio}
  \emph {et~al.}}]{Wang2020Efficient}%
  \BibitemOpen
  \bibfield  {author} {\bibinfo {author} {\bibfnamefont {C.~S.}\ \bibnamefont
  {Wang}}, \bibinfo {author} {\bibfnamefont {J.~C.}\ \bibnamefont {Curtis}},
  \bibinfo {author} {\bibfnamefont {B.~J.}\ \bibnamefont {Lester}}, \bibinfo
  {author} {\bibfnamefont {Y.}~\bibnamefont {Zhang}}, \bibinfo {author}
  {\bibfnamefont {Y.~Y.}\ \bibnamefont {Gao}}, \bibinfo {author} {\bibfnamefont
  {J.}~\bibnamefont {Freeze}}, \bibinfo {author} {\bibfnamefont {V.~S.}\
  \bibnamefont {Batista}}, \bibinfo {author} {\bibfnamefont {P.~H.}\
  \bibnamefont {Vaccaro}}, \bibinfo {author} {\bibfnamefont {I.~L.}\
  \bibnamefont {Chuang}}, \bibinfo {author} {\bibfnamefont {L.}~\bibnamefont
  {Frunzio}},  \emph {et~al.},\ }\bibfield  {title} {\enquote {\bibinfo {title}
  {Efficient multiphoton sampling of molecular vibronic spectra on a
  superconducting bosonic processor},}\ }\href {\doibase
  10.1103/PhysRevX.10.021060} {\bibfield  {journal} {\bibinfo  {journal} {Phys.
  Rev. X}\ }\textbf {\bibinfo {volume} {10}},\ \bibinfo {pages} {021060}
  (\bibinfo {year} {2020})}\BibitemShut {NoStop}%
\bibitem [{\citenamefont {Wang}\ \emph
  {et~al.}(2019{\natexlab{a}})\citenamefont {Wang}, \citenamefont {Wu},
  \citenamefont {Ma}, \citenamefont {Cai}, \citenamefont {Hu}, \citenamefont
  {Mu}, \citenamefont {Xu}, \citenamefont {Chen}, \citenamefont {Wang},
  \citenamefont {Song} \emph {et~al.}}]{Wang2019Heisenberglimited}%
  \BibitemOpen
  \bibfield  {author} {\bibinfo {author} {\bibfnamefont {W.}~\bibnamefont
  {Wang}}, \bibinfo {author} {\bibfnamefont {Y.}~\bibnamefont {Wu}}, \bibinfo
  {author} {\bibfnamefont {Y.}~\bibnamefont {Ma}}, \bibinfo {author}
  {\bibfnamefont {W.}~\bibnamefont {Cai}}, \bibinfo {author} {\bibfnamefont
  {L.}~\bibnamefont {Hu}}, \bibinfo {author} {\bibfnamefont {X.}~\bibnamefont
  {Mu}}, \bibinfo {author} {\bibfnamefont {Y.}~\bibnamefont {Xu}}, \bibinfo
  {author} {\bibfnamefont {Z.-J.}\ \bibnamefont {Chen}}, \bibinfo {author}
  {\bibfnamefont {H.}~\bibnamefont {Wang}}, \bibinfo {author} {\bibfnamefont
  {Y.~P.}\ \bibnamefont {Song}},  \emph {et~al.},\ }\bibfield  {title}
  {\enquote {\bibinfo {title} {Heisenberg-limited single-mode quantum metrology
  in a superconducting circuit},}\ }\href {\doibase 10.1038/s41467-019-12290-7}
  {\bibfield  {journal} {\bibinfo  {journal} {Nat. Commun.}\ }\textbf {\bibinfo
  {volume} {10}},\ \bibinfo {pages} {4382} (\bibinfo {year}
  {2019}{\natexlab{a}})}\BibitemShut {NoStop}%
\bibitem [{\citenamefont {Deng}\ \emph {et~al.}(2024)\citenamefont {Deng},
  \citenamefont {Li}, \citenamefont {Chen}, \citenamefont {Ni}, \citenamefont
  {Cai}, \citenamefont {Mai}, \citenamefont {Zhang}, \citenamefont {Zheng},
  \citenamefont {Yu}, \citenamefont {Zou} \emph
  {et~al.}}]{Deng2024metrology100Fock}%
  \BibitemOpen
  \bibfield  {author} {\bibinfo {author} {\bibfnamefont {X.}~\bibnamefont
  {Deng}}, \bibinfo {author} {\bibfnamefont {S.}~\bibnamefont {Li}}, \bibinfo
  {author} {\bibfnamefont {Z.-J.}\ \bibnamefont {Chen}}, \bibinfo {author}
  {\bibfnamefont {Z.}~\bibnamefont {Ni}}, \bibinfo {author} {\bibfnamefont
  {Y.}~\bibnamefont {Cai}}, \bibinfo {author} {\bibfnamefont {J.}~\bibnamefont
  {Mai}}, \bibinfo {author} {\bibfnamefont {L.}~\bibnamefont {Zhang}}, \bibinfo
  {author} {\bibfnamefont {P.}~\bibnamefont {Zheng}}, \bibinfo {author}
  {\bibfnamefont {H.}~\bibnamefont {Yu}}, \bibinfo {author} {\bibfnamefont
  {C.-L.}\ \bibnamefont {Zou}},  \emph {et~al.},\ }\bibfield  {title} {\enquote
  {\bibinfo {title} {Quantum-enhanced metrology with large fock states},}\
  }\href {\doibase 10.1038/s41567-024-02619-5} {\bibfield  {journal} {\bibinfo
  {journal} {Nat. Phys.}\ } (\bibinfo {year} {2024}),\
  10.1038/s41567-024-02619-5}\BibitemShut {NoStop}%
\bibitem [{\citenamefont {Michael}\ \emph {et~al.}(2016)\citenamefont
  {Michael}, \citenamefont {Silveri}, \citenamefont {Brierley}, \citenamefont
  {Albert}, \citenamefont {Salmilehto}, \citenamefont {Jiang},\ and\
  \citenamefont {Girvin}}]{Michael2016NewClass}%
  \BibitemOpen
  \bibfield  {author} {\bibinfo {author} {\bibfnamefont {M.~H.}\ \bibnamefont
  {Michael}}, \bibinfo {author} {\bibfnamefont {M.}~\bibnamefont {Silveri}},
  \bibinfo {author} {\bibfnamefont {R.~T.}\ \bibnamefont {Brierley}}, \bibinfo
  {author} {\bibfnamefont {V.~V.}\ \bibnamefont {Albert}}, \bibinfo {author}
  {\bibfnamefont {J.}~\bibnamefont {Salmilehto}}, \bibinfo {author}
  {\bibfnamefont {L.}~\bibnamefont {Jiang}}, \ and\ \bibinfo {author}
  {\bibfnamefont {S.~M.}\ \bibnamefont {Girvin}},\ }\bibfield  {title}
  {\enquote {\bibinfo {title} {New class of quantum error-correcting codes for
  a bosonic mode},}\ }\href {\doibase 10.1103/PhysRevX.6.031006} {\bibfield
  {journal} {\bibinfo  {journal} {Phys. Rev. X}\ }\textbf {\bibinfo {volume}
  {6}},\ \bibinfo {pages} {031006} (\bibinfo {year} {2016})}\BibitemShut
  {NoStop}%
\bibitem [{\citenamefont {Leghtas}\ \emph {et~al.}(2013)\citenamefont
  {Leghtas}, \citenamefont {Kirchmair}, \citenamefont {Vlastakis},
  \citenamefont {Schoelkopf}, \citenamefont {Devoret},\ and\ \citenamefont
  {Mirrahimi}}]{Leghtas2013Hardware-Efficient}%
  \BibitemOpen
  \bibfield  {author} {\bibinfo {author} {\bibfnamefont {Z.}~\bibnamefont
  {Leghtas}}, \bibinfo {author} {\bibfnamefont {G.}~\bibnamefont {Kirchmair}},
  \bibinfo {author} {\bibfnamefont {B.}~\bibnamefont {Vlastakis}}, \bibinfo
  {author} {\bibfnamefont {R.~J.}\ \bibnamefont {Schoelkopf}}, \bibinfo
  {author} {\bibfnamefont {M.~H.}\ \bibnamefont {Devoret}}, \ and\ \bibinfo
  {author} {\bibfnamefont {M.}~\bibnamefont {Mirrahimi}},\ }\bibfield  {title}
  {\enquote {\bibinfo {title} {Hardware-efficient autonomous quantum memory
  protection},}\ }\href {\doibase 10.1103/PhysRevLett.111.120501} {\bibfield
  {journal} {\bibinfo  {journal} {Phys. Rev. Lett.}\ }\textbf {\bibinfo
  {volume} {111}},\ \bibinfo {pages} {120501} (\bibinfo {year}
  {2013})}\BibitemShut {NoStop}%
\bibitem [{\citenamefont {Gottesman}\ \emph {et~al.}(2001)\citenamefont
  {Gottesman}, \citenamefont {Kitaev},\ and\ \citenamefont
  {Preskill}}]{GKP2001Encoding}%
  \BibitemOpen
  \bibfield  {author} {\bibinfo {author} {\bibfnamefont {D.}~\bibnamefont
  {Gottesman}}, \bibinfo {author} {\bibfnamefont {A.}~\bibnamefont {Kitaev}}, \
  and\ \bibinfo {author} {\bibfnamefont {J.}~\bibnamefont {Preskill}},\
  }\bibfield  {title} {\enquote {\bibinfo {title} {Encoding a qubit in an
  oscillator},}\ }\href {\doibase 10.1103/PhysRevA.64.012310} {\bibfield
  {journal} {\bibinfo  {journal} {Phys. Rev. A}\ }\textbf {\bibinfo {volume}
  {64}},\ \bibinfo {pages} {012310} (\bibinfo {year} {2001})}\BibitemShut
  {NoStop}%
\bibitem [{\citenamefont {Hu}\ \emph {et~al.}(2019)\citenamefont {Hu},
  \citenamefont {Ma}, \citenamefont {Cai}, \citenamefont {Mu}, \citenamefont
  {Xu}, \citenamefont {Wang}, \citenamefont {Wu}, \citenamefont {Wang},
  \citenamefont {Song}, \citenamefont {Zou} \emph {et~al.}}]{Hu2019binomial}%
  \BibitemOpen
  \bibfield  {author} {\bibinfo {author} {\bibfnamefont {L.}~\bibnamefont
  {Hu}}, \bibinfo {author} {\bibfnamefont {Y.}~\bibnamefont {Ma}}, \bibinfo
  {author} {\bibfnamefont {W.}~\bibnamefont {Cai}}, \bibinfo {author}
  {\bibfnamefont {X.}~\bibnamefont {Mu}}, \bibinfo {author} {\bibfnamefont
  {Y.}~\bibnamefont {Xu}}, \bibinfo {author} {\bibfnamefont {W.}~\bibnamefont
  {Wang}}, \bibinfo {author} {\bibfnamefont {Y.}~\bibnamefont {Wu}}, \bibinfo
  {author} {\bibfnamefont {H.}~\bibnamefont {Wang}}, \bibinfo {author}
  {\bibfnamefont {Y.~P.}\ \bibnamefont {Song}}, \bibinfo {author}
  {\bibfnamefont {C.~L.}\ \bibnamefont {Zou}},  \emph {et~al.},\ }\bibfield
  {title} {\enquote {\bibinfo {title} {Quantum error correction and universal
  gate set operation on a binomial bosonic logical qubit},}\ }\href {\doibase
  10.1038/s41567-018-0414-3} {\bibfield  {journal} {\bibinfo  {journal} {Nat.
  Phys.}\ }\textbf {\bibinfo {volume} {15}},\ \bibinfo {pages} {503} (\bibinfo
  {year} {2019})}\BibitemShut {NoStop}%
\bibitem [{\citenamefont {Ma}\ \emph {et~al.}(2020)\citenamefont {Ma},
  \citenamefont {Xu}, \citenamefont {Mu}, \citenamefont {Cai}, \citenamefont
  {Hu}, \citenamefont {Wang}, \citenamefont {Pan}, \citenamefont {Wang},
  \citenamefont {Song}, \citenamefont {Zou} \emph {et~al.}}]{Ma2020PASS}%
  \BibitemOpen
  \bibfield  {author} {\bibinfo {author} {\bibfnamefont {Y.}~\bibnamefont
  {Ma}}, \bibinfo {author} {\bibfnamefont {Y.}~\bibnamefont {Xu}}, \bibinfo
  {author} {\bibfnamefont {X.}~\bibnamefont {Mu}}, \bibinfo {author}
  {\bibfnamefont {W.}~\bibnamefont {Cai}}, \bibinfo {author} {\bibfnamefont
  {L.}~\bibnamefont {Hu}}, \bibinfo {author} {\bibfnamefont {W.}~\bibnamefont
  {Wang}}, \bibinfo {author} {\bibfnamefont {X.}~\bibnamefont {Pan}}, \bibinfo
  {author} {\bibfnamefont {H.}~\bibnamefont {Wang}}, \bibinfo {author}
  {\bibfnamefont {Y.~P.}\ \bibnamefont {Song}}, \bibinfo {author}
  {\bibfnamefont {C.~L.}\ \bibnamefont {Zou}},  \emph {et~al.},\ }\bibfield
  {title} {\enquote {\bibinfo {title} {Error-transparent operations on a
  logical qubit protected by quantum error correction},}\ }\href {\doibase
  10.1038/s41567-020-0893-x} {\bibfield  {journal} {\bibinfo  {journal} {Nat.
  Phys.}\ }\textbf {\bibinfo {volume} {16}},\ \bibinfo {pages} {827} (\bibinfo
  {year} {2020})}\BibitemShut {NoStop}%
\bibitem [{\citenamefont {Gertler}\ \emph {et~al.}(2021)\citenamefont
  {Gertler}, \citenamefont {Baker}, \citenamefont {Li}, \citenamefont {Shirol},
  \citenamefont {Koch},\ and\ \citenamefont {Wang}}]{Gertler2021BosonicAQEC}%
  \BibitemOpen
  \bibfield  {author} {\bibinfo {author} {\bibfnamefont {J.~M.}\ \bibnamefont
  {Gertler}}, \bibinfo {author} {\bibfnamefont {B.}~\bibnamefont {Baker}},
  \bibinfo {author} {\bibfnamefont {J.}~\bibnamefont {Li}}, \bibinfo {author}
  {\bibfnamefont {S.}~\bibnamefont {Shirol}}, \bibinfo {author} {\bibfnamefont
  {J.}~\bibnamefont {Koch}}, \ and\ \bibinfo {author} {\bibfnamefont
  {C.}~\bibnamefont {Wang}},\ }\bibfield  {title} {\enquote {\bibinfo {title}
  {Protecting a bosonic qubit with autonomous quantum error correction},}\
  }\href {\doibase 10.1038/s41586-021-03257-0} {\bibfield  {journal} {\bibinfo
  {journal} {Nature}\ }\textbf {\bibinfo {volume} {590}},\ \bibinfo {pages}
  {243} (\bibinfo {year} {2021})}\BibitemShut {NoStop}%
\bibitem [{\citenamefont {Reinhold}\ \emph {et~al.}(2020)\citenamefont
  {Reinhold}, \citenamefont {Rosenblum}, \citenamefont {Ma}, \citenamefont
  {Frunzio}, \citenamefont {Jiang},\ and\ \citenamefont
  {Schoelkopf}}]{Reinhold2020ErrorcorrectedGate}%
  \BibitemOpen
  \bibfield  {author} {\bibinfo {author} {\bibfnamefont {P.}~\bibnamefont
  {Reinhold}}, \bibinfo {author} {\bibfnamefont {S.}~\bibnamefont {Rosenblum}},
  \bibinfo {author} {\bibfnamefont {W.-L.}\ \bibnamefont {Ma}}, \bibinfo
  {author} {\bibfnamefont {L.}~\bibnamefont {Frunzio}}, \bibinfo {author}
  {\bibfnamefont {L.}~\bibnamefont {Jiang}}, \ and\ \bibinfo {author}
  {\bibfnamefont {R.~J.}\ \bibnamefont {Schoelkopf}},\ }\bibfield  {title}
  {\enquote {\bibinfo {title} {Error-corrected gates on an encoded qubit},}\
  }\href {\doibase 10.1038/s41567-020-0931-8} {\bibfield  {journal} {\bibinfo
  {journal} {Nat. Phys.}\ }\textbf {\bibinfo {volume} {16}},\ \bibinfo {pages}
  {822} (\bibinfo {year} {2020})}\BibitemShut {NoStop}%
\bibitem [{\citenamefont {Ofek}\ \emph {et~al.}(2016)\citenamefont {Ofek},
  \citenamefont {Petrenko}, \citenamefont {Heeres}, \citenamefont {Reinhold},
  \citenamefont {Leghtas}, \citenamefont {Vlastakis}, \citenamefont {Liu},
  \citenamefont {Frunzio}, \citenamefont {Girvin}, \citenamefont {Jiang} \emph
  {et~al.}}]{Ofek2016Extending}%
  \BibitemOpen
  \bibfield  {author} {\bibinfo {author} {\bibfnamefont {N.}~\bibnamefont
  {Ofek}}, \bibinfo {author} {\bibfnamefont {A.}~\bibnamefont {Petrenko}},
  \bibinfo {author} {\bibfnamefont {R.}~\bibnamefont {Heeres}}, \bibinfo
  {author} {\bibfnamefont {P.}~\bibnamefont {Reinhold}}, \bibinfo {author}
  {\bibfnamefont {Z.}~\bibnamefont {Leghtas}}, \bibinfo {author} {\bibfnamefont
  {B.}~\bibnamefont {Vlastakis}}, \bibinfo {author} {\bibfnamefont
  {Y.}~\bibnamefont {Liu}}, \bibinfo {author} {\bibfnamefont {L.}~\bibnamefont
  {Frunzio}}, \bibinfo {author} {\bibfnamefont {S.~M.}\ \bibnamefont {Girvin}},
  \bibinfo {author} {\bibfnamefont {L.}~\bibnamefont {Jiang}},  \emph
  {et~al.},\ }\bibfield  {title} {\enquote {\bibinfo {title} {Extending the
  lifetime of a quantum bit with error correction in superconducting
  circuits},}\ }\href {\doibase 10.1038/nature18949} {\bibfield  {journal}
  {\bibinfo  {journal} {Nature}\ }\textbf {\bibinfo {volume} {536}},\ \bibinfo
  {pages} {441} (\bibinfo {year} {2016})}\BibitemShut {NoStop}%
\bibitem [{\citenamefont {Ni}\ \emph {et~al.}(2023)\citenamefont {Ni},
  \citenamefont {Li}, \citenamefont {Deng}, \citenamefont {Cai}, \citenamefont
  {Zhang}, \citenamefont {Wang}, \citenamefont {Yang}, \citenamefont {Yu},
  \citenamefont {Yan}, \citenamefont {Liu} \emph {et~al.}}]{Ni2023Beating}%
  \BibitemOpen
  \bibfield  {author} {\bibinfo {author} {\bibfnamefont {Z.}~\bibnamefont
  {Ni}}, \bibinfo {author} {\bibfnamefont {S.}~\bibnamefont {Li}}, \bibinfo
  {author} {\bibfnamefont {X.}~\bibnamefont {Deng}}, \bibinfo {author}
  {\bibfnamefont {Y.}~\bibnamefont {Cai}}, \bibinfo {author} {\bibfnamefont
  {L.}~\bibnamefont {Zhang}}, \bibinfo {author} {\bibfnamefont
  {W.}~\bibnamefont {Wang}}, \bibinfo {author} {\bibfnamefont {Z.-B.}\
  \bibnamefont {Yang}}, \bibinfo {author} {\bibfnamefont {H.}~\bibnamefont
  {Yu}}, \bibinfo {author} {\bibfnamefont {F.}~\bibnamefont {Yan}}, \bibinfo
  {author} {\bibfnamefont {S.}~\bibnamefont {Liu}},  \emph {et~al.},\
  }\bibfield  {title} {\enquote {\bibinfo {title} {Beating the break-even point
  with a discrete-variable-encoded logical qubit},}\ }\href {\doibase
  10.1038/s41586-023-05784-4} {\bibfield  {journal} {\bibinfo  {journal}
  {Nature}\ }\textbf {\bibinfo {volume} {616}},\ \bibinfo {pages} {56}
  (\bibinfo {year} {2023})}\BibitemShut {NoStop}%
\bibitem [{\citenamefont {Sivak}\ \emph {et~al.}(2023)\citenamefont {Sivak},
  \citenamefont {Eickbusch}, \citenamefont {Royer}, \citenamefont {Singh},
  \citenamefont {Tsioutsios}, \citenamefont {Ganjam}, \citenamefont {Miano},
  \citenamefont {Brock}, \citenamefont {Ding}, \citenamefont {Frunzio} \emph
  {et~al.}}]{Sivak2023Realtime}%
  \BibitemOpen
  \bibfield  {author} {\bibinfo {author} {\bibfnamefont {V.~V.}\ \bibnamefont
  {Sivak}}, \bibinfo {author} {\bibfnamefont {A.}~\bibnamefont {Eickbusch}},
  \bibinfo {author} {\bibfnamefont {B.}~\bibnamefont {Royer}}, \bibinfo
  {author} {\bibfnamefont {S.}~\bibnamefont {Singh}}, \bibinfo {author}
  {\bibfnamefont {I.}~\bibnamefont {Tsioutsios}}, \bibinfo {author}
  {\bibfnamefont {S.}~\bibnamefont {Ganjam}}, \bibinfo {author} {\bibfnamefont
  {A.}~\bibnamefont {Miano}}, \bibinfo {author} {\bibfnamefont {B.~L.}\
  \bibnamefont {Brock}}, \bibinfo {author} {\bibfnamefont {A.~Z.}\ \bibnamefont
  {Ding}}, \bibinfo {author} {\bibfnamefont {L.}~\bibnamefont {Frunzio}},
  \emph {et~al.},\ }\bibfield  {title} {\enquote {\bibinfo {title} {Real-time
  quantum error correction beyond break-even},}\ }\href {\doibase
  10.1038/s41586-023-05782-6} {\bibfield  {journal} {\bibinfo  {journal}
  {Nature}\ }\textbf {\bibinfo {volume} {616}},\ \bibinfo {pages} {50}
  (\bibinfo {year} {2023})}\BibitemShut {NoStop}%
\bibitem [{\citenamefont {Mirrahimi}\ \emph {et~al.}(2014)\citenamefont
  {Mirrahimi}, \citenamefont {Leghtas}, \citenamefont {Albert}, \citenamefont
  {Touzard}, \citenamefont {Schoelkopf}, \citenamefont {Jiang},\ and\
  \citenamefont {Devoret}}]{Mirrahimi2014Dynamically}%
  \BibitemOpen
  \bibfield  {author} {\bibinfo {author} {\bibfnamefont {M.}~\bibnamefont
  {Mirrahimi}}, \bibinfo {author} {\bibfnamefont {Z.}~\bibnamefont {Leghtas}},
  \bibinfo {author} {\bibfnamefont {V.~V.}\ \bibnamefont {Albert}}, \bibinfo
  {author} {\bibfnamefont {S.}~\bibnamefont {Touzard}}, \bibinfo {author}
  {\bibfnamefont {R.~J.}\ \bibnamefont {Schoelkopf}}, \bibinfo {author}
  {\bibfnamefont {L.}~\bibnamefont {Jiang}}, \ and\ \bibinfo {author}
  {\bibfnamefont {M.~H.}\ \bibnamefont {Devoret}},\ }\bibfield  {title}
  {\enquote {\bibinfo {title} {Dynamically protected cat-qubits: a new paradigm
  for universal quantum computation},}\ }\href {\doibase
  10.1088/1367-2630/16/4/045014} {\bibfield  {journal} {\bibinfo  {journal}
  {New Journal of Physics}\ }\textbf {\bibinfo {volume} {16}},\ \bibinfo
  {pages} {045014} (\bibinfo {year} {2014})}\BibitemShut {NoStop}%
\bibitem [{\citenamefont {Teoh}\ \emph {et~al.}(2023)\citenamefont {Teoh},
  \citenamefont {Winkel}, \citenamefont {Babla}, \citenamefont {Chapman},
  \citenamefont {Claes}, \citenamefont {de~Graaf}, \citenamefont {Garmon},
  \citenamefont {Kalfus}, \citenamefont {Lu}, \citenamefont {Maiti} \emph
  {et~al.}}]{Teoh2023Dualrail}%
  \BibitemOpen
  \bibfield  {author} {\bibinfo {author} {\bibfnamefont {J.~D.}\ \bibnamefont
  {Teoh}}, \bibinfo {author} {\bibfnamefont {P.}~\bibnamefont {Winkel}},
  \bibinfo {author} {\bibfnamefont {H.~K.}\ \bibnamefont {Babla}}, \bibinfo
  {author} {\bibfnamefont {B.~J.}\ \bibnamefont {Chapman}}, \bibinfo {author}
  {\bibfnamefont {J.}~\bibnamefont {Claes}}, \bibinfo {author} {\bibfnamefont
  {S.~J.}\ \bibnamefont {de~Graaf}}, \bibinfo {author} {\bibfnamefont
  {J.~W.~O.}\ \bibnamefont {Garmon}}, \bibinfo {author} {\bibfnamefont {W.~D.}\
  \bibnamefont {Kalfus}}, \bibinfo {author} {\bibfnamefont {Y.}~\bibnamefont
  {Lu}}, \bibinfo {author} {\bibfnamefont {A.}~\bibnamefont {Maiti}},  \emph
  {et~al.},\ }\bibfield  {title} {\enquote {\bibinfo {title} {Dual-rail
  encoding with superconducting cavities},}\ }\href {\doibase
  doi:10.1073/pnas.2221736120} {\bibfield  {journal} {\bibinfo  {journal}
  {Proceedings of the National Academy of Sciences}\ }\textbf {\bibinfo
  {volume} {120}},\ \bibinfo {pages} {e2221736120} (\bibinfo {year}
  {2023})}\BibitemShut {NoStop}%
\bibitem [{\citenamefont {Grimm}\ \emph {et~al.}(2020)\citenamefont {Grimm},
  \citenamefont {Frattini}, \citenamefont {Puri}, \citenamefont {Mundhada},
  \citenamefont {Touzard}, \citenamefont {Mirrahimi}, \citenamefont {Girvin},
  \citenamefont {Shankar},\ and\ \citenamefont
  {Devoret}}]{Grimm2020Stabilization}%
  \BibitemOpen
  \bibfield  {author} {\bibinfo {author} {\bibfnamefont {A.}~\bibnamefont
  {Grimm}}, \bibinfo {author} {\bibfnamefont {N.~E.}\ \bibnamefont {Frattini}},
  \bibinfo {author} {\bibfnamefont {S.}~\bibnamefont {Puri}}, \bibinfo {author}
  {\bibfnamefont {S.~O.}\ \bibnamefont {Mundhada}}, \bibinfo {author}
  {\bibfnamefont {S.}~\bibnamefont {Touzard}}, \bibinfo {author} {\bibfnamefont
  {M.}~\bibnamefont {Mirrahimi}}, \bibinfo {author} {\bibfnamefont {S.~M.}\
  \bibnamefont {Girvin}}, \bibinfo {author} {\bibfnamefont {S.}~\bibnamefont
  {Shankar}}, \ and\ \bibinfo {author} {\bibfnamefont {M.~H.}\ \bibnamefont
  {Devoret}},\ }\bibfield  {title} {\enquote {\bibinfo {title} {Stabilization
  and operation of a kerr-cat qubit},}\ }\href {\doibase
  10.1038/s41586-020-2587-z} {\bibfield  {journal} {\bibinfo  {journal}
  {Nature}\ }\textbf {\bibinfo {volume} {584}},\ \bibinfo {pages} {205}
  (\bibinfo {year} {2020})}\BibitemShut {NoStop}%
\bibitem [{\citenamefont {Lescanne}\ \emph {et~al.}(2020)\citenamefont
  {Lescanne}, \citenamefont {Villiers}, \citenamefont {Peronnin}, \citenamefont
  {Sarlette}, \citenamefont {Delbecq}, \citenamefont {Huard}, \citenamefont
  {Kontos}, \citenamefont {Mirrahimi},\ and\ \citenamefont
  {Leghtas}}]{Lescanne2020Exponential}%
  \BibitemOpen
  \bibfield  {author} {\bibinfo {author} {\bibfnamefont {R.}~\bibnamefont
  {Lescanne}}, \bibinfo {author} {\bibfnamefont {M.}~\bibnamefont {Villiers}},
  \bibinfo {author} {\bibfnamefont {T.}~\bibnamefont {Peronnin}}, \bibinfo
  {author} {\bibfnamefont {A.}~\bibnamefont {Sarlette}}, \bibinfo {author}
  {\bibfnamefont {M.}~\bibnamefont {Delbecq}}, \bibinfo {author} {\bibfnamefont
  {B.}~\bibnamefont {Huard}}, \bibinfo {author} {\bibfnamefont
  {T.}~\bibnamefont {Kontos}}, \bibinfo {author} {\bibfnamefont
  {M.}~\bibnamefont {Mirrahimi}}, \ and\ \bibinfo {author} {\bibfnamefont
  {Z.}~\bibnamefont {Leghtas}},\ }\bibfield  {title} {\enquote {\bibinfo
  {title} {Exponential suppression of bit-flips in a qubit encoded in an
  oscillator},}\ }\href {\doibase 10.1038/s41567-020-0824-x} {\bibfield
  {journal} {\bibinfo  {journal} {Nat. Phys.}\ }\textbf {\bibinfo {volume}
  {16}},\ \bibinfo {pages} {509} (\bibinfo {year} {2020})}\BibitemShut
  {NoStop}%
\bibitem [{\citenamefont {Berdou}\ \emph {et~al.}(2023)\citenamefont {Berdou},
  \citenamefont {Murani}, \citenamefont {R\'{e}glade}, \citenamefont {Smith},
  \citenamefont {Villiers}, \citenamefont {Palomo}, \citenamefont {Rosticher},
  \citenamefont {Denis}, \citenamefont {Morfin}, \citenamefont {Delbecq} \emph
  {et~al.}}]{Berdou2023OneHundred}%
  \BibitemOpen
  \bibfield  {author} {\bibinfo {author} {\bibfnamefont {C.}~\bibnamefont
  {Berdou}}, \bibinfo {author} {\bibfnamefont {A.}~\bibnamefont {Murani}},
  \bibinfo {author} {\bibfnamefont {U.}~\bibnamefont {R\'{e}glade}}, \bibinfo
  {author} {\bibfnamefont {W.~C.}\ \bibnamefont {Smith}}, \bibinfo {author}
  {\bibfnamefont {M.}~\bibnamefont {Villiers}}, \bibinfo {author}
  {\bibfnamefont {J.}~\bibnamefont {Palomo}}, \bibinfo {author} {\bibfnamefont
  {M.}~\bibnamefont {Rosticher}}, \bibinfo {author} {\bibfnamefont
  {A.}~\bibnamefont {Denis}}, \bibinfo {author} {\bibfnamefont
  {P.}~\bibnamefont {Morfin}}, \bibinfo {author} {\bibfnamefont
  {M.}~\bibnamefont {Delbecq}},  \emph {et~al.},\ }\bibfield  {title} {\enquote
  {\bibinfo {title} {One hundred second bit-flip time in a two-photon
  dissipative oscillator},}\ }\href {\doibase 10.1103/PRXQuantum.4.020350}
  {\bibfield  {journal} {\bibinfo  {journal} {PRX Quantum}\ }\textbf {\bibinfo
  {volume} {4}},\ \bibinfo {pages} {020350} (\bibinfo {year}
  {2023})}\BibitemShut {NoStop}%
\bibitem [{\citenamefont {R{\'e}glade}\ \emph {et~al.}(2024)\citenamefont
  {R{\'e}glade}, \citenamefont {Bocquet}, \citenamefont {Gautier},
  \citenamefont {Cohen}, \citenamefont {Marquet}, \citenamefont {Albertinale},
  \citenamefont {Pankratova}, \citenamefont {Hall\'{e}n}, \citenamefont
  {Rautschke}, \citenamefont {Sellem} \emph
  {et~al.}}]{Reglade2024NatureDCat10s}%
  \BibitemOpen
  \bibfield  {author} {\bibinfo {author} {\bibfnamefont {U.}~\bibnamefont
  {R{\'e}glade}}, \bibinfo {author} {\bibfnamefont {A.}~\bibnamefont
  {Bocquet}}, \bibinfo {author} {\bibfnamefont {R.}~\bibnamefont {Gautier}},
  \bibinfo {author} {\bibfnamefont {J.}~\bibnamefont {Cohen}}, \bibinfo
  {author} {\bibfnamefont {A.}~\bibnamefont {Marquet}}, \bibinfo {author}
  {\bibfnamefont {E.}~\bibnamefont {Albertinale}}, \bibinfo {author}
  {\bibfnamefont {N.}~\bibnamefont {Pankratova}}, \bibinfo {author}
  {\bibfnamefont {M.}~\bibnamefont {Hall\'{e}n}}, \bibinfo {author}
  {\bibfnamefont {F.}~\bibnamefont {Rautschke}}, \bibinfo {author}
  {\bibfnamefont {L.~A.}\ \bibnamefont {Sellem}},  \emph {et~al.},\ }\bibfield
  {title} {\enquote {\bibinfo {title} {Quantum control of a cat qubit with
  bit-flip times exceeding ten seconds},}\ }\href {\doibase
  10.1038/s41586-024-07294-3} {\bibfield  {journal} {\bibinfo  {journal}
  {Nature}\ }\textbf {\bibinfo {volume} {629}},\ \bibinfo {pages} {778}
  (\bibinfo {year} {2024})}\BibitemShut {NoStop}%
\bibitem [{\citenamefont {Guillaud}\ and\ \citenamefont
  {Mirrahimi}(2019)}]{Guillaud2019Repetition}%
  \BibitemOpen
  \bibfield  {author} {\bibinfo {author} {\bibfnamefont {J.}~\bibnamefont
  {Guillaud}}\ and\ \bibinfo {author} {\bibfnamefont {M.}~\bibnamefont
  {Mirrahimi}},\ }\bibfield  {title} {\enquote {\bibinfo {title} {Repetition
  cat qubits for fault-tolerant quantum computation},}\ }\href {\doibase
  10.1103/PhysRevX.9.041053} {\bibfield  {journal} {\bibinfo  {journal} {Phys.
  Rev. X}\ }\textbf {\bibinfo {volume} {9}},\ \bibinfo {pages} {041053}
  (\bibinfo {year} {2019})}\BibitemShut {NoStop}%
\bibitem [{\citenamefont {Puri}\ \emph {et~al.}(2020)\citenamefont {Puri},
  \citenamefont {St-Jean}, \citenamefont {Gross}, \citenamefont {Grimm},
  \citenamefont {Frattini}, \citenamefont {Iyer}, \citenamefont {Krishna},
  \citenamefont {Touzard}, \citenamefont {Jiang}, \citenamefont {Blais} \emph
  {et~al.}}]{Puri2020Bias-preserving}%
  \BibitemOpen
  \bibfield  {author} {\bibinfo {author} {\bibfnamefont {S.}~\bibnamefont
  {Puri}}, \bibinfo {author} {\bibfnamefont {L.}~\bibnamefont {St-Jean}},
  \bibinfo {author} {\bibfnamefont {J.~A.}\ \bibnamefont {Gross}}, \bibinfo
  {author} {\bibfnamefont {A.}~\bibnamefont {Grimm}}, \bibinfo {author}
  {\bibfnamefont {N.~E.}\ \bibnamefont {Frattini}}, \bibinfo {author}
  {\bibfnamefont {P.~S.}\ \bibnamefont {Iyer}}, \bibinfo {author}
  {\bibfnamefont {A.}~\bibnamefont {Krishna}}, \bibinfo {author} {\bibfnamefont
  {S.}~\bibnamefont {Touzard}}, \bibinfo {author} {\bibfnamefont
  {L.}~\bibnamefont {Jiang}}, \bibinfo {author} {\bibfnamefont
  {A.}~\bibnamefont {Blais}},  \emph {et~al.},\ }\bibfield  {title} {\enquote
  {\bibinfo {title} {Bias-preserving gates with stabilized cat qubits},}\
  }\href {\doibase doi:10.1126/sciadv.aay5901} {\bibfield  {journal} {\bibinfo
  {journal} {Science Advances}\ }\textbf {\bibinfo {volume} {6}},\ \bibinfo
  {pages} {eaay5901} (\bibinfo {year} {2020})}\BibitemShut {NoStop}%
\bibitem [{\citenamefont {Guillaud}\ and\ \citenamefont
  {Mirrahimi}(2021)}]{Guillaud2021Error}%
  \BibitemOpen
  \bibfield  {author} {\bibinfo {author} {\bibfnamefont {J.}~\bibnamefont
  {Guillaud}}\ and\ \bibinfo {author} {\bibfnamefont {M.}~\bibnamefont
  {Mirrahimi}},\ }\bibfield  {title} {\enquote {\bibinfo {title} {Error rates
  and resource overheads of repetition cat qubits},}\ }\href {\doibase
  10.1103/PhysRevA.103.042413} {\bibfield  {journal} {\bibinfo  {journal}
  {Phys. Rev. A}\ }\textbf {\bibinfo {volume} {103}},\ \bibinfo {pages}
  {042413} (\bibinfo {year} {2021})}\BibitemShut {NoStop}%
\bibitem [{\citenamefont {Putterman}\ \emph {et~al.}(2024)\citenamefont
  {Putterman}, \citenamefont {Noh}, \citenamefont {Hann}, \citenamefont
  {MacCabe}, \citenamefont {Aghaeimeibodi}, \citenamefont {Patel},
  \citenamefont {Lee}, \citenamefont {Jones}, \citenamefont {Moradinejad},\
  and\ \citenamefont {Rodriguez}}]{Putterman2024ConcatenatedCat}%
  \BibitemOpen
  \bibfield  {author} {\bibinfo {author} {\bibfnamefont {H.}~\bibnamefont
  {Putterman}}, \bibinfo {author} {\bibfnamefont {K.}~\bibnamefont {Noh}},
  \bibinfo {author} {\bibfnamefont {C.~T.}\ \bibnamefont {Hann}}, \bibinfo
  {author} {\bibfnamefont {G.~S.}\ \bibnamefont {MacCabe}}, \bibinfo {author}
  {\bibfnamefont {S.}~\bibnamefont {Aghaeimeibodi}}, \bibinfo {author}
  {\bibfnamefont {R.~N.}\ \bibnamefont {Patel}}, \bibinfo {author}
  {\bibfnamefont {M.}~\bibnamefont {Lee}}, \bibinfo {author} {\bibfnamefont
  {W.~M.}\ \bibnamefont {Jones}}, \bibinfo {author} {\bibfnamefont
  {H.}~\bibnamefont {Moradinejad}}, \ and\ \bibinfo {author} {\bibfnamefont
  {R.}~\bibnamefont {Rodriguez}},\ }\bibfield  {title} {\enquote {\bibinfo
  {title} {Hardware-efficient quantum error correction using concatenated
  bosonic qubits},}\ }\href@noop {} {\bibfield  {journal} {\bibinfo  {journal}
  {arXiv preprint arXiv:2409.13025}\ } (\bibinfo {year} {2024})}\BibitemShut
  {NoStop}%
\bibitem [{\citenamefont {Darmawan}\ \emph {et~al.}(2021)\citenamefont
  {Darmawan}, \citenamefont {Brown}, \citenamefont {Grimsmo}, \citenamefont
  {Tuckett},\ and\ \citenamefont {Puri}}]{Darmawan2021Practical}%
  \BibitemOpen
  \bibfield  {author} {\bibinfo {author} {\bibfnamefont {A.~S.}\ \bibnamefont
  {Darmawan}}, \bibinfo {author} {\bibfnamefont {B.~J.}\ \bibnamefont {Brown}},
  \bibinfo {author} {\bibfnamefont {A.~L.}\ \bibnamefont {Grimsmo}}, \bibinfo
  {author} {\bibfnamefont {D.~K.}\ \bibnamefont {Tuckett}}, \ and\ \bibinfo
  {author} {\bibfnamefont {S.}~\bibnamefont {Puri}},\ }\bibfield  {title}
  {\enquote {\bibinfo {title} {Practical quantum error correction with the xzzx
  code and kerr-cat qubits},}\ }\href {\doibase 10.1103/PRXQuantum.2.030345}
  {\bibfield  {journal} {\bibinfo  {journal} {PRX Quantum}\ }\textbf {\bibinfo
  {volume} {2}},\ \bibinfo {pages} {030345} (\bibinfo {year}
  {2021})}\BibitemShut {NoStop}%
\bibitem [{\citenamefont {Chamberland}\ \emph {et~al.}(2022)\citenamefont
  {Chamberland}, \citenamefont {Noh}, \citenamefont {Arrangoiz-Arriola},
  \citenamefont {Campbell}, \citenamefont {Hann}, \citenamefont {Iverson},
  \citenamefont {Putterman}, \citenamefont {Bohdanowicz}, \citenamefont
  {Flammia}, \citenamefont {Keller} \emph {et~al.}}]{Chamberland2022Building}%
  \BibitemOpen
  \bibfield  {author} {\bibinfo {author} {\bibfnamefont {C.}~\bibnamefont
  {Chamberland}}, \bibinfo {author} {\bibfnamefont {K.}~\bibnamefont {Noh}},
  \bibinfo {author} {\bibfnamefont {P.}~\bibnamefont {Arrangoiz-Arriola}},
  \bibinfo {author} {\bibfnamefont {E.~T.}\ \bibnamefont {Campbell}}, \bibinfo
  {author} {\bibfnamefont {C.~T.}\ \bibnamefont {Hann}}, \bibinfo {author}
  {\bibfnamefont {J.}~\bibnamefont {Iverson}}, \bibinfo {author} {\bibfnamefont
  {H.}~\bibnamefont {Putterman}}, \bibinfo {author} {\bibfnamefont {T.~C.}\
  \bibnamefont {Bohdanowicz}}, \bibinfo {author} {\bibfnamefont {S.~T.}\
  \bibnamefont {Flammia}}, \bibinfo {author} {\bibfnamefont {A.}~\bibnamefont
  {Keller}},  \emph {et~al.},\ }\bibfield  {title} {\enquote {\bibinfo {title}
  {Building a fault-tolerant quantum computer using concatenated cat codes},}\
  }\href {\doibase 10.1103/PRXQuantum.3.010329} {\bibfield  {journal} {\bibinfo
   {journal} {PRX Quantum}\ }\textbf {\bibinfo {volume} {3}},\ \bibinfo {pages}
  {010329} (\bibinfo {year} {2022})}\BibitemShut {NoStop}%
\bibitem [{\citenamefont {Aliferis}\ and\ \citenamefont
  {Preskill}(2008)}]{Aliferis2008Fault-tolerent}%
  \BibitemOpen
  \bibfield  {author} {\bibinfo {author} {\bibfnamefont {P.}~\bibnamefont
  {Aliferis}}\ and\ \bibinfo {author} {\bibfnamefont {J.}~\bibnamefont
  {Preskill}},\ }\bibfield  {title} {\enquote {\bibinfo {title} {Fault-tolerant
  quantum computation against biased noise},}\ }\href {\doibase
  10.1103/PhysRevA.78.052331} {\bibfield  {journal} {\bibinfo  {journal} {Phys.
  Rev. A}\ }\textbf {\bibinfo {volume} {78}},\ \bibinfo {pages} {052331}
  (\bibinfo {year} {2008})}\BibitemShut {NoStop}%
\bibitem [{\citenamefont {Tuckett}\ \emph {et~al.}(2018)\citenamefont
  {Tuckett}, \citenamefont {Bartlett},\ and\ \citenamefont
  {Flammia}}]{Tuckett2018Ultrahigh}%
  \BibitemOpen
  \bibfield  {author} {\bibinfo {author} {\bibfnamefont {D.~K.}\ \bibnamefont
  {Tuckett}}, \bibinfo {author} {\bibfnamefont {S.~D.}\ \bibnamefont
  {Bartlett}}, \ and\ \bibinfo {author} {\bibfnamefont {S.~T.}\ \bibnamefont
  {Flammia}},\ }\bibfield  {title} {\enquote {\bibinfo {title} {Ultrahigh error
  threshold for surface codes with biased noise},}\ }\href {\doibase
  10.1103/PhysRevLett.120.050505} {\bibfield  {journal} {\bibinfo  {journal}
  {Phys. Rev. Lett.}\ }\textbf {\bibinfo {volume} {120}},\ \bibinfo {pages}
  {050505} (\bibinfo {year} {2018})}\BibitemShut {NoStop}%
\bibitem [{\citenamefont {Bonilla~Ataides}\ \emph {et~al.}(2021)\citenamefont
  {Bonilla~Ataides}, \citenamefont {Tuckett}, \citenamefont {Bartlett},
  \citenamefont {Flammia},\ and\ \citenamefont {Brown}}]{Bonilla2021XZZX}%
  \BibitemOpen
  \bibfield  {author} {\bibinfo {author} {\bibfnamefont {J.~P.}\ \bibnamefont
  {Bonilla~Ataides}}, \bibinfo {author} {\bibfnamefont {D.~K.}\ \bibnamefont
  {Tuckett}}, \bibinfo {author} {\bibfnamefont {S.~D.}\ \bibnamefont
  {Bartlett}}, \bibinfo {author} {\bibfnamefont {S.~T.}\ \bibnamefont
  {Flammia}}, \ and\ \bibinfo {author} {\bibfnamefont {B.~J.}\ \bibnamefont
  {Brown}},\ }\bibfield  {title} {\enquote {\bibinfo {title} {The xzzx surface
  code},}\ }\href {\doibase 10.1038/s41467-021-22274-1} {\bibfield  {journal}
  {\bibinfo  {journal} {Nat. Commun.}\ }\textbf {\bibinfo {volume} {12}},\
  \bibinfo {pages} {2172} (\bibinfo {year} {2021})}\BibitemShut {NoStop}%
\bibitem [{\citenamefont {Blais}\ \emph {et~al.}(2004)\citenamefont {Blais},
  \citenamefont {Huang}, \citenamefont {Wallraff}, \citenamefont {Girvin},\
  and\ \citenamefont {Schoelkopf}}]{Blais2004CavityQED}%
  \BibitemOpen
  \bibfield  {author} {\bibinfo {author} {\bibfnamefont {A.}~\bibnamefont
  {Blais}}, \bibinfo {author} {\bibfnamefont {R.-S.}\ \bibnamefont {Huang}},
  \bibinfo {author} {\bibfnamefont {A.}~\bibnamefont {Wallraff}}, \bibinfo
  {author} {\bibfnamefont {S.~M.}\ \bibnamefont {Girvin}}, \ and\ \bibinfo
  {author} {\bibfnamefont {R.~J.}\ \bibnamefont {Schoelkopf}},\ }\bibfield
  {title} {\enquote {\bibinfo {title} {Cavity quantum electrodynamics for
  superconducting electrical circuits: An architecture for quantum
  computation},}\ }\href {\doibase 10.1103/PhysRevA.69.062320} {\bibfield
  {journal} {\bibinfo  {journal} {Phys. Rev. A}\ }\textbf {\bibinfo {volume}
  {69}},\ \bibinfo {pages} {062320} (\bibinfo {year} {2004})}\BibitemShut
  {NoStop}%
\bibitem [{\citenamefont {Khaneja}\ \emph {et~al.}(2005)\citenamefont
  {Khaneja}, \citenamefont {Reiss}, \citenamefont {Kehlet}, \citenamefont
  {Schulte-Herbr{\"u}ggen},\ and\ \citenamefont {Glaser}}]{Khaneja2005GEAPE}%
  \BibitemOpen
  \bibfield  {author} {\bibinfo {author} {\bibfnamefont {N.}~\bibnamefont
  {Khaneja}}, \bibinfo {author} {\bibfnamefont {T.}~\bibnamefont {Reiss}},
  \bibinfo {author} {\bibfnamefont {C.}~\bibnamefont {Kehlet}}, \bibinfo
  {author} {\bibfnamefont {T.}~\bibnamefont {Schulte-Herbr{\"u}ggen}}, \ and\
  \bibinfo {author} {\bibfnamefont {S.~J.}\ \bibnamefont {Glaser}},\ }\bibfield
   {title} {\enquote {\bibinfo {title} {Optimal control of coupled spin
  dynamics: design of nmr pulse sequences by gradient ascent algorithms},}\
  }\href {\doibase https://doi.org/10.1016/j.jmr.2004.11.004} {\bibfield
  {journal} {\bibinfo  {journal} {Journal of Magnetic Resonance}\ }\textbf
  {\bibinfo {volume} {172}},\ \bibinfo {pages} {296} (\bibinfo {year}
  {2005})}\BibitemShut {NoStop}%
\bibitem [{\citenamefont {Eickbusch}\ \emph {et~al.}(2022)\citenamefont
  {Eickbusch}, \citenamefont {Sivak}, \citenamefont {Ding}, \citenamefont
  {Elder}, \citenamefont {Jha}, \citenamefont {Venkatraman}, \citenamefont
  {Royer}, \citenamefont {Girvin}, \citenamefont {Schoelkopf},\ and\
  \citenamefont {Devoret}}]{Eickbusch2022ECD}%
  \BibitemOpen
  \bibfield  {author} {\bibinfo {author} {\bibfnamefont {A.}~\bibnamefont
  {Eickbusch}}, \bibinfo {author} {\bibfnamefont {V.}~\bibnamefont {Sivak}},
  \bibinfo {author} {\bibfnamefont {A.~Z.}\ \bibnamefont {Ding}}, \bibinfo
  {author} {\bibfnamefont {S.~S.}\ \bibnamefont {Elder}}, \bibinfo {author}
  {\bibfnamefont {S.~R.}\ \bibnamefont {Jha}}, \bibinfo {author} {\bibfnamefont
  {J.}~\bibnamefont {Venkatraman}}, \bibinfo {author} {\bibfnamefont
  {B.}~\bibnamefont {Royer}}, \bibinfo {author} {\bibfnamefont {S.~M.}\
  \bibnamefont {Girvin}}, \bibinfo {author} {\bibfnamefont {R.~J.}\
  \bibnamefont {Schoelkopf}}, \ and\ \bibinfo {author} {\bibfnamefont {M.~H.}\
  \bibnamefont {Devoret}},\ }\bibfield  {title} {\enquote {\bibinfo {title}
  {Fast universal control of an oscillator with weak dispersive coupling to a
  qubit},}\ }\href {\doibase 10.1038/s41567-022-01776-9} {\bibfield  {journal}
  {\bibinfo  {journal} {Nat. Phys.}\ }\textbf {\bibinfo {volume} {18}},\
  \bibinfo {pages} {1464} (\bibinfo {year} {2022})}\BibitemShut {NoStop}%
\bibitem [{\citenamefont {Ma}\ \emph {et~al.}(2021)\citenamefont {Ma},
  \citenamefont {Puri}, \citenamefont {Schoelkopf}, \citenamefont {Devoret},
  \citenamefont {Girvin},\ and\ \citenamefont {Jiang}}]{Ma2021QuantumControl}%
  \BibitemOpen
  \bibfield  {author} {\bibinfo {author} {\bibfnamefont {W.-L.}\ \bibnamefont
  {Ma}}, \bibinfo {author} {\bibfnamefont {S.}~\bibnamefont {Puri}}, \bibinfo
  {author} {\bibfnamefont {R.~J.}\ \bibnamefont {Schoelkopf}}, \bibinfo
  {author} {\bibfnamefont {M.~H.}\ \bibnamefont {Devoret}}, \bibinfo {author}
  {\bibfnamefont {S.~M.}\ \bibnamefont {Girvin}}, \ and\ \bibinfo {author}
  {\bibfnamefont {L.}~\bibnamefont {Jiang}},\ }\bibfield  {title} {\enquote
  {\bibinfo {title} {Quantum control of bosonic modes with superconducting
  circuits},}\ }\href {\doibase https://doi.org/10.1016/j.scib.2021.05.024}
  {\bibfield  {journal} {\bibinfo  {journal} {Science Bulletin}\ }\textbf
  {\bibinfo {volume} {66}},\ \bibinfo {pages} {1789} (\bibinfo {year}
  {2021})}\BibitemShut {NoStop}%
\bibitem [{\citenamefont {Chapman}\ \emph {et~al.}(2023)\citenamefont
  {Chapman}, \citenamefont {de~Graaf}, \citenamefont {Xue}, \citenamefont
  {Zhang}, \citenamefont {Teoh}, \citenamefont {Curtis}, \citenamefont
  {Tsunoda}, \citenamefont {Eickbusch}, \citenamefont {Read}, \citenamefont
  {Koottandavida} \emph {et~al.}}]{Chapman2023SNAILBeamSplitter}%
  \BibitemOpen
  \bibfield  {author} {\bibinfo {author} {\bibfnamefont {B.~J.}\ \bibnamefont
  {Chapman}}, \bibinfo {author} {\bibfnamefont {S.~J.}\ \bibnamefont
  {de~Graaf}}, \bibinfo {author} {\bibfnamefont {S.~H.}\ \bibnamefont {Xue}},
  \bibinfo {author} {\bibfnamefont {Y.}~\bibnamefont {Zhang}}, \bibinfo
  {author} {\bibfnamefont {J.}~\bibnamefont {Teoh}}, \bibinfo {author}
  {\bibfnamefont {J.~C.}\ \bibnamefont {Curtis}}, \bibinfo {author}
  {\bibfnamefont {T.}~\bibnamefont {Tsunoda}}, \bibinfo {author} {\bibfnamefont
  {A.}~\bibnamefont {Eickbusch}}, \bibinfo {author} {\bibfnamefont {A.~P.}\
  \bibnamefont {Read}}, \bibinfo {author} {\bibfnamefont {A.}~\bibnamefont
  {Koottandavida}},  \emph {et~al.},\ }\bibfield  {title} {\enquote {\bibinfo
  {title} {High-on-off-ratio beam-splitter interaction for gates on bosonically
  encoded qubits},}\ }\href {\doibase 10.1103/PRXQuantum.4.020355} {\bibfield
  {journal} {\bibinfo  {journal} {PRX Quantum}\ }\textbf {\bibinfo {volume}
  {4}},\ \bibinfo {pages} {020355} (\bibinfo {year} {2023})}\BibitemShut
  {NoStop}%
\bibitem [{\citenamefont {Lu}\ \emph {et~al.}(2023)\citenamefont {Lu},
  \citenamefont {Maiti}, \citenamefont {Garmon}, \citenamefont {Ganjam},
  \citenamefont {Zhang}, \citenamefont {Claes}, \citenamefont {Frunzio},
  \citenamefont {Girvin},\ and\ \citenamefont
  {Schoelkopf}}]{Yao2023Highfidelity}%
  \BibitemOpen
  \bibfield  {author} {\bibinfo {author} {\bibfnamefont {Y.}~\bibnamefont
  {Lu}}, \bibinfo {author} {\bibfnamefont {A.}~\bibnamefont {Maiti}}, \bibinfo
  {author} {\bibfnamefont {J.~W.~O.}\ \bibnamefont {Garmon}}, \bibinfo {author}
  {\bibfnamefont {S.}~\bibnamefont {Ganjam}}, \bibinfo {author} {\bibfnamefont
  {Y.}~\bibnamefont {Zhang}}, \bibinfo {author} {\bibfnamefont
  {J.}~\bibnamefont {Claes}}, \bibinfo {author} {\bibfnamefont
  {L.}~\bibnamefont {Frunzio}}, \bibinfo {author} {\bibfnamefont {S.~M.}\
  \bibnamefont {Girvin}}, \ and\ \bibinfo {author} {\bibfnamefont {R.~J.}\
  \bibnamefont {Schoelkopf}},\ }\bibfield  {title} {\enquote {\bibinfo {title}
  {High-fidelity parametric beamsplitting with a parity-protected converter},}\
  }\href {\doibase 10.1038/s41467-023-41104-0} {\bibfield  {journal} {\bibinfo
  {journal} {Nat. Commun.}\ }\textbf {\bibinfo {volume} {14}},\ \bibinfo
  {pages} {5767} (\bibinfo {year} {2023})}\BibitemShut {NoStop}%
\bibitem [{\citenamefont {Koch}\ \emph {et~al.}(2007)\citenamefont {Koch},
  \citenamefont {Yu}, \citenamefont {Gambetta}, \citenamefont {Houck},
  \citenamefont {Schuster}, \citenamefont {Majer}, \citenamefont {Blais},
  \citenamefont {Devoret}, \citenamefont {Girvin},\ and\ \citenamefont
  {Schoelkopf}}]{Koch2007Transmon}%
  \BibitemOpen
  \bibfield  {author} {\bibinfo {author} {\bibfnamefont {J.}~\bibnamefont
  {Koch}}, \bibinfo {author} {\bibfnamefont {T.~M.}\ \bibnamefont {Yu}},
  \bibinfo {author} {\bibfnamefont {J.}~\bibnamefont {Gambetta}}, \bibinfo
  {author} {\bibfnamefont {A.~A.}\ \bibnamefont {Houck}}, \bibinfo {author}
  {\bibfnamefont {D.~I.}\ \bibnamefont {Schuster}}, \bibinfo {author}
  {\bibfnamefont {J.}~\bibnamefont {Majer}}, \bibinfo {author} {\bibfnamefont
  {A.}~\bibnamefont {Blais}}, \bibinfo {author} {\bibfnamefont {M.~H.}\
  \bibnamefont {Devoret}}, \bibinfo {author} {\bibfnamefont {S.~M.}\
  \bibnamefont {Girvin}}, \ and\ \bibinfo {author} {\bibfnamefont {R.~J.}\
  \bibnamefont {Schoelkopf}},\ }\bibfield  {title} {\enquote {\bibinfo {title}
  {Charge-insensitive qubit design derived from the cooper pair box},}\ }\href
  {\doibase 10.1103/PhysRevA.76.042319} {\bibfield  {journal} {\bibinfo
  {journal} {Phys. Rev. A}\ }\textbf {\bibinfo {volume} {76}},\ \bibinfo
  {pages} {042319} (\bibinfo {year} {2007})}\BibitemShut {NoStop}%
\bibitem [{\citenamefont {Leghtas}\ \emph {et~al.}(2015)\citenamefont
  {Leghtas}, \citenamefont {Touzard}, \citenamefont {Pop}, \citenamefont {Kou},
  \citenamefont {Vlastakis}, \citenamefont {Petrenko}, \citenamefont {Sliwa},
  \citenamefont {Narla}, \citenamefont {Shankar}, \citenamefont {Hatridge}
  \emph {et~al.}}]{Leghtas2015Confining}%
  \BibitemOpen
  \bibfield  {author} {\bibinfo {author} {\bibfnamefont {Z.}~\bibnamefont
  {Leghtas}}, \bibinfo {author} {\bibfnamefont {S.}~\bibnamefont {Touzard}},
  \bibinfo {author} {\bibfnamefont {I.~M.}\ \bibnamefont {Pop}}, \bibinfo
  {author} {\bibfnamefont {A.}~\bibnamefont {Kou}}, \bibinfo {author}
  {\bibfnamefont {B.}~\bibnamefont {Vlastakis}}, \bibinfo {author}
  {\bibfnamefont {A.}~\bibnamefont {Petrenko}}, \bibinfo {author}
  {\bibfnamefont {K.~M.}\ \bibnamefont {Sliwa}}, \bibinfo {author}
  {\bibfnamefont {A.}~\bibnamefont {Narla}}, \bibinfo {author} {\bibfnamefont
  {S.}~\bibnamefont {Shankar}}, \bibinfo {author} {\bibfnamefont {M.~J.}\
  \bibnamefont {Hatridge}},  \emph {et~al.},\ }\bibfield  {title} {\enquote
  {\bibinfo {title} {Confining the state of light to a quantum manifold by
  engineered two-photon loss},}\ }\href {\doibase doi:10.1126/science.aaa2085}
  {\bibfield  {journal} {\bibinfo  {journal} {Science}\ }\textbf {\bibinfo
  {volume} {347}},\ \bibinfo {pages} {853} (\bibinfo {year}
  {2015})}\BibitemShut {NoStop}%
\bibitem [{\citenamefont {Touzard}\ \emph {et~al.}(2018)\citenamefont
  {Touzard}, \citenamefont {Grimm}, \citenamefont {Leghtas}, \citenamefont
  {Mundhada}, \citenamefont {Reinhold}, \citenamefont {Axline}, \citenamefont
  {Reagor}, \citenamefont {Chou}, \citenamefont {Blumoff}, \citenamefont
  {Sliwa} \emph {et~al.}}]{Touzard2018Coherent}%
  \BibitemOpen
  \bibfield  {author} {\bibinfo {author} {\bibfnamefont {S.}~\bibnamefont
  {Touzard}}, \bibinfo {author} {\bibfnamefont {A.}~\bibnamefont {Grimm}},
  \bibinfo {author} {\bibfnamefont {Z.}~\bibnamefont {Leghtas}}, \bibinfo
  {author} {\bibfnamefont {S.~O.}\ \bibnamefont {Mundhada}}, \bibinfo {author}
  {\bibfnamefont {P.}~\bibnamefont {Reinhold}}, \bibinfo {author}
  {\bibfnamefont {C.}~\bibnamefont {Axline}}, \bibinfo {author} {\bibfnamefont
  {M.}~\bibnamefont {Reagor}}, \bibinfo {author} {\bibfnamefont
  {K.}~\bibnamefont {Chou}}, \bibinfo {author} {\bibfnamefont {J.}~\bibnamefont
  {Blumoff}}, \bibinfo {author} {\bibfnamefont {K.~M.}\ \bibnamefont {Sliwa}},
  \emph {et~al.},\ }\bibfield  {title} {\enquote {\bibinfo {title} {Coherent
  oscillations inside a quantum manifold stabilized by dissipation},}\ }\href
  {\doibase 10.1103/PhysRevX.8.021005} {\bibfield  {journal} {\bibinfo
  {journal} {Phys. Rev. X}\ }\textbf {\bibinfo {volume} {8}},\ \bibinfo {pages}
  {021005} (\bibinfo {year} {2018})}\BibitemShut {NoStop}%
\bibitem [{\citenamefont {Wang}\ \emph
  {et~al.}(2019{\natexlab{b}})\citenamefont {Wang}, \citenamefont {Pechal},
  \citenamefont {Wollack}, \citenamefont {Arrangoiz-Arriola}, \citenamefont
  {Gao}, \citenamefont {Lee},\ and\ \citenamefont
  {Safavi-Naeini}}]{Wang2019Quantum}%
  \BibitemOpen
  \bibfield  {author} {\bibinfo {author} {\bibfnamefont {Z.}~\bibnamefont
  {Wang}}, \bibinfo {author} {\bibfnamefont {M.}~\bibnamefont {Pechal}},
  \bibinfo {author} {\bibfnamefont {E.~A.}\ \bibnamefont {Wollack}}, \bibinfo
  {author} {\bibfnamefont {P.}~\bibnamefont {Arrangoiz-Arriola}}, \bibinfo
  {author} {\bibfnamefont {M.}~\bibnamefont {Gao}}, \bibinfo {author}
  {\bibfnamefont {N.~R.}\ \bibnamefont {Lee}}, \ and\ \bibinfo {author}
  {\bibfnamefont {A.~H.}\ \bibnamefont {Safavi-Naeini}},\ }\bibfield  {title}
  {\enquote {\bibinfo {title} {Quantum dynamics of a few-photon parametric
  oscillator},}\ }\href {\doibase 10.1103/PhysRevX.9.021049} {\bibfield
  {journal} {\bibinfo  {journal} {Phys. Rev. X}\ }\textbf {\bibinfo {volume}
  {9}},\ \bibinfo {pages} {021049} (\bibinfo {year}
  {2019}{\natexlab{b}})}\BibitemShut {NoStop}%
\bibitem [{\citenamefont {Andersen}\ \emph {et~al.}(2020)\citenamefont
  {Andersen}, \citenamefont {Kamal}, \citenamefont {Masluk}, \citenamefont
  {Pop}, \citenamefont {Blais},\ and\ \citenamefont
  {Devoret}}]{Andersen2020Quantum}%
  \BibitemOpen
  \bibfield  {author} {\bibinfo {author} {\bibfnamefont {C.~K.}\ \bibnamefont
  {Andersen}}, \bibinfo {author} {\bibfnamefont {A.}~\bibnamefont {Kamal}},
  \bibinfo {author} {\bibfnamefont {N.~A.}\ \bibnamefont {Masluk}}, \bibinfo
  {author} {\bibfnamefont {I.~M.}\ \bibnamefont {Pop}}, \bibinfo {author}
  {\bibfnamefont {A.}~\bibnamefont {Blais}}, \ and\ \bibinfo {author}
  {\bibfnamefont {M.~H.}\ \bibnamefont {Devoret}},\ }\bibfield  {title}
  {\enquote {\bibinfo {title} {Quantum versus classical switching dynamics of
  driven dissipative kerr resonators},}\ }\href {\doibase
  10.1103/PhysRevApplied.13.044017} {\bibfield  {journal} {\bibinfo  {journal}
  {Phys. Rev. Appl.}\ }\textbf {\bibinfo {volume} {13}},\ \bibinfo {pages}
  {044017} (\bibinfo {year} {2020})}\BibitemShut {NoStop}%
\bibitem [{\citenamefont {Yamaji}\ \emph {et~al.}(2022)\citenamefont {Yamaji},
  \citenamefont {Kagami}, \citenamefont {Yamaguchi}, \citenamefont {Satoh},
  \citenamefont {Koshino}, \citenamefont {Goto}, \citenamefont {Lin},
  \citenamefont {Nakamura},\ and\ \citenamefont
  {Yamamoto}}]{Yamaji2022Spectroscopic}%
  \BibitemOpen
  \bibfield  {author} {\bibinfo {author} {\bibfnamefont {T.}~\bibnamefont
  {Yamaji}}, \bibinfo {author} {\bibfnamefont {S.}~\bibnamefont {Kagami}},
  \bibinfo {author} {\bibfnamefont {A.}~\bibnamefont {Yamaguchi}}, \bibinfo
  {author} {\bibfnamefont {T.}~\bibnamefont {Satoh}}, \bibinfo {author}
  {\bibfnamefont {K.}~\bibnamefont {Koshino}}, \bibinfo {author} {\bibfnamefont
  {H.}~\bibnamefont {Goto}}, \bibinfo {author} {\bibfnamefont {Z.~R.}\
  \bibnamefont {Lin}}, \bibinfo {author} {\bibfnamefont {Y.}~\bibnamefont
  {Nakamura}}, \ and\ \bibinfo {author} {\bibfnamefont {T.}~\bibnamefont
  {Yamamoto}},\ }\bibfield  {title} {\enquote {\bibinfo {title} {Spectroscopic
  observation of the crossover from a classical duffing oscillator to a kerr
  parametric oscillator},}\ }\href {\doibase 10.1103/PhysRevA.105.023519}
  {\bibfield  {journal} {\bibinfo  {journal} {Phys. Rev. A}\ }\textbf {\bibinfo
  {volume} {105}},\ \bibinfo {pages} {023519} (\bibinfo {year}
  {2022})}\BibitemShut {NoStop}%
\bibitem [{\citenamefont {Yamaji}\ \emph {et~al.}(2023)\citenamefont {Yamaji},
  \citenamefont {Masuda}, \citenamefont {Yamaguchi}, \citenamefont {Satoh},
  \citenamefont {Morioka}, \citenamefont {Igarashi}, \citenamefont {Shirane},\
  and\ \citenamefont {Yamamoto}}]{Yamaji2023Correlated}%
  \BibitemOpen
  \bibfield  {author} {\bibinfo {author} {\bibfnamefont {T.}~\bibnamefont
  {Yamaji}}, \bibinfo {author} {\bibfnamefont {S.}~\bibnamefont {Masuda}},
  \bibinfo {author} {\bibfnamefont {A.}~\bibnamefont {Yamaguchi}}, \bibinfo
  {author} {\bibfnamefont {T.}~\bibnamefont {Satoh}}, \bibinfo {author}
  {\bibfnamefont {A.}~\bibnamefont {Morioka}}, \bibinfo {author} {\bibfnamefont
  {Y.}~\bibnamefont {Igarashi}}, \bibinfo {author} {\bibfnamefont
  {M.}~\bibnamefont {Shirane}}, \ and\ \bibinfo {author} {\bibfnamefont
  {T.}~\bibnamefont {Yamamoto}},\ }\bibfield  {title} {\enquote {\bibinfo
  {title} {Correlated oscillations in kerr parametric oscillators with tunable
  effective coupling},}\ }\href {\doibase 10.1103/PhysRevApplied.20.014057}
  {\bibfield  {journal} {\bibinfo  {journal} {Phys. Rev. Appl.}\ }\textbf
  {\bibinfo {volume} {20}},\ \bibinfo {pages} {014057} (\bibinfo {year}
  {2023})}\BibitemShut {NoStop}%
\bibitem [{\citenamefont {Iyama}\ \emph {et~al.}(2024)\citenamefont {Iyama},
  \citenamefont {Kamiya}, \citenamefont {Fujii}, \citenamefont {Mukai},
  \citenamefont {Zhou}, \citenamefont {Nagase}, \citenamefont {Tomonaga},
  \citenamefont {Wang}, \citenamefont {Xue}, \citenamefont {Watabe} \emph
  {et~al.}}]{Iyama2024Observation}%
  \BibitemOpen
  \bibfield  {author} {\bibinfo {author} {\bibfnamefont {D.}~\bibnamefont
  {Iyama}}, \bibinfo {author} {\bibfnamefont {T.}~\bibnamefont {Kamiya}},
  \bibinfo {author} {\bibfnamefont {S.}~\bibnamefont {Fujii}}, \bibinfo
  {author} {\bibfnamefont {H.}~\bibnamefont {Mukai}}, \bibinfo {author}
  {\bibfnamefont {Y.}~\bibnamefont {Zhou}}, \bibinfo {author} {\bibfnamefont
  {T.}~\bibnamefont {Nagase}}, \bibinfo {author} {\bibfnamefont
  {A.}~\bibnamefont {Tomonaga}}, \bibinfo {author} {\bibfnamefont
  {R.}~\bibnamefont {Wang}}, \bibinfo {author} {\bibfnamefont {J.-J.}\
  \bibnamefont {Xue}}, \bibinfo {author} {\bibfnamefont {S.}~\bibnamefont
  {Watabe}},  \emph {et~al.},\ }\bibfield  {title} {\enquote {\bibinfo {title}
  {Observation and manipulation of quantum interference in a superconducting
  kerr parametric oscillator},}\ }\href {\doibase 10.1038/s41467-023-44496-1}
  {\bibfield  {journal} {\bibinfo  {journal} {Nat. Commun.}\ }\textbf {\bibinfo
  {volume} {15}},\ \bibinfo {pages} {86} (\bibinfo {year} {2024})}\BibitemShut
  {NoStop}%
\bibitem [{\citenamefont {Hoshi}\ \emph {et~al.}(2024)\citenamefont {Hoshi},
  \citenamefont {Nagase}, \citenamefont {Kwon}, \citenamefont {Iyama},
  \citenamefont {Kamiya}, \citenamefont {Fujii}, \citenamefont {Mukai},
  \citenamefont {Ahmed}, \citenamefont {Kockum},\ and\ \citenamefont
  {Watabe}}]{Hoshi2024EntanglingCat}%
  \BibitemOpen
  \bibfield  {author} {\bibinfo {author} {\bibfnamefont {D.}~\bibnamefont
  {Hoshi}}, \bibinfo {author} {\bibfnamefont {T.}~\bibnamefont {Nagase}},
  \bibinfo {author} {\bibfnamefont {S.}~\bibnamefont {Kwon}}, \bibinfo {author}
  {\bibfnamefont {D.}~\bibnamefont {Iyama}}, \bibinfo {author} {\bibfnamefont
  {T.}~\bibnamefont {Kamiya}}, \bibinfo {author} {\bibfnamefont
  {S.}~\bibnamefont {Fujii}}, \bibinfo {author} {\bibfnamefont
  {H.}~\bibnamefont {Mukai}}, \bibinfo {author} {\bibfnamefont
  {S.}~\bibnamefont {Ahmed}}, \bibinfo {author} {\bibfnamefont {A.~F.}\
  \bibnamefont {Kockum}}, \ and\ \bibinfo {author} {\bibfnamefont
  {S.}~\bibnamefont {Watabe}},\ }\bibfield  {title} {\enquote {\bibinfo {title}
  {Entangling schr\"odinger's cat states by seeding a bell state or swapping
  the cats},}\ }\href@noop {} {\bibfield  {journal} {\bibinfo  {journal} {arXiv
  preprint arXiv:2406.17999}\ } (\bibinfo {year} {2024})}\BibitemShut {NoStop}%
\bibitem [{\citenamefont {Frattini}\ \emph {et~al.}(2017)\citenamefont
  {Frattini}, \citenamefont {Vool}, \citenamefont {Shankar}, \citenamefont
  {Narla}, \citenamefont {Sliwa},\ and\ \citenamefont
  {Devoret}}]{Frattini2017SNAIL}%
  \BibitemOpen
  \bibfield  {author} {\bibinfo {author} {\bibfnamefont {N.~E.}\ \bibnamefont
  {Frattini}}, \bibinfo {author} {\bibfnamefont {U.}~\bibnamefont {Vool}},
  \bibinfo {author} {\bibfnamefont {S.}~\bibnamefont {Shankar}}, \bibinfo
  {author} {\bibfnamefont {A.}~\bibnamefont {Narla}}, \bibinfo {author}
  {\bibfnamefont {K.~M.}\ \bibnamefont {Sliwa}}, \ and\ \bibinfo {author}
  {\bibfnamefont {M.~H.}\ \bibnamefont {Devoret}},\ }\bibfield  {title}
  {\enquote {\bibinfo {title} {3-wave mixing josephson dipole element},}\
  }\href {\doibase 10.1063/1.4984142} {\bibfield  {journal} {\bibinfo
  {journal} {Appl. Phys. Lett.}\ }\textbf {\bibinfo {volume} {110}},\ \bibinfo
  {pages} {222603} (\bibinfo {year} {2017})}\BibitemShut {NoStop}%
\bibitem [{\citenamefont {Frattini}\ \emph {et~al.}(2024)\citenamefont
  {Frattini}, \citenamefont {Corti\~{n}as}, \citenamefont {Venkatraman},
  \citenamefont {Xiao}, \citenamefont {Su}, \citenamefont {Lei}, \citenamefont
  {Chapman}, \citenamefont {Joshi}, \citenamefont {Girvin}, \citenamefont
  {Schoelkopf} \emph {et~al.}}]{Frattini2024Observation}%
  \BibitemOpen
  \bibfield  {author} {\bibinfo {author} {\bibfnamefont {N.~E.}\ \bibnamefont
  {Frattini}}, \bibinfo {author} {\bibfnamefont {R.~G.}\ \bibnamefont
  {Corti\~{n}as}}, \bibinfo {author} {\bibfnamefont {J.}~\bibnamefont
  {Venkatraman}}, \bibinfo {author} {\bibfnamefont {X.}~\bibnamefont {Xiao}},
  \bibinfo {author} {\bibfnamefont {Q.}~\bibnamefont {Su}}, \bibinfo {author}
  {\bibfnamefont {C.~U.}\ \bibnamefont {Lei}}, \bibinfo {author} {\bibfnamefont
  {B.~J.}\ \bibnamefont {Chapman}}, \bibinfo {author} {\bibfnamefont {V.~R.}\
  \bibnamefont {Joshi}}, \bibinfo {author} {\bibfnamefont {S.~M.}\ \bibnamefont
  {Girvin}}, \bibinfo {author} {\bibfnamefont {R.~J.}\ \bibnamefont
  {Schoelkopf}},  \emph {et~al.},\ }\bibfield  {title} {\enquote {\bibinfo
  {title} {Observation of pairwise level degeneracies and the quantum regime of
  the arrhenius law in a double-well parametric oscillator},}\ }\href {\doibase
  10.1103/PhysRevX.14.031040} {\bibfield  {journal} {\bibinfo  {journal} {Phys.
  Rev. X}\ }\textbf {\bibinfo {volume} {14}},\ \bibinfo {pages} {031040}
  (\bibinfo {year} {2024})}\BibitemShut {NoStop}%
\bibitem [{\citenamefont {Hajr}\ \emph {et~al.}(2024)\citenamefont {Hajr},
  \citenamefont {Qing}, \citenamefont {Wang}, \citenamefont {Koolstra},
  \citenamefont {Pedramrazi}, \citenamefont {Kang}, \citenamefont {Chen},
  \citenamefont {Nguyen}, \citenamefont {Junger}, \citenamefont {Goss} \emph
  {et~al.}}]{Hajr2024TaKerrCat}%
  \BibitemOpen
  \bibfield  {author} {\bibinfo {author} {\bibfnamefont {A.}~\bibnamefont
  {Hajr}}, \bibinfo {author} {\bibfnamefont {B.}~\bibnamefont {Qing}}, \bibinfo
  {author} {\bibfnamefont {K.}~\bibnamefont {Wang}}, \bibinfo {author}
  {\bibfnamefont {G.}~\bibnamefont {Koolstra}}, \bibinfo {author}
  {\bibfnamefont {Z.}~\bibnamefont {Pedramrazi}}, \bibinfo {author}
  {\bibfnamefont {Z.}~\bibnamefont {Kang}}, \bibinfo {author} {\bibfnamefont
  {L.}~\bibnamefont {Chen}}, \bibinfo {author} {\bibfnamefont {L.~B.}\
  \bibnamefont {Nguyen}}, \bibinfo {author} {\bibfnamefont {C.}~\bibnamefont
  {Junger}}, \bibinfo {author} {\bibfnamefont {N.}~\bibnamefont {Goss}},  \emph
  {et~al.},\ }\bibfield  {title} {\enquote {\bibinfo {title} {High-coherence
  kerr-cat qubit in 2d architecture},}\ }\href {\doibase
  10.48550/arXiv.2404.16697} {\ ,\ \bibinfo {pages} {arXiv:2404.16697}
  (\bibinfo {year} {2024})}\BibitemShut {NoStop}%
\bibitem [{\citenamefont {Venkatraman}\ \emph {et~al.}(2024)\citenamefont
  {Venkatraman}, \citenamefont {Corti{\~n}as}, \citenamefont {Frattini},
  \citenamefont {Xiao},\ and\ \citenamefont
  {Devoret}}]{Venkatraman2024DetunedKerrCat}%
  \BibitemOpen
  \bibfield  {author} {\bibinfo {author} {\bibfnamefont {J.}~\bibnamefont
  {Venkatraman}}, \bibinfo {author} {\bibfnamefont {R.~G.}\ \bibnamefont
  {Corti{\~n}as}}, \bibinfo {author} {\bibfnamefont {N.~E.}\ \bibnamefont
  {Frattini}}, \bibinfo {author} {\bibfnamefont {X.}~\bibnamefont {Xiao}}, \
  and\ \bibinfo {author} {\bibfnamefont {M.~H.}\ \bibnamefont {Devoret}},\
  }\bibfield  {title} {\enquote {\bibinfo {title} {A driven kerr oscillator
  with two-fold degeneracies for qubit protection},}\ }\href {\doibase
  doi:10.1073/pnas.2311241121} {\bibfield  {journal} {\bibinfo  {journal}
  {Proceedings of the National Academy of Sciences}\ }\textbf {\bibinfo
  {volume} {121}},\ \bibinfo {pages} {e2311241121} (\bibinfo {year}
  {2024})}\BibitemShut {NoStop}%
\bibitem [{\citenamefont {Deppe}\ \emph {et~al.}(2004)\citenamefont {Deppe},
  \citenamefont {Saito}, \citenamefont {Tanaka},\ and\ \citenamefont
  {Takayanagi}}]{Deppe2004JJCapacitances}%
  \BibitemOpen
  \bibfield  {author} {\bibinfo {author} {\bibfnamefont {F.}~\bibnamefont
  {Deppe}}, \bibinfo {author} {\bibfnamefont {S.}~\bibnamefont {Saito}},
  \bibinfo {author} {\bibfnamefont {H.}~\bibnamefont {Tanaka}}, \ and\ \bibinfo
  {author} {\bibfnamefont {H.}~\bibnamefont {Takayanagi}},\ }\bibfield  {title}
  {\enquote {\bibinfo {title} {Determination of the capacitance of nm scale
  josephson junctions},}\ }\href {\doibase 10.1063/1.1645673} {\bibfield
  {journal} {\bibinfo  {journal} {J. Appl. Phys.}\ }\textbf {\bibinfo {volume}
  {95}},\ \bibinfo {pages} {2607} (\bibinfo {year} {2004})}\BibitemShut
  {NoStop}%
\bibitem [{\citenamefont {Hillmann}\ and\ \citenamefont
  {Quijandr{\'\i}a}(2022)}]{Hillmann2022Designing}%
  \BibitemOpen
  \bibfield  {author} {\bibinfo {author} {\bibfnamefont {T.}~\bibnamefont
  {Hillmann}}\ and\ \bibinfo {author} {\bibfnamefont {F.}~\bibnamefont
  {Quijandr{\'\i}a}},\ }\bibfield  {title} {\enquote {\bibinfo {title}
  {Designing kerr interactions for quantum information processing via
  counterrotating terms of asymmetric josephson-junction loops},}\ }\href
  {\doibase 10.1103/PhysRevApplied.17.064018} {\bibfield  {journal} {\bibinfo
  {journal} {Phys. Rev. Appl.}\ }\textbf {\bibinfo {volume} {17}},\ \bibinfo
  {pages} {064018} (\bibinfo {year} {2022})}\BibitemShut {NoStop}%
\bibitem [{\citenamefont {Blais}\ \emph {et~al.}(2021)\citenamefont {Blais},
  \citenamefont {Grimsmo}, \citenamefont {Girvin},\ and\ \citenamefont
  {Wallraff}}]{Blais2021Circuit}%
  \BibitemOpen
  \bibfield  {author} {\bibinfo {author} {\bibfnamefont {A.}~\bibnamefont
  {Blais}}, \bibinfo {author} {\bibfnamefont {A.~L.}\ \bibnamefont {Grimsmo}},
  \bibinfo {author} {\bibfnamefont {S.~M.}\ \bibnamefont {Girvin}}, \ and\
  \bibinfo {author} {\bibfnamefont {A.}~\bibnamefont {Wallraff}},\ }\bibfield
  {title} {\enquote {\bibinfo {title} {Circuit quantum electrodynamics},}\
  }\href {\doibase 10.1103/RevModPhys.93.025005} {\bibfield  {journal}
  {\bibinfo  {journal} {Rev. Mod. Phys.}\ }\textbf {\bibinfo {volume} {93}},\
  \bibinfo {pages} {025005} (\bibinfo {year} {2021})}\BibitemShut {NoStop}%
\bibitem [{\citenamefont {Huang}\ \emph {et~al.}(2023)\citenamefont {Huang},
  \citenamefont {Li}, \citenamefont {Chen}, \citenamefont {Zhang},
  \citenamefont {Zou}, \citenamefont {Guo},\ and\ \citenamefont
  {Zou}}]{Huang2023Residual}%
  \BibitemOpen
  \bibfield  {author} {\bibinfo {author} {\bibfnamefont {Y.-X.}\ \bibnamefont
  {Huang}}, \bibinfo {author} {\bibfnamefont {M.}~\bibnamefont {Li}}, \bibinfo
  {author} {\bibfnamefont {Z.-J.}\ \bibnamefont {Chen}}, \bibinfo {author}
  {\bibfnamefont {Y.-L.}\ \bibnamefont {Zhang}}, \bibinfo {author}
  {\bibfnamefont {X.-B.}\ \bibnamefont {Zou}}, \bibinfo {author} {\bibfnamefont
  {G.-C.}\ \bibnamefont {Guo}}, \ and\ \bibinfo {author} {\bibfnamefont
  {C.-L.}\ \bibnamefont {Zou}},\ }\bibfield  {title} {\enquote {\bibinfo
  {title} {Residual quantum effects beyond mean-field treatment in quantum
  optics systems},}\ }\href {\doibase https://doi.org/10.1002/lpor.202200599}
  {\bibfield  {journal} {\bibinfo  {journal} {Laser \& Photonics Reviews}\
  }\textbf {\bibinfo {volume} {17}},\ \bibinfo {pages} {2200599} (\bibinfo
  {year} {2023})}\BibitemShut {NoStop}%
\bibitem [{\citenamefont {Zhang}\ \emph {et~al.}(2019)\citenamefont {Zhang},
  \citenamefont {Lester}, \citenamefont {Gao}, \citenamefont {Jiang},
  \citenamefont {Schoelkopf},\ and\ \citenamefont
  {Girvin}}]{Yaxing2019Engineering}%
  \BibitemOpen
  \bibfield  {author} {\bibinfo {author} {\bibfnamefont {Y.}~\bibnamefont
  {Zhang}}, \bibinfo {author} {\bibfnamefont {B.~J.}\ \bibnamefont {Lester}},
  \bibinfo {author} {\bibfnamefont {Y.~Y.}\ \bibnamefont {Gao}}, \bibinfo
  {author} {\bibfnamefont {L.}~\bibnamefont {Jiang}}, \bibinfo {author}
  {\bibfnamefont {R.~J.}\ \bibnamefont {Schoelkopf}}, \ and\ \bibinfo {author}
  {\bibfnamefont {S.~M.}\ \bibnamefont {Girvin}},\ }\bibfield  {title}
  {\enquote {\bibinfo {title} {Engineering bilinear mode coupling in circuit
  qed: Theory and experiment},}\ }\href {\doibase 10.1103/PhysRevA.99.012314}
  {\bibfield  {journal} {\bibinfo  {journal} {Phys. Rev. A}\ }\textbf {\bibinfo
  {volume} {99}},\ \bibinfo {pages} {012314} (\bibinfo {year}
  {2019})}\BibitemShut {NoStop}%
\bibitem [{\citenamefont {Miano}\ \emph {et~al.}(2023)\citenamefont {Miano},
  \citenamefont {Joshi}, \citenamefont {Liu}, \citenamefont {Dai},
  \citenamefont {Parakh}, \citenamefont {Frunzio},\ and\ \citenamefont
  {Devoret}}]{Miano2023Hamiltonian}%
  \BibitemOpen
  \bibfield  {author} {\bibinfo {author} {\bibfnamefont {A.}~\bibnamefont
  {Miano}}, \bibinfo {author} {\bibfnamefont {V.~R.}\ \bibnamefont {Joshi}},
  \bibinfo {author} {\bibfnamefont {G.}~\bibnamefont {Liu}}, \bibinfo {author}
  {\bibfnamefont {W.}~\bibnamefont {Dai}}, \bibinfo {author} {\bibfnamefont
  {P.~D.}\ \bibnamefont {Parakh}}, \bibinfo {author} {\bibfnamefont
  {L.}~\bibnamefont {Frunzio}}, \ and\ \bibinfo {author} {\bibfnamefont
  {M.~H.}\ \bibnamefont {Devoret}},\ }\bibfield  {title} {\enquote {\bibinfo
  {title} {Hamiltonian extrema of an arbitrary flux-biased josephson
  circuit},}\ }\href {\doibase 10.1103/PRXQuantum.4.030324} {\bibfield
  {journal} {\bibinfo  {journal} {PRX Quantum}\ }\textbf {\bibinfo {volume}
  {4}},\ \bibinfo {pages} {030324} (\bibinfo {year} {2023})}\BibitemShut
  {NoStop}%
\bibitem [{\citenamefont {Minev}\ \emph {et~al.}(2021)\citenamefont {Minev},
  \citenamefont {Leghtas}, \citenamefont {Mundhada}, \citenamefont
  {Christakis}, \citenamefont {Pop},\ and\ \citenamefont
  {Devoret}}]{Minev2021pyEPR}%
  \BibitemOpen
  \bibfield  {author} {\bibinfo {author} {\bibfnamefont {Z.~K.}\ \bibnamefont
  {Minev}}, \bibinfo {author} {\bibfnamefont {Z.}~\bibnamefont {Leghtas}},
  \bibinfo {author} {\bibfnamefont {S.~O.}\ \bibnamefont {Mundhada}}, \bibinfo
  {author} {\bibfnamefont {L.}~\bibnamefont {Christakis}}, \bibinfo {author}
  {\bibfnamefont {I.~M.}\ \bibnamefont {Pop}}, \ and\ \bibinfo {author}
  {\bibfnamefont {M.~H.}\ \bibnamefont {Devoret}},\ }\bibfield  {title}
  {\enquote {\bibinfo {title} {Energy-participation quantization of josephson
  circuits},}\ }\href {\doibase 10.1038/s41534-021-00461-8} {\bibfield
  {journal} {\bibinfo  {journal} {npj Quantum Inf.}\ }\textbf {\bibinfo
  {volume} {7}},\ \bibinfo {pages} {131} (\bibinfo {year} {2021})}\BibitemShut
  {NoStop}%
\bibitem [{\citenamefont {Riwar}\ and\ \citenamefont
  {DiVincenzo}(2022)}]{Riwar2022Circuit}%
  \BibitemOpen
  \bibfield  {author} {\bibinfo {author} {\bibfnamefont {R.~P.}\ \bibnamefont
  {Riwar}}\ and\ \bibinfo {author} {\bibfnamefont {D.~P.}\ \bibnamefont
  {DiVincenzo}},\ }\bibfield  {title} {\enquote {\bibinfo {title} {Circuit
  quantization with time-dependent magnetic fields for realistic geometries},}\
  }\href {\doibase 10.1038/s41534-022-00539-x} {\bibfield  {journal} {\bibinfo
  {journal} {npj Quantum Inf.}\ }\textbf {\bibinfo {volume} {8}},\ \bibinfo
  {pages} {36} (\bibinfo {year} {2022})}\BibitemShut {NoStop}%
\bibitem [{\citenamefont {You}\ \emph {et~al.}(2019)\citenamefont {You},
  \citenamefont {Sauls},\ and\ \citenamefont {Koch}}]{You2019Circuit}%
  \BibitemOpen
  \bibfield  {author} {\bibinfo {author} {\bibfnamefont {X.}~\bibnamefont
  {You}}, \bibinfo {author} {\bibfnamefont {J.~A.}\ \bibnamefont {Sauls}}, \
  and\ \bibinfo {author} {\bibfnamefont {J.}~\bibnamefont {Koch}},\ }\bibfield
  {title} {\enquote {\bibinfo {title} {Circuit quantization in the presence of
  time-dependent external flux},}\ }\href {\doibase 10.1103/PhysRevB.99.174512}
  {\bibfield  {journal} {\bibinfo  {journal} {Phys. Rev. B.}\ }\textbf
  {\bibinfo {volume} {99}},\ \bibinfo {pages} {174512} (\bibinfo {year}
  {2019})}\BibitemShut {NoStop}%
\bibitem [{\citenamefont {Bryon}\ \emph {et~al.}(2023)\citenamefont {Bryon},
  \citenamefont {Weiss}, \citenamefont {You}, \citenamefont {Sussman},
  \citenamefont {Croot}, \citenamefont {Huang}, \citenamefont {Koch},\ and\
  \citenamefont {Houck}}]{Bryon2023FluxDriveFluxonium}%
  \BibitemOpen
  \bibfield  {author} {\bibinfo {author} {\bibfnamefont {J.}~\bibnamefont
  {Bryon}}, \bibinfo {author} {\bibfnamefont {D.~K.}\ \bibnamefont {Weiss}},
  \bibinfo {author} {\bibfnamefont {X.}~\bibnamefont {You}}, \bibinfo {author}
  {\bibfnamefont {S.}~\bibnamefont {Sussman}}, \bibinfo {author} {\bibfnamefont
  {X.}~\bibnamefont {Croot}}, \bibinfo {author} {\bibfnamefont
  {Z.}~\bibnamefont {Huang}}, \bibinfo {author} {\bibfnamefont
  {J.}~\bibnamefont {Koch}}, \ and\ \bibinfo {author} {\bibfnamefont {A.~A.}\
  \bibnamefont {Houck}},\ }\bibfield  {title} {\enquote {\bibinfo {title}
  {Time-dependent magnetic flux in devices for circuit quantum
  electrodynamics},}\ }\href {\doibase 10.1103/PhysRevApplied.19.034031}
  {\bibfield  {journal} {\bibinfo  {journal} {Phys. Rev. Appl.}\ }\textbf
  {\bibinfo {volume} {19}},\ \bibinfo {pages} {034031} (\bibinfo {year}
  {2023})}\BibitemShut {NoStop}%
\bibitem [{\citenamefont {Manucharyan}\ \emph {et~al.}(2009)\citenamefont
  {Manucharyan}, \citenamefont {Koch}, \citenamefont {Glazman},\ and\
  \citenamefont {Devoret}}]{Manucharyan2009Fluxonium}%
  \BibitemOpen
  \bibfield  {author} {\bibinfo {author} {\bibfnamefont {V.~E.}\ \bibnamefont
  {Manucharyan}}, \bibinfo {author} {\bibfnamefont {J.}~\bibnamefont {Koch}},
  \bibinfo {author} {\bibfnamefont {L.~I.}\ \bibnamefont {Glazman}}, \ and\
  \bibinfo {author} {\bibfnamefont {M.~H.}\ \bibnamefont {Devoret}},\
  }\bibfield  {title} {\enquote {\bibinfo {title} {Fluxonium: single
  cooper-pair circuit free of charge offsets},}\ }\href {\doibase
  doi:10.1126/science.1175552} {\bibfield  {journal} {\bibinfo  {journal}
  {Science}\ }\textbf {\bibinfo {volume} {326}},\ \bibinfo {pages} {113}
  (\bibinfo {year} {2009})}\BibitemShut {NoStop}%
\bibitem [{\citenamefont {Pop}\ \emph {et~al.}(2014)\citenamefont {Pop},
  \citenamefont {Geerlings}, \citenamefont {Catelani}, \citenamefont
  {Schoelkopf}, \citenamefont {Glazman},\ and\ \citenamefont
  {Devoret}}]{Pop2014Fluxonium}%
  \BibitemOpen
  \bibfield  {author} {\bibinfo {author} {\bibfnamefont {I.~M.}\ \bibnamefont
  {Pop}}, \bibinfo {author} {\bibfnamefont {K.}~\bibnamefont {Geerlings}},
  \bibinfo {author} {\bibfnamefont {G.}~\bibnamefont {Catelani}}, \bibinfo
  {author} {\bibfnamefont {R.~J.}\ \bibnamefont {Schoelkopf}}, \bibinfo
  {author} {\bibfnamefont {L.~I.}\ \bibnamefont {Glazman}}, \ and\ \bibinfo
  {author} {\bibfnamefont {M.~H.}\ \bibnamefont {Devoret}},\ }\bibfield
  {title} {\enquote {\bibinfo {title} {Coherent suppression of electromagnetic
  dissipation due to superconducting quasiparticles},}\ }\href {\doibase
  10.1038/nature13017} {\bibfield  {journal} {\bibinfo  {journal} {Nature}\
  }\textbf {\bibinfo {volume} {508}},\ \bibinfo {pages} {369} (\bibinfo {year}
  {2014})}\BibitemShut {NoStop}%
\bibitem [{\citenamefont {Matveev}\ \emph {et~al.}(2002)\citenamefont
  {Matveev}, \citenamefont {Larkin},\ and\ \citenamefont
  {Glazman}}]{Matveev2002Persistent}%
  \BibitemOpen
  \bibfield  {author} {\bibinfo {author} {\bibfnamefont {K.~A.}\ \bibnamefont
  {Matveev}}, \bibinfo {author} {\bibfnamefont {A.~I.}\ \bibnamefont {Larkin}},
  \ and\ \bibinfo {author} {\bibfnamefont {L.~I.}\ \bibnamefont {Glazman}},\
  }\bibfield  {title} {\enquote {\bibinfo {title} {Persistent current in
  superconducting nanorings},}\ }\href {\doibase 10.1103/PhysRevLett.89.096802}
  {\bibfield  {journal} {\bibinfo  {journal} {Phys. Rev. Lett.}\ }\textbf
  {\bibinfo {volume} {89}},\ \bibinfo {pages} {096802} (\bibinfo {year}
  {2002})}\BibitemShut {NoStop}%
\bibitem [{\citenamefont {Astafiev}\ \emph {et~al.}(2012)\citenamefont
  {Astafiev}, \citenamefont {Ioffe}, \citenamefont {Kafanov}, \citenamefont
  {Pashkin}, \citenamefont {Arutyunov}, \citenamefont {Shahar}, \citenamefont
  {Cohen},\ and\ \citenamefont {Tsai}}]{Astafiev2012Coherent}%
  \BibitemOpen
  \bibfield  {author} {\bibinfo {author} {\bibfnamefont {O.~V.}\ \bibnamefont
  {Astafiev}}, \bibinfo {author} {\bibfnamefont {L.~B.}\ \bibnamefont {Ioffe}},
  \bibinfo {author} {\bibfnamefont {S.}~\bibnamefont {Kafanov}}, \bibinfo
  {author} {\bibfnamefont {Y.~A.}\ \bibnamefont {Pashkin}}, \bibinfo {author}
  {\bibfnamefont {K.~Y.}\ \bibnamefont {Arutyunov}}, \bibinfo {author}
  {\bibfnamefont {D.}~\bibnamefont {Shahar}}, \bibinfo {author} {\bibfnamefont
  {O.}~\bibnamefont {Cohen}}, \ and\ \bibinfo {author} {\bibfnamefont {J.~S.}\
  \bibnamefont {Tsai}},\ }\bibfield  {title} {\enquote {\bibinfo {title}
  {Coherent quantum phase slip},}\ }\href {\doibase 10.1038/nature10930}
  {\bibfield  {journal} {\bibinfo  {journal} {Nature}\ }\textbf {\bibinfo
  {volume} {484}},\ \bibinfo {pages} {355} (\bibinfo {year}
  {2012})}\BibitemShut {NoStop}%
\bibitem [{\citenamefont {Manucharyan}\ \emph {et~al.}(2012)\citenamefont
  {Manucharyan}, \citenamefont {Masluk}, \citenamefont {Kamal}, \citenamefont
  {Koch}, \citenamefont {Glazman},\ and\ \citenamefont
  {Devoret}}]{Manucharyan2012Evidence}%
  \BibitemOpen
  \bibfield  {author} {\bibinfo {author} {\bibfnamefont {V.~E.}\ \bibnamefont
  {Manucharyan}}, \bibinfo {author} {\bibfnamefont {N.~A.}\ \bibnamefont
  {Masluk}}, \bibinfo {author} {\bibfnamefont {A.}~\bibnamefont {Kamal}},
  \bibinfo {author} {\bibfnamefont {J.}~\bibnamefont {Koch}}, \bibinfo {author}
  {\bibfnamefont {L.~I.}\ \bibnamefont {Glazman}}, \ and\ \bibinfo {author}
  {\bibfnamefont {M.~H.}\ \bibnamefont {Devoret}},\ }\bibfield  {title}
  {\enquote {\bibinfo {title} {Evidence for coherent quantum phase slips across
  a josephson junction array},}\ }\href {\doibase 10.1103/PhysRevB.85.024521}
  {\bibfield  {journal} {\bibinfo  {journal} {Phys. Rev. B.}\ }\textbf
  {\bibinfo {volume} {85}},\ \bibinfo {pages} {024521} (\bibinfo {year}
  {2012})}\BibitemShut {NoStop}%
\bibitem [{\citenamefont {Kwon}\ \emph {et~al.}(2022)\citenamefont {Kwon},
  \citenamefont {Watabe},\ and\ \citenamefont {Tsai}}]{Kwon2022Autonomous}%
  \BibitemOpen
  \bibfield  {author} {\bibinfo {author} {\bibfnamefont {S.}~\bibnamefont
  {Kwon}}, \bibinfo {author} {\bibfnamefont {S.}~\bibnamefont {Watabe}}, \ and\
  \bibinfo {author} {\bibfnamefont {J.-S.}\ \bibnamefont {Tsai}},\ }\bibfield
  {title} {\enquote {\bibinfo {title} {Autonomous quantum error correction in a
  four-photon kerr parametric oscillator},}\ }\href {\doibase
  10.1038/s41534-022-00553-z} {\bibfield  {journal} {\bibinfo  {journal} {npj
  Quantum Inf.}\ }\textbf {\bibinfo {volume} {8}},\ \bibinfo {pages} {40}
  (\bibinfo {year} {2022})}\BibitemShut {NoStop}%
\bibitem [{\citenamefont {Bhandari}\ \emph {et~al.}(2024)\citenamefont
  {Bhandari}, \citenamefont {Huang}, \citenamefont {Hajr}, \citenamefont
  {Yanik}, \citenamefont {Qing}, \citenamefont {Wang}, \citenamefont
  {Santiago}, \citenamefont {Dressel}, \citenamefont {Siddiqi},\ and\
  \citenamefont {Jordan}}]{Bhandari2024STS}%
  \BibitemOpen
  \bibfield  {author} {\bibinfo {author} {\bibfnamefont {B.}~\bibnamefont
  {Bhandari}}, \bibinfo {author} {\bibfnamefont {I.}~\bibnamefont {Huang}},
  \bibinfo {author} {\bibfnamefont {A.}~\bibnamefont {Hajr}}, \bibinfo {author}
  {\bibfnamefont {K.}~\bibnamefont {Yanik}}, \bibinfo {author} {\bibfnamefont
  {B.}~\bibnamefont {Qing}}, \bibinfo {author} {\bibfnamefont {K.}~\bibnamefont
  {Wang}}, \bibinfo {author} {\bibfnamefont {D.~I.}\ \bibnamefont {Santiago}},
  \bibinfo {author} {\bibfnamefont {J.}~\bibnamefont {Dressel}}, \bibinfo
  {author} {\bibfnamefont {I.}~\bibnamefont {Siddiqi}}, \ and\ \bibinfo
  {author} {\bibfnamefont {A.~N.}\ \bibnamefont {Jordan}},\ }\bibfield  {title}
  {\enquote {\bibinfo {title} {Symmetrically threaded squids as next generation
  kerr-cat qubits},}\ }\href {\doibase 10.48550/arXiv.2405.11375} {\ ,\
  \bibinfo {pages} {arXiv:2405.11375} (\bibinfo {year} {2024})}\BibitemShut
  {NoStop}%
\bibitem [{\citenamefont {Marquet}\ \emph {et~al.}(2024)\citenamefont
  {Marquet}, \citenamefont {Essig}, \citenamefont {Cohen}, \citenamefont
  {Cottet}, \citenamefont {Murani}, \citenamefont {Albertinale}, \citenamefont
  {Dupouy}, \citenamefont {Bienfait}, \citenamefont {Peronnin}, \citenamefont
  {Jezouin} \emph {et~al.}}]{Marquet2024Autoparametric}%
  \BibitemOpen
  \bibfield  {author} {\bibinfo {author} {\bibfnamefont {A.}~\bibnamefont
  {Marquet}}, \bibinfo {author} {\bibfnamefont {A.}~\bibnamefont {Essig}},
  \bibinfo {author} {\bibfnamefont {J.}~\bibnamefont {Cohen}}, \bibinfo
  {author} {\bibfnamefont {N.}~\bibnamefont {Cottet}}, \bibinfo {author}
  {\bibfnamefont {A.}~\bibnamefont {Murani}}, \bibinfo {author} {\bibfnamefont
  {E.}~\bibnamefont {Albertinale}}, \bibinfo {author} {\bibfnamefont
  {S.}~\bibnamefont {Dupouy}}, \bibinfo {author} {\bibfnamefont
  {A.}~\bibnamefont {Bienfait}}, \bibinfo {author} {\bibfnamefont
  {T.}~\bibnamefont {Peronnin}}, \bibinfo {author} {\bibfnamefont
  {S.}~\bibnamefont {Jezouin}},  \emph {et~al.},\ }\bibfield  {title} {\enquote
  {\bibinfo {title} {Autoparametric resonance extending the bit-flip time of a
  cat qubit up to 0.3 s},}\ }\href {\doibase 10.1103/PhysRevX.14.021019}
  {\bibfield  {journal} {\bibinfo  {journal} {Phys. Rev. X}\ }\textbf {\bibinfo
  {volume} {14}},\ \bibinfo {pages} {021019} (\bibinfo {year}
  {2024})}\BibitemShut {NoStop}%
\end{thebibliography}%

\end{document}